\documentclass[aps,12pt]{revtex4}
\usepackage[dvips]{graphicx,psfrag}
\usepackage{dcolumn}
\usepackage{bm}
\usepackage{color}

\usepackage{amsmath,amssymb,mathrsfs,braket}

\def\simle{\mathrel{\rlap{\raise 0.511ex \hbox{$<$}}{\lower 0.511ex \hbox{$\sim$}}}}
\def\simge{\mathrel{ \rlap{\raise 0.511ex \hbox{$>$}}{\lower 0.511ex \hbox{$\sim$}}}}
 
\def\q{{\boldsymbol q}}

\newcommand \beq{\begin{eqnarray}}
\newcommand \eeq{\end{eqnarray}}
\def\simle{\mathrel{\rlap{\raise 0.511ex \hbox{$<$}}{\lower 0.511ex 
\hbox{$\sim$}}}}
\def\simge{\mathrel{ \rlap{\raise 0.511ex 
\hbox{$>$}}{\lower 0.511ex \hbox{$\sim$}}}}
\newcommand{\del}{\partial}

\newcommand{\Slash}[1]{{\ooalign{\hfil/\hfil\crcr$#1$}}} 

\def\A{{\boldsymbol A}}
\def\F{{\boldsymbol F}}
\def\E{{\boldsymbol E}}
\def\B{{\boldsymbol B}}
\def\x{{\boldsymbol x}}
\def\k{{\boldsymbol k}}
\def\bnabla{{\boldsymbol \nabla}}

\def\r{{\boldsymbol r}}

\def\x{{\boldsymbol x}}

\def\k{{\boldsymbol k}}
\def\q{{\boldsymbol q}}
\def\p{{\boldsymbol p}}
\def\0{{\boldsymbol 0}}
\def\v{{\boldsymbol v}}

\begin{document}

\title{Quantum Fields at Finite Temperature\\
``from tera to nano Kelvin''}

\author{Jean-Paul Blaizot}
\email{jean-paul.blaizot@cea.fr}
\affiliation{Institut de Physique Th\'eorique, CEA-Saclay, 91191 Gif-sur-Yvette Cedex, France.}

\date{\today}

\begin{abstract}

These lectures introduce techniques that are used in the description of systems of particles and fields at high temperature (or density).  These methods   have a broad range of physical applications. We shall discuss two specific applications: one related to hot and dense matter composed of quarks and gluons, with  temperatures in the tera Kelvin range, the other related to Bose-Einstein condensation in ultra-cold gases, with temperatures in the nano Kelvin range. As we shall see, in both systems, long wavelength collective phenomena   lead to similar features, in spite of the huge difference in orders of magnitude of the respective energy scales. 

\end{abstract}

\maketitle



\section{Lecture 1}
\section*{Introduction}

Let me start with a few words of explanation about the title of these lectures, ``Quantum fields at finite temperature, from tera to nano Kelvin''.  {\it Tera} is a Greek word which means $10^{12}$, while {\it nano} means $10^{-9}$ (this latter word is presumably very familiar to you because of the ``{\it nano} technologies'').
The Kelvin is the unit of temperature. In dealing with systems with so vastly different orders of magnitude it is useful  to think in terms of the corresponding energy scales.
Energies are conveniently measured in electron-volt. To convert Kelvin into electron-volt, recall that 300 K is about $1/40$ eV or, if you wish, 1 eV is about 120000 K. So $10^{12}$ K translates approximately into 100 MeV. This is an energy scale typical of (high energy) nuclear processes. At the opposite end, the nano Kelvin corresponds to a subatomic energy scale. It is for instance realized in experiments with ultra cold atoms, where the phenomenon of Bose-Einstein condensation has been observed. The lectures will discuss theoretical techniques that are relevant to the calculation of the properties of matter in these two extreme energy regimes. The beauty of theoretical physics is that the same techniques are indeed capable to provide an adequate description of some important aspects of these vastly different systems. 

That quantum field theory appears  as an essential tool in the description of hot and dense matter, composed of quarks and gluons, is a priori natural: the dynamics of quarks and gluons is governed by  a quantum field theory, called Quantum Chromodynamics (QCD), of which I shall say more in today's lecture. What is perhaps more surprising is that field theory is also useful for understanding the behavior of cold atoms. 
Atoms are objects which we can study  in isolation, and their dynamics obey non relativistic many-body quantum mechanics. However,  collections of atoms can undergo collective, long wavelength oscillations.
By  long wavelength, I mean a wavelength much larger than the typical distance between the atoms. Such  long wavelength oscillations necessarily involve collectively many  atoms. And these collective excitations can be described by (classical) field theory. These long wavelength phenomena provide  the connection between the two topics that I plan to discuss, hot and dense matter composed of quarks and gluons, and the Bose-Einstein condensation of weakly interacting cold atoms

A central theme of our discussion will be that,  in both systems, the effect of the interaction can be large, 
although  the strength of the interaction between the elementary constituents is small. At first sight, this looks like a paradox. But the clue to resolve this apparent puzzle has just been mentioned: collective, long wavelength phenomena involve many degrees of freedom, and the cooperation of these degrees of freedom compensates for the weakness of the coupling. Technically, this feature shows up in infrared divergences in the Feynman diagrams of perturbation theory. Because of these divergences, perturbation theory in fact breaks down, and other techniques have to be developed to perform calculations in the weak coupling regime. As we shall see, in both systems that we shall consider, a simple effective theory will allow us to overcome the difficulties met in perturbation theory (albeit only partially in the case of QCD).

After this brief and general introduction, and the explanation of the title, let me say a few words about the plan of the lectures.  
There will be six lectures. The first four lectures will be mostly devoted to the physics of  hot and dense matter, with in mind the
quark gluon plasma. That will give me the opportunity to introduce techniques of quantum field theory at finite temperature which can be used in many other contexts. In these lectures, I shall be mainly concerned with the calculations of thermodynamical quantities (like the pressure). I shall use the scalar field as a prototype of a quantum field theory in order to illustrate the main difficulties that one encounters in perturbative calculations at finite temperature. Many of these difficulties are common to other field theories, in particular QCD. In the latter case further complications arise, that I shall briefly indicate. I shall also introduce a simple effective field theory that allows us to handle the infrared divergences of perturbation theory. This effective theory can be extended to the case of QCD, but this will not be covered in the lectures. We shall rather find a direct application of this effective theory in the study of Bose-Einstein condensation, to which the last two lectures will be devoted.  Bose Einstein condensation is a phase transition
which takes place in the ideal  Bose gas, that is,  without interaction. The question that I want to address is how the interactions between the atoms modify this remarkable phenomenon. More specifically, I shall be interested by the shift in the transition temperature $T_c$ caused by a very small repulsion between the atoms. Naively, since the interaction between the atoms can be chosen as small as desired, you may think of using perturbation theory in order to calculate the shift in $T_c$. But we shall discover that perturbation theory is meaningless, it is plagued by infrared divergences. And we shall see that the effective theory introduced in the first part of these lectures can be used to obtain an elegant solution to this problem. 

Finite temperature aspects of many-body physics or quantum field theory are presented in a number of textbooks, for instance \cite{KB62,Abrikosov63,Fetter71,BR86,Kapusta,Kapusta:2006pm,LeBellac96}. Complements to the present lectures can be found in several lecture notes or review articles. Thus, the lectures 1-4 are based more specifically on \cite{Blaizot:2001su,Blaizot:2003tw,Blaizot:1997dy,Blaizot:2001yq,Blaizot:2001nr}. Lectures 5-6 are based on \cite{Blaizot:2008xx}. These papers should be consulted for systematic references to the original literature. Further references will be given in the text about specific results that will be used or referred to.

\subsection{A brief introduction to QCD and its symmetries}

Let me now begin the discussion of hot and dense matter. 
At sufficiently high temperature and/or density, one expects  nuclear matter --the matter that makes atomic nuclei-- to turn into a plasma of quarks and gluons, whose interactions are described by Quantum Chromodynamics (QCD). I shall then  remind you a few basic properties of QCD that are important to understand the bulk features of the quark-gluon plasma.

\subsubsection{Quantum Chromodynamics}

As I just said, QCD is the theory that governs the dynamics of quarks and gluons. Quarks are spin 1/2 fermions, that I shall  represent by a field $\Psi_f(x)$. The index $f$ refers to the so-called ``flavor''. There  exists six such flavors, denoted $u,d, s,c, t, b$, and the corresponding quarks have different masses: the masses of $u,d$ are very small,  of  order  2-4 MeV, that of the strange quark is in the hundred MeV range, that of 
the charm quark is of the order of a GeV, the same for the bottom, 4-5 GeV, while  the top quark is much heavier, $\sim$ 170 GeV. We shall be mostly concerned with matter that results from the ``melting'' of neutrons and protons, that is,  matter made of up and down quarks. In addition to flavor, quarks carry another internal quantum number, color. There are $N_c=3$ different colors for quarks.

Gluons are somewhat similar to photons. They are the modes of a vector gauge field $A_\mu$. As photons, they are massless bosons, with spin 1 (and two polarization states). In contrast to photons, which are electrically neutral, gluons carry color charge (gluons exist in $N_c^2-1=8$ colors), and interact directly among themselves.

The interaction between quarks and gluons is coded in the QCD lagrangian which takes the following form.   
\begin{equation}\label{lagrangianQCD}
\mathscr{L}=-\frac{1}{4}F^{\mu\nu}_a F^a_{\mu\nu}+\Psi_f(i\Slash{D}-m_f)\Psi_f,
\end{equation}
where $m_f$ is the mass of the quark with flavor $f$, $\Slash{D}\equiv\gamma^\mu D_{\mu}$, with $D_\mu$ the covariant derivative 
$D_\mu=\partial_{\mu}+igA_{\mu}$,
with $A_{\mu}$  the gauge field. This is a non-Abelian gauge field, i.e., a color matrix,  $A_{\mu}=A_{\mu}^a t^a$, where $t^a$ is a generator of the gauge group $SU(3)$ (in the fundamental representation). The field strength tensor reads
\begin{equation}\label{Fmunu}
F^{\mu\nu}_a=\partial^{\mu}A^{\nu}_a-\partial^{\nu}A^{\mu}_a+gf_{abc}A^{\mu}_bA^{\nu}_c.
\end{equation}
The first two terms are just those you would get in  Quantum Electrodynamics (QED). 
In a non-Abelian gauge theory you have the additional term, quadratic term in $A$, which is responsible for the interactions among the gluons ($f_{abc}$ are the 
structure constants of the gauge group).

\begin{figure}
	\centering
	\includegraphics[scale=0.75,clip]{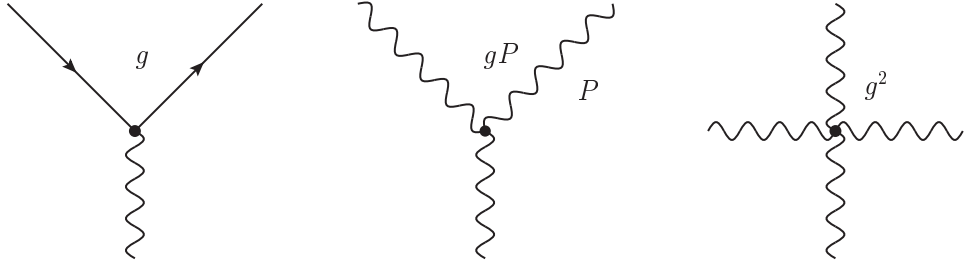} \\
	\caption{Feynman diagrams describing the elementary QCD interactions.}
	\label{Fig_Lec1:FeynmanDiagram}
\end{figure}

Now the main thing that I want to do here is to remind you how the interactions among quarks and gluons, and among gluons themselves, can be red off the QCD lagrangian.  Let's look first at the quarks. Their interactions with the gluon field is contained in the term $\bar{\Psi}\gamma^\mu A_\mu \Psi$, and I shall represent this interaction by the first diagram in Fig.~\ref{Fig_Lec1:FeynmanDiagram}. The strength of the coupling is $g$. Let us now turn to the gluons. If one would ignore the last piece of the field strength tensor (\ref{Fmunu}),  i.e., set $g=0$, 
then $(F_{\mu\nu})^2$ would be quadratic in the gauge potential. And  a lagrangian which is quadratic in the field
describes only normal modes, or free particles. These modes are what we call the gluons. When $g\ne 0$, the gluons interact, and the interaction vertices can be obtained by analyzing the $(F^{\mu\nu}_a)^2$, which, aside from the part quadratic in the gauge potential, contains also  the product of $\partial^\mu A_a^\nu$ and $gf_{abc}A^\mu_b A^\nu_c$, which generates a three gluon vertex, proportional to $g$  and to the derivative of the field (so that the strength of the three-gluon interaction is proportional to the momentum of one of the gluons). We have also a four gluon vertex  that is proportional to $g^2$. These vertices are displayed in Fig.~\ref{Fig_Lec1:FeynmanDiagram}.

\subsubsection{Symmetries of QCD}

The symmetries of QCD play an important role in the analysis of the phase diagram of hot and dense matter. I shall briefly review here these symmetries. 

There is of course the color symmetry, a local gauge symmetry. The QCD lagrangian is invariant under (local) color rotations of the quark field, accompanied by an appropriate transformation of the gauge potential:
\beq
A_\mu\to hA_\mu h^{-1}-\frac{i}{g}h\del_\mu h^{-1}\qquad \psi\to h\psi\qquad D_\mu\psi\to hD_\mu\Psi,
\eeq
with
\beq
h(x)={\rm e}^{i\theta^a(x)t^a}.
\eeq
This gauge symmetry is actually the guiding principle behind the whole construction of the QCD lagrangian (\ref{lagrangianQCD}). 

Color symmetry is responsible for  color confinement, 
 the fact that color charges cannot be isolated. The quarks  in  Nature combine to form color singlet states, the  hadrons. However, when matter is at extremely high temperature  confinement ``disappears'', 
during the so called confinement/deconfinement transition. In these lectures, we shall be interested in the thermodynamics of this deconfined  phase, called the quark-gluon plasma.

When the mass of the quark is strictly zero, there is another symmetry, called  chiral symmetry. Let me digress here on an issue that will appear at various occasions.  The mass of the quark is something which is not determined by QCD, but by physics at a higher energy scale, by the electro-weak physics of the standard model. So the mass of the quark is something
which is not ``negotiable''. However as a theorist, I can  play games and  consider a theory which is not quite QCD, but which differs from it only by the fact that  $m_f=0$. If I allow myself to do that, I observe that the lagrangian possesses another symmetry, chiral symmetry, which corresponds to the independent rotation of the left and right quarks. It is a global symmetry.

What I did for the quark masses can be done for all the other parameters which are around. I have told you that $N_c=3$, and indeed in  Nature
there are three colors of quarks and eight colors of gluons. That is an experimental fact. But it is sometimes instructive to consider  theories which look
like QCD, but in which $N_c$ can be varied. In particular you can sometimes obtain interesting insights by considering the limit where $N_c$ is infinite.
The theory corresponding to $N_c=\infty$ has many features similar to those of the theory with $N_c=3$.
But letting $N_c$ going to infinity allows you to do  calculations which you cannot do if $N_c$ is finite.
 We can also play with the number of flavors, etc.

Returning to chiral symmetry, we note that it is 
spontaneously broken in the vacuum. This feature has many  consequences, for  low energy nuclear physics in particular:  pions, the lightest hadrons, can be understood as (pesudo) Goldstone bosons.
Chiral symmetry is also important in characterizing the bulk properties of hot and dense matter, as it  is  restored by thermal fluctuations at high temperature. Thus, the quark condensate $\braket{\bar{\Psi}\Psi}$, an order parameter for chiral symmetry,  is non vanishing in the vacuum, but   vanishes at  high temperature.  

One last symmetry which I want to mention, again without going into  too much detail,   is   scale invariance.
 The QCD lagrangian (\ref{lagrangianQCD}) is left invariant in the rescaling of the coordinates $x\to\lambda x$, where $\lambda$ is an arbitrary number,  accompanied by a corresponding rescaling of the fields ($A_\mu\to \lambda^{-1} A_\mu$).  This symmetry (which holds for massless quarks) is easily verified by noticing that the gauge field $A_\mu$ has mass dimension 1, and that there are no dimensionful parameters  in the lagrangian. In particular the coupling constant $g$ is dimensionless. This is in fact a rather remarkable property because, as you know,  the QCD lagrangian  is supposed to allow you to calculate the  mass of the proton for instance. So how are you going to get the mass of the proton in GeV, if there is nothing in the lagrangian that ``knows'' about GeV ? 

At this point,  I need to remind you a few things about  renormalization and the running coupling constant. When we go beyond the classical level, and calculate physical processes, we need to take into account  the effects of
the short wavelength fluctuations and that usually leads to divergent quantities. 
In order to control these divergences,   you need to introduce some  cutoff, and this is where an energy scale enters.
In the case of QCD this energy scale is called $\Lambda_{QCD}$. It's value is of  the order of 250 MeV. The key point now is that the original QCD coupling constant  becomes  a ``running coupling constant'', i.e., it depends on the scale of the processes that one considers, or more precisely on the ratio of that scale to $\Lambda_{QCD}$, according to the (one-loop) formula 
\begin{equation}
g^2(\mu)=\frac{8\pi^2}{b_0\ln(\mu/\Lambda_{QCD})},\qquad b_0=\frac{11 N_c}{3}-\frac{2}{3}N_f.
\end{equation}
This formula has a remarkable  consequence,  called asymptotic freedom:
 when $\mu$ is much bigger than $\Lambda_{QCD}$, $g(\mu)$ goes to zero.
In the high temperature quark-gluon plasma,  the typical energy scale is of the order of the
temperature. Thus, when the temperature is large compare to  $\Lambda_{QCD}$  the coupling constant becomes  small. This is essentially the argument that leads one to expect that the quark-gluon plasma is a weakly interacting system at high temperature.

Now returning to the scale invariance, one notes that the the symmetry that exists at the level of the classical lagrangian is broken at the quantum level. This manifests itself in particular in a so-called  ``quantum anomaly'', that can be measured by the  trace of the energy momentum tensor. That trace, which  should normally vanish for a system of massless particles (reflecting the scale invariance), is given (at one-loop) by 
\begin{equation}
T^\mu_{\ \mu}=\frac{\beta(g)}{g^2}\mbox{Tr}(F^{\mu\nu}F_{\mu\nu}).
\end{equation}
The function $\beta(g)$ is called the beta function. It describes the variation of the running coupling with the scale $\mu$. It is given by 
\begin{equation}
\beta(g)=\mu\frac{dg}{d\mu}=-\frac{b_0}{16\pi^2} g^3.
\end{equation}
At finite temperature, (and after subtraction of the vacuum contribution), one can write the trace of the energy momentum tensor as $\epsilon-3P$, where $\epsilon $ is the energy density and $P$ the pressure . This quantity is non vanishing above the deconfinement transition, and (slowly) goes to zero with increasing temperature, in agreement with asymptotic freedom. 

I shall end here the discussion of the symmetries of QCD, and how these can help to characterize the bulk properties of dense matter. I shall return briefly to the phase diagram later today. At this point, the main message that I want to leave you with is that, because of asymptotic freedom, one expects matter at high temperature to be simple: a weakly interacting system of quarks and gluons. Since the interactions are weak it is natural to try and calculate their effects using perturbation theory. We shall learn in these lectures that the situation is in fact more subtle. But anyway, before calculating the effects of the interactions, it is important to recall some well-known properties of the non interacting system.

\subsection{Thermodynamics of relativistic particles}

Let me then remind you about the thermodynamics of free relativistic particles. This will also offer us the opportunity of a short reminder of basics of statistical mechanics that will be useful later at various points in the lectures. 

\subsubsection{ Some reminders}

 As you know, the statistical description of quantum systems involve the  so-called density operator ${\cal D}$, which, for systems in equilibrium at temperature $T=1/\beta$, is of the form
\begin{equation}
{\cal D}=\frac{1}{{\cal Z}}e^{-\beta(H-\mu Q)},
\end{equation}
where $H$ is the hamiltonian, $Q$ is the conserved charge which can be the electric charge, the baryon number, the strangeness, etc,
and $\mu$ is the associated chemical potential. The fact that $Q$ is a conserved charge means that $H$ commutes with Q, $[H,Q]=0$. It follows that the 
the eigenstates of the hamiltonian can be classified according to the eigenstates of $Q$. In other words one can write
\begin{equation}
H\ket{\psi_n}=E_n\ket{\psi_n},\quad Q\ket{\psi_n}=q_n\ket{\psi_n}.
\end{equation}
We can also rewrite the density operator as
\begin{equation}
{\cal D}=\sum_n\ket{\psi_n}p_n\bra{\psi_n},
\end{equation}
where $p_n$ is the probability to find the system in the particular eigenstate $\ket{\psi_n}$ of the hamiltonian. We have $\sum_np_n=1$.

The partition function 
\begin{equation}
{\cal Z}=\mbox{Tr}\ e^{-\beta(H-\mu Q)}
\end{equation}
is  the central object of most calculations, since most thermodynamical functions can be obtained from ${\cal Z}$. In particular, the thermodynamic potential reads
\begin{equation}
\Omega=-T\ln {\cal Z}.
\end{equation}
(I am using  natural units where $k_B=1$. In other words, I am measuring the temperature in unit of energy or mass.) The thermodynamic potential is also
\begin{equation}
\Omega=E-TS-\mu N,
\end{equation}
where 
\beq
E=\braket{H}=\frac{1}{{\cal Z}}\mbox{Tr}\left(He^{-\beta(H-\mu Q)}\right),
\eeq
and similarly $N$ is the expectation value of $Q$, $N=\braket{Q}$, which is calculated in the same way.
$S$ is the entropy.
\begin{equation}
S=-k_B\mbox{Tr}\ {\cal D}\ln {\cal D}=-k_B\sum_n p_n\ln p_n.
\end{equation}
It is a positive quantity because $p_n$ is a positive number smaller than 1.
You can verify that
\begin{equation}
E=-\left.\frac{\partial\ln {\cal Z}}{\partial\beta}\right|_{\beta\mu},\quad N=-\left.\frac{\partial\ln {\cal Z}}{\partial(\beta\mu)}\right|_{\beta}.
\end{equation}
The relation between the pressure and the thermodynamics potential is ($V$ is the volume)
\begin{equation}
P=-\left.\frac{\partial\Omega}{\partial V}\right|_{T,\mu}.
\end{equation}
$\Omega$ is the a function of temperature, chemical potential and volume. It  is proportional to the volume
\begin{equation}
\Omega(T,\mu,V)=V\omega(\mu,T), 
\end{equation}
from which one deduces that  $\Omega=-PV$.

\subsubsection{Free particles}

Now I would like to consider free particles. Let me write a typical free particle hamiltonian in second quantization 
\begin{equation}
H_0=\sum_\p\varepsilon_\p a^\dag_\p a_\p.
\end{equation}
Here $\p$ represents the quantum numbers for one particle. It is typically  the momentum, but I am not going to separate the momentum, the spin,
the color, etc, and just use $\p$ as a generic symbol for the set of all the quantum numbers that are needed to characterize entirely the state of a single particle. I am assuming that the hamiltonian is diagonal in this representation. And $\varepsilon_\p$ is
just the energy for a single particle in the state which is labeled by $\p$. The operators  $a^\dag_\p$ and $a_\p$ are creation and annihilation operator and these satisfy  commutation or
anticommutation relations, depending on whether the particles are bosons or fermions. That is, for bosons we have
\begin{equation}
a_\p a^\dag_\p - a^\dag_\p a_\p=1, 
\end{equation}
and for fermions
\begin{equation}
a_\p a^\dag_\p + a^\dag_\p a_\p=1.
\end{equation}

Now we want to calculate the partition function.
\begin{equation}
{\cal Z}_0=\mbox{Tr}\ e^{-\beta H_0}=\mbox{Tr}\left(e^{-\beta\sum_\p\varepsilon_\p a^\dag_\p a_\p}\right).
\end{equation}
I have to tell you a bit more about what is this trace here. We deal with free particles. The partition function involves a sum over states that have  arbitrary numbers of particles:
the state with zero particle (the vacuum), the states with one particle, with two particles, and so on.
How do you characterize the state with $n$ (identical) particles? You can look at the different states of one particle. Assume that the values of $\p$ are 
all discrete: $\p_0, \p_1, \p_2, \ldots, \p_k, \ldots $ is the list of all the possible states that a single particle can occupy.
To characterize a state with a large number of identical particles, it is enough to tell what is the number of particles which occupy each single particle state  $\p$. 
The operator $\hat{n}_\p\equiv a^\dag_\p a_\p$ is the operator which counts the number of particle in the state $\p$. Now, I can easily rewrite  ${\cal Z}_0$ in terms of this operator.
I get
\begin{equation}
{\cal Z}_0=\sum_{\{n_\p\}}\left(\prod_\p e^{-\beta\varepsilon_\p n_\p}\right)=\prod_\p \left (   \sum_{n_\p}e^{-\beta\varepsilon_\p n_\p}\right)
\end{equation}
where $\sum_{\{n_\p\} }$ is the sum over all ``configurations'', that is all sets of possible numbers $n_\p$ (eigenvalues of the operator $\hat{n}_\p$). 
I have to distinguish two cases, the fermions and the bosons. For fermions, you cannot put more than one particle in a given state,  so $n_\p=0, 1$.
The sum is then very easy:
\begin{equation}
{\cal Z}_0=\prod_\p\left(1+e^{-\beta\varepsilon_\p}\right)
\end{equation}
For bosons, $n_\p$ can be any integer, and we get
\begin{equation}
{\cal Z}_0=\prod_\p\left(1+e^{-\beta\varepsilon_\p}+e^{-2\beta\varepsilon_\p}+\cdots\right)=\prod_\p\frac{1}{1-e^{-\beta\varepsilon_\p}}.
\end{equation}

Knowing the partition function, you can calculate all thermodynamic quantities according to the formulae recalled above. In particular let me call
$f_\p\equiv\braket{\hat{n}_\p}$. This is usually referred to as the occupation number. In  equilibrium, there is not an exact number
of particles in each individual state, but each single particle state is occupied with some probability,
and there are fluctuations. The occupation number can be obtained from the formula
\begin{equation}
f_\p=-\frac{\partial \ln {\cal Z}}{\partial(\beta\varepsilon_\p)}=\frac{1}{e^{\beta\varepsilon_\p}\mp 1}, 
\end{equation}
with $-$ for bosons and $+$ for fermions.

\subsection{The quark-hadron transition in the bag model.}

I now return, as promised, to the phase diagram of hot and dense matter, with a short digression on a simple model that mainly exploits the formulae that we have just recalled. Further details on this model may be found in \cite{Blaizot:1996ns}. 

\begin{figure}
\includegraphics[width=.5\textwidth]{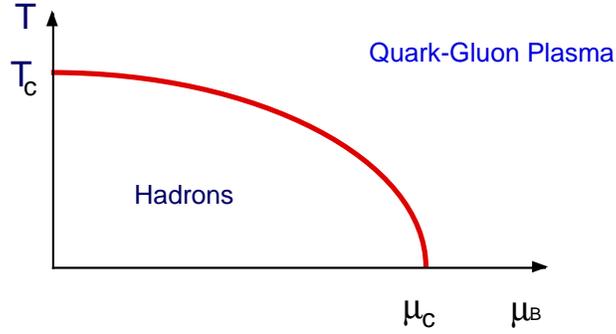}
\caption{The expected phase diagram of hot and dense hadronic matter 
        in the plane $\left(\mu_ B,T \right) $, where $T $ 
        is the temperature and $ \mu_ B $ the baryon  chemical potential}
\label{fig:phase_diag}
\end{figure}

The phase diagram of dense hadronic matter has the expected shape indicated
in Fig.~\ref{fig:phase_diag}.  There is a low density, low
temperature region, corresponding to the world of ordinary hadrons,
and a high  density, high temperature region, where the dominant
degrees of freedom are  quarks and gluons. The
precise determination of the transition line  requires
elaborate non perturbative techniques, such as those of lattice gauge
theories.  But one can get rough orders of
magnitude for the transition temperature and  density using a simple 
model dealing
mostly with non-interacting particles.

Let us first consider  the transition  in the
  case  where $ \mu_B =0$.
At low temperature  this baryon free matter is composed of the lightest mesons,
i.e. mostly the pions. At sufficiently  high temperature one should also
take into account heavier mesons, but in  the present discussion
this is an inessential complication. We shall even  make a further 
approximation by treating the pion as a
massless particle. At very  high temperature, we shall consider that 
hadronic matter is composed
only of  quarks and antiquarks (in equal numbers), and gluons,
forming a quark-gluon  plasma\index{quark-gluon  plasma}. In both the high
temperature and the low temperature phases,  interactions are neglected
(except for the  bag constant to be introduced below).
The description of the transition will therefore  be dominated by
entropy\index{entropy} considerations, i.e. by counting  the degrees of 
freedom.

The energy density
$ \varepsilon $ and the pressure $ P $ of a
gas of massless pions are given by:
\beq
  \varepsilon  = 3\cdot  {\pi^ 2 \over 30} T^4\, ,\ \ \ \ \ \ P
= 3\cdot{ \pi^ 2 \over 90} T^4, \eeq where the factors 3 account
for the 3 types of pions $ (\pi^ +, $ $ \pi^ -, $
$ \pi^ 0). $

The energy density and pressure of the quark-gluon plasma are given
by similar  formulae:
\beq
  \varepsilon &  =& 37\cdot{ \pi^ 2 \over 30} T^4+B, \nonumber\\ P & =
& 37\cdot{ \pi^ 2 \over 90} T^4 - B,
  \eeq where $ 37 = 2\times 8+{7 \over 8}\times 2\times 2\times
2\times 3 $ is the effective number of degrees of freedom\index{degrees of
freedom}  of gluons (8 colors, 2 spin states) and quarks (3 colors, 2 spins, 2
flavors, $ q $ and $ \bar q) $. The quantity $ B $, which is  added to the
energy density, and subtracted from the pressure, summarizes
interaction effects which are  responsible for a change in the
vacuum structure between the low temperature  and the high
temperature phases. It was introduced first in the \lq\lq bag
model\rq\rq\  of hadron structure as a restoring force needed to
equilibrate the pressure  generated by the kinetic energy of the
quarks inside the bag\index{bag model}. Roughly, the  energy of
the bag is
\beq
  E(R) = {4\pi \over 3} R^3B + {C \over R},
\eeq
where $ C/R$ is the kinetic energy of massless quarks.
Minimizing with respect to $ R, $ one finds that the energy at
equilibrium is
$ E \left(R_0 \right)=4B V_0, $  where $ V_0 = 4\pi R_0^3/3 $ is
the equilibrium volume. For a proton with $ E_0
\approx  1$ GeV and
$ R_0\approx 0.7$ fm, one finds $ E_0/V_0\simeq 0.7$
GeV/fm$^3 $, which corresponds to a \lq\lq bag constant\rq\rq\ $ B
\approx  175$ MeV/fm$^3$, or $ B^{1/4} \approx  192$ MeV.

\begin{figure}
\includegraphics[width=.5\textwidth]{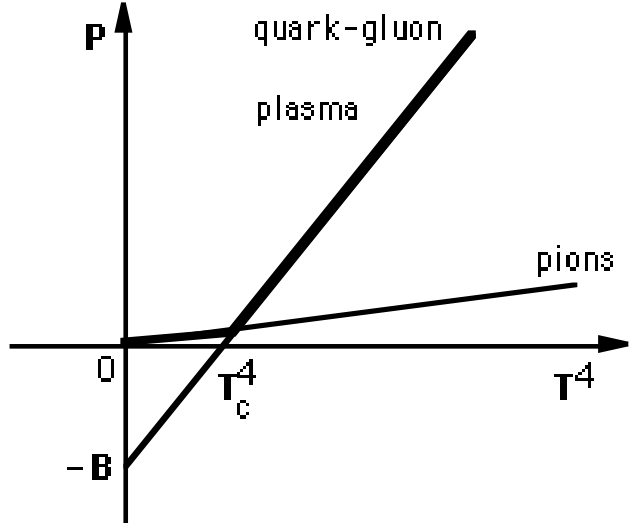}
\caption{The pressure of the massless pion gas compared to that of a 
quark-gluon
plasma, showing the transition temperature $T_c$.}
\label{fig:pression}
\end{figure}

We can now compare the two phases as a function of the temperature.
Fig.~\ref{fig:pression} shows how $ P $ varies as a
function of $ T^4. $ One sees that there exists  a transition
temperature
\beq\label{Tc}
  T_c= \left({45 \over 17\pi^ 2} \right)^{1/4}\ B^{1/4} \approx
0.72\ B^{1/4}, \eeq beyond which the quark-gluon plasma is
thermodynamically favored (has largest  pressure) compared to the
pion gas. For $ B^{1/4}\approx 200$ MeV, $ T_c \approx
150$ MeV.

\begin{figure}
\includegraphics[width=.5\textwidth]{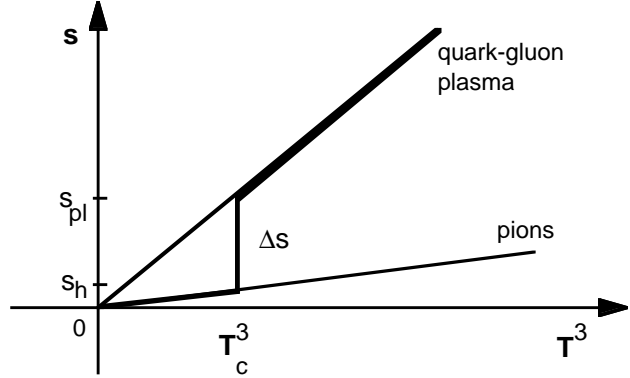}
\caption{The entropy density. The jump $ \Delta s $ at the transition is
proportional to the increase in the number of active degrees of freedom }
\label{fig:entropden}
\end{figure}

The variation of the entropy
density\index{entropy density} $ s =
\partial P /\partial T $  as a function of the temperature  is
displayed in Fig.~\ref{fig:entropden}. Note that  the bag constant $ 
B $ does not
enter explicitly the expression of the entropy. However, $ B $
is involved in  Fig.~\ref{fig:entropden} indirectly, via the
temperature $ T_c $ where the  discontinuity $ \Delta s $ occurs.
One verifies easily that the jump in entropy density
$ \Delta s = \Delta \varepsilon /T_c $ is directly proportional to
the change in the number of active  degrees of freedom when $ T $
crosses $ T_c $.

In order to extend these considerations to the case where
$\mu_B\ne 0$, we note that the transition is taking place when
the total pressure approximately vanishes, that is when the kinetic
pressure of quarks and gluons approximately equilibrates the bag
pressure. Taking as a criterion for the phase transition the
condition $P=0$, one replaces the value (\ref{Tc}) for $T_c$ by the
value $(90/37\pi^2)^{1/4}B^{1/4}\approx 0.70 B^{1/4}$, which is
nearly identical to (\ref{Tc}). We shall then assume that for any
value of $\mu_B$ and $T$, the phase transition occurs when
$P(\mu_B,T)=B$, where $B$ is the bag constant and $P(\mu_B,T)$
is the kinetic pressure of quarks and gluons:
\beq
P(\mu_B,T)=\frac{37}{90}\pi^2T^4+\frac{\mu_B^2}{9}(T^2+\frac{\mu_B^2}{9\pi^2}).
\eeq
The transition line is then given by $P(\mu_c,T_c)=B$, and it has
indeed the shape illustrated in Fig.~\ref{fig:phase_diag}.

The model that we have just described reproduces some of the bulk
features of  the equation of state obtained through lattice gauge
calculations. In particular, it exhibits the
characteristic increase of the entropy  density\index{entropy density} at the
transition which corresponds to the emergence of a large  number of new
degrees of freedom associated with quarks and gluons.   One should be
cautious, however, and not attempt to draw too detailed conclusions
about the nature of the phase transitions from such a simple model.  In 
particular
  this model  predicts
(by construction!) a discontinuous
transition; but this prediction should not be trusted  \cite{Blaizot:1996ns}.

\subsection{Perturbative evaluation of the partition function}

I now begin the formal part of the lectures, and start introducing the techniques that will allow us to treat the effects of the interactions.

The direct evaluation of the partition function of an interacting system  is
rarely possible as this requires a complete knowledge of the
spectrum of the hamiltonian $H$. Various approximation schemes have therefore
been devised to calculate ${\cal Z}$. We briefly describe one of them, the perturbative
expansion. We assume that we can split the hamiltonian
into 
$ H=H_0+H_1 $
with $H_1\ll H_0$, and that the spectrum of $H_0$ is known
$H_0\ket{ \psi_n^0}=E_n^0\ket{\psi_n^0}$. For convenience, we  assume that the states
$\ket{\psi_n^0}$ are also eigenstates of the charge $Q$, which allows us to
treat $Q$ on the same footing as $H_0$. Thus, in the following, we shall assume
that the term $-\mu Q$ is included in $H_0$. 

\subsubsection{The imaginary time formalism}

We define the following
``evolution operator'':   
\beq
U(\tau)=\exp\left( -\tau H\right) \equiv U_0(\tau) U_I(\tau) ,
\eeq
where $U_0(\tau)\equiv\exp(-H_0\tau)$. The operator $U(\tau)$ is analogous to
the familiar evolution operator of quantum mechanics,
$\exp\left(-iHt\right)$. It differs from it solely by the replacement of the 
time $t$ by  $-i\tau$. Because of this analogy, we refer to $\tau$ as 
the ``imaginary time'' ($\tau$ is real!). It has no 
direct physical interpretation: its role is to properly keep track of the ordering
of operators in the perturbative expansion (indeed in a ``classical'' 
approximation where the operators are allowed to commute the time 
dependence disappears). The operator $U_I(\tau)=\exp\left(
\tau H_0 \right)\exp\left( -\tau H \right)$ is called
the {\it interaction representation} of $U$. We also define the interaction
representation of the perturbation $H_1$:
\beq
 H_1(\tau)=e^{\tau H_0}H_1 e^{-\tau
H_0}, 
\eeq
 and similarly for other operators. 
It is easily verified that $U_I(\tau)$ satisfies the following equation
\beq
\frac{{\rm d} U_I(\tau)}{{\rm d}\tau}=-H_1(\tau) U_I(\tau), 
\eeq 
with initial condition $U_I(0)=1$. By solving this equation one establishes the following important formula
\beq
{\rm e}^{-\beta H} ={\rm e}^{-\beta H_0}\,{\rm T}\exp\left\{-\int_0^\beta d\tau
H_1(\tau)\right\}, 
\eeq
where the symbol T implies an ordering of the operators on its right,  from left to right in
decreasing order of their time arguments.
Using this formula one can 
rewrite ${\cal Z}$ in the form  
\beq\label{Z_pert}
{\cal Z}={\cal Z}_0\,\langle {\rm T}\exp\left\{-\int_0^\beta d\tau
H_1(\tau)\right\}\rangle_0,
\eeq
where, for any operator $\cal O$,
\beq\label{operator_ev}
\langle {\cal O} \rangle_0\equiv{\rm Tr}\left({e^{-\beta
H_0}\over {\cal Z}_0}{\cal O}\right)
.\eeq 
The operator $H_1$ is usually expressed in terms of
creation and annihilation operators $a$ and $a^\dagger$.
Then the  calculation of ${\cal Z}$ reduces to that of the expectation values of 
time ordered products of such operators. Whenever $H_0$ is a quadratic function
of these operators, Wick's theorem applies, and all expectation values can be
expressed in terms of single particle propagators. A diagrammatic expansion can be worked
out following standard techniques. The partition function can be written as
${\cal Z}={\cal Z}_0\exp(\Gamma_c)$ where $\Gamma_c$ represents the sum of all
connected diagrams.

\subsubsection{Free propagators}

  The study of the free propagators will give us the opportunity to add a few
remarks on the structure of perturbation theory at finite temperature.
Let us consider a system with unperturbed hamiltonian: 
\beq
H_0=\sum_\p\epsilon_\p a^\dagger_\p a_\p, 
\eeq
which commutes with the particle number operator
$Q=\sum_\p a^\dagger_\p a_\p$. We define time dependent creation and
annihilation operators in the interaction picture:  
\beq\label{a_de_tau}
a^\dagger_\p(\tau) &\equiv & e^{\tau H_0}a^\dagger_\p
 e^{-\tau H_0}=e^{\epsilon_\p\tau}a^\dagger_\p\nonumber\\
a_\p(\tau) &\equiv & e^{\tau H_0}a_\p
 e^{-\tau H_0}=e^{-\epsilon_\p\tau}a_\p
.\eeq
The last equalities follow simply from the equation of motion $d a^{(\dagger)}(\tau)/d\tau=[H_0,a^{(\dagger)}(\tau)]$, and the commutation relations:
\beq
[H_0,a_\p^\dagger]=\epsilon_\p a_\p^\dagger\qquad\qquad
[H_0,a_\p]=-\epsilon_\p a_\p
\eeq
which hold for bosons and fermions. 
The single particle propagator can then be obtained by a direct calculation:
\beq
G_\p(\tau_1-\tau_2) &=& \langle{\rm
T}a_\p(\tau_1)a_\p^\dagger(\tau_2)\rangle_0\nonumber\\
&=& e^{-\epsilon_\p(\tau_1-\tau_2)}
\left[\theta(\tau_1-\tau_2)(1\pm
n_\p)\pm n_\p\theta(\tau_2-\tau_1)\right], 
\label{freeG}
\eeq
where:
\beq
n_\p\equiv \langle a^\dagger_\p
a_\p\rangle_0=\frac{1}{e^{\beta\epsilon_\p}\mp 1}, 
\eeq
and the upper (lower) sign is for bosons (fermions).
The fact that $G$ is a function of $\tau_1-\tau_2$
alone may be viewed as a consequence of the fact that $H_0$ is independent of (imaginary) time.  Note
that since $0<\tau_1,\tau_2<\beta$, one has  $-\beta<\tau_1-\tau_2<\beta$. One can verify on the expression
(\ref{freeG}) that, in this interval, 
$G_\p(\tau)$ is a periodic (boson) or antiperiodic (fermion) function of
$\tau$:
\beq\label{periodicity}
G_\p(\tau-\beta)=\pm G_\p(\tau)\qquad\qquad(0\le\tau\le \beta).
\eeq
To show this, note the following useful relation:
 \beq
 {\rm e}^{\beta\epsilon_\p}\,n_\p=1\pm n_\p.
 \eeq
Thanks to its periodicity (\ref{periodicity}),  the propagator can be represented by a Fourier series
\beq\label{eq:Gdetau}
 G_\p(\tau)=\frac{1}{\beta}\sum_\nu
e^{-i\omega_\nu\tau}G_\p(i\omega_\nu),
\eeq
where the $\omega_\nu$'s are called the Matsubara frequencies:
\beq
\begin{array}{llll}
\omega_\nu &=& 2\nu\pi/\beta & \qquad{\rm bosons,} \\
\omega_\nu &=& (2\nu+1)\pi/\beta & \qquad{\rm fermions.}
\end{array}
\eeq
The inverse transform is given by
\beq\label{G_de_omega}
G(i\omega_\nu) = \int_0^\beta {\rm d}\tau\, e^{i\omega_\nu\tau}G(\tau)
= \frac{1}{H_0-i\omega_\nu}.
\eeq
Using the property
\beq
\delta(\tau)={1\over\beta}\sum_\nu e^{-i\omega_\nu\tau}\qquad\qquad -\beta<\tau<\beta
\eeq
and Eqs.~(\ref{eq:Gdetau},\ref{G_de_omega}), it is easily
seen that $G(\tau)$ satisfies the differential equation
\beq\label{equa_diff_G}
(\del_\tau+H_0)\,G(\tau)=\delta(\tau)
,\eeq
which may be also verified directly from Eq.~(\ref{freeG}). 
Alternatively, the single propagator at finite temperature may be obtained as
the solution of this equation (\ref{equa_diff_G}) with periodic (bosons) or antiperiodic (fermions) boundary conditions.

\noindent{\bf Remark 1.} The periodicity or antiperiodicity that we have uncovered on the
explicit form of the unperturbed propagator is, in fact, a general property of the
propagators of a many-body system in thermal equilibrium. It is a consequence of the
commutation relations of the creation and annihilation operators and the cyclic
invariance of the trace. I leave it to you as an exercise to establish this general property.

\noindent{\bf Remark 2.} The statistical
factor $n_\k$ can be obtained from the relation $n_\k= \pm G(\tau=0^-,\k)$.
In the approximation in which the sum over Matsubara
frequencies is limited to the term $\omega_n=0$, one gets from (\ref{eq:Gdetau}):
$n_\k\approx \frac{T}{\epsilon_\k}$.
Such an approximation corresponds to a ``classical field'' approximation 
 valid when 
the occupation factors are large. This approximation, typically valid for long
wavelength (small $k$), should not be confused with  the
classical limit reached when the thermal wavelength of the particles
becomes small compared to their average separation distance. In this
limit, the occupation of the single particle states becomes small, and
the statistical factors can be approximated by their Boltzmann form:
\beq
\frac{1}{{\rm e}^{\beta(\epsilon_\p-\mu)}\pm 1}\approx {\rm
e}^{-\beta(\epsilon_\p-\mu)}\ll 1\qquad {\rm e}^{-\beta\mu}\gg 1.
\eeq

\section{Lecture 2}

I briefly remind you where we stand.
Last time I started to explain to you
how to do calculations in  field theory
at finite temperature,  
and I told you that we were going to use two languages.   
One is the operator formalism that  in many cases offers the most direct physical interpretation. This is a convenient formalism if you know the Hamiltonian.  
But (relativistic) quantum field theory is more often formulated  
in terms of a lagrangian rather than a hamiltonian,  
and the most appropriate formalism is then that of path integrals.  
I shall remind you how the calculations proceed in the two formalisms.   

Then, what I intend to do today is to apply these formalisms to the calculation of the  thermodynamics of scalar fields.   
I would like to show you,  on the simple example of the scalar field,  
how we go on calculating various Feynman diagrams, and alert you on the   
difficulties which emerge in such calculations. 
That will take us slowly towards the problems specific to QCD. 

\subsection{Calculation of the partition function}
\subsubsection{Operator formalism (reminder from last lecture)}

We established last time  
the basic formula for the calculation of the partition function,
${\cal Z} ={\rm Tr}(e^{-\beta H})$.
We showed that it can be written as an
expectation value of a time-ordered exponential:
\begin{align} 
{\cal Z} ={\rm Tr} e^{-\beta H}
= 
{\cal Z}_0 
\left< {\rm T} 
\exp  
\left \{ -\int_0^\beta d\tau H_1(\tau) \right \} 
\right>_0 
\label{eq:Z} 
\; , 
\end{align} 
where $\beta =1/T$ is the inverse temperature, 
and the expectation value is taken with the density operator associated with the free hamiltonian $H_0$. 
That is,  for any operator $O$, we have
\begin{align} 
\left< 
O 
\right>_0 
= 
\frac{1}{{\cal Z}_0} 
{\rm Tr}[ e^{-\beta H_0} O]
\; . 
\end{align} 
Obviously, if you choose $O$ to be the  identity,
 this formula  tells you that 
$1=\left <1\right>_0= \frac{1}{{\cal Z}_0} {\rm tr}e^{-\beta  H_0}$,
which indicates that ${\cal Z}_0={\rm tr}e^{-\beta  H_0}$: ${\cal Z}_0$ is the partition function corresponding to the hamiltonian $H_0$.
I have written $H = H_0 +H_1$,  
assuming that $H_1 \ll H_0$, in some sense that I shall
 specify  later on.   
I remind you that $H_1(\tau )$ is the interaction representation of   
$H_1$,  i.e.,
$H_1(\tau)=e^{\tau H_0} H_1 e^{-\tau H_0}$. 
The expression (\ref{eq:Z}) suggests a way to calculate ${\cal Z}$, as an expansion  in powers of $H_1$.
This is perturbation theory in the operator formalism.

\begin{figure}[tb]
\begin{center}
\hfil
\includegraphics[width=4cm]{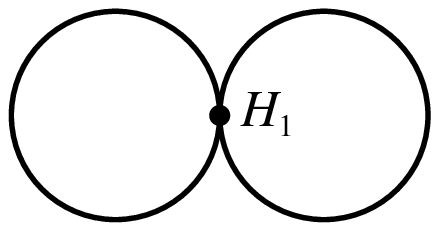}
\hfil
\includegraphics[width=4cm]{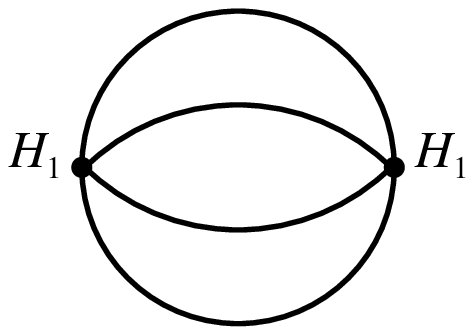}
\hfil
 \caption{
Examples of  Feynman diagrams in $\phi^4$ theory.}
\label{fig:fey1}
\end{center}
\end{figure}

Now, what does it mean expanding in powers of $H_1$?   
After expanding the exponential in Eq.~(\ref{eq:Z}), 
you have to take the  expectation value.  
If $H_0$, as most often, is a quadratic form of  
creation and annihilation operators, for instance 
$ H_0 = \sum_\p \epsilon_\p a^\dagger_\p a_\p$,   
then the calculation that you have to do  can be   
expressed in terms of  Feynman diagrams.  
The lowest order Feynman diagrams for the $\phi^4$ theory that we shall discuss soon are displayed in Fig.~\ref{fig:fey1},  
with $H_1$ entering the vertices.
So the left diagram  will be of first order in $H_1$,  
and the right one will be of second order in $H_1$.   
The lines which join two vertices at times $\tau_1$ and $\tau_2$ are free propagators  $G^0_\p (\tau_1-\tau_2)$.  Such a propagator, as we saw last time, has two pieces depending on whether $\tau_1$ is
bigger or smaller than $\tau_2$:   
\begin{align} 
G^0_\p (\tau_1-\tau_2) 
= 
e^{-\epsilon_\p (\tau_1 - \tau_2)} 
\left[ 
\theta (\tau_1 - \tau_2)(1\pm n_\p) 
\pm 
\theta (\tau_2 - \tau_1)n_\p 
\right] 
\; , 
\end{align}  
where $n_\p$ is the statistical factor  
(occupation number) for bosons or fermions,  
and the plus or minus sign corresponds to each of these two possible
situations.
Remember also that $G^0_\p (\tau)$ with $\tau\equiv \tau_1-\tau_2$,  
is a periodic function of $\tau$.  
Thus, it can be expanded as  a Fourier series
\begin{align} 
G^0_\p (\tau) 
= 
\frac{1}{\beta} 
\sum_n 
e^{-i\omega_n \tau} 
G_\p (i\omega_n) 
,\qquad G_\p(i\omega_n)  
=  
\frac{1}{\epsilon_\p - i\omega_n}  
\; .
\end{align} 
where $\omega_n=2n\pi/\beta$ is  a Matsubara frequency.

\subsubsection{Path integral  formalism}

Now, I would like to move to the path integral formalism.   
The path integral formalism is based on the famous Feynman formula   
 for probability amplitudes.
Let me first discuss the case of ordinary quantum mechanics.
Consider a particle in one dimension, which is located at point $x_1$ a the initial time $t=0$.
Its motion is described by a hamiltonian $H$, typically of the form  $H=p^2/2m+V(x)$.  
At  time $t$,  
the state of the particle is given by 
\begin{align} 
\left .
e^{-i H t} 
| 
x_1 \right> 
\; .  
\end{align} 
The probability amplitude that at time $t$, the particle is located at point  $x_2$ is
\begin{align} 
\left<x_2 | 
e^{-i H t} 
| 
x_1 \right> 
\; .  
\end{align} 
Now what Feynman told us is that  
this can be written as an integral 
over paths $x(t)$  
such that $x(0)=x_1$ and $x(t)=x_2$:
\begin{align} 
\left<x_2 | 
e^{-i H t} 
| 
x_1 \right> 
= 
\int_{x(0)=x_1}^{x(t)=x_2} {\cal D}(x(t)) 
\;
e^{i\int_0^t (\frac{1}{2}m\dot{x}^2-V(x))dt'} 
\; , 
\label{eq:pathint_realtime} 
\end{align} 
where we recognize in the exponent the action integral. Note that the notation in Eq.~(\ref{eq:pathint_realtime}) is symbolic: most of  the paths involved in the sum are not smooth paths. 
But I'm  assuming that most of you have some familiarity with this expression.

What I shall do now is to use this  expression 
(\ref{eq:pathint_realtime}) 
in order to calculate the partition function $Z$.   
I shall rely on the analogy that I already pointed out last time
-- the analogy between the evolution operator 
$e^{-i H t}$ and the density operator $e^{-\beta H}$
of statistical mechanics.   
So I'm going to substitute  $it\to\tau$, with  $\tau$ real.    
Then I shall calculate the matrix elements 
$\left<x_2 | e^{- H \tau} | x_1 \right> $
according to the rule we used in order to write the expression
(\ref{eq:pathint_realtime}),
and find 
a path integral from $x_1$ to $x_2$, where  now the path is a function of
$\tau$. 
Watch out how things get modified 
in the action integral:  
$it$ is replaced by $\tau$ , $idt$ is  replaced by
$d \tau$, and $\del_t$ by $i\del\tau$, so that  the kinetic energy acquires a minus sign.   
A minus sign exists also in front of the potential energy, and 
I pull this overall minus sign out of the integrand.
Then I get
\begin{align} 
\left<x_2 | 
e^{- H \tau} 
| 
x_1 \right> 
= 
\int_{x(0)=x_1}^{x(\tau)=x_2} {\cal D}(x(\tau)) 
\;
e^{- \int_0^\tau (\frac{1}{2} m \dot{x}^2 + V(x))d\tau'} 
\; .  
\label{eq:pathint_imagtime} 
\end{align} 
where now $\dot{x} = dx/d\tau$. This is the formula that will be useful to calculate the partition function. Pay attention to the difference with Eq.~(\ref{eq:pathint_realtime}): 
the factor  $i$  in front of the action in the real time path integral
(\ref{eq:pathint_realtime}) has been replaced by a minus sign. 
And there is a {\it plus} sign in front of the potential
instead of minus sign.   

Now I return to the partition function
${\cal Z}={\rm Tr} e^{- \beta H}$, which can be calculated as 
\begin{align} 
{\rm Tr} e^{- \beta H} 
= 
\int dx 
\left<x | 
e^{- \beta H } 
| 
x \right> 
.\end{align} 
The matrix element is given by the path integral
(\ref{eq:pathint_imagtime}), with $x(0)=x(\beta)=x$, and we shall write, symbolically,
\begin{align} 
{\rm Tr} e^{- \beta H} 
= 
\int dx 
\left< x | 
e^{- \beta H } 
|x \right> 
= 
\int \limits_{{x(0)=x(\beta)}} {\cal D}(x(\tau)) \; e^{-S_E} 
\; , 
\label{eq:pathint_qm} 
\end{align} 
where I have introduced the notation $S_E$ for the ``Euclidean action''
\begin{align} 
S_E
=
\int_0^{\beta} \left ( \frac{1}{2} m \dot{x}^2 + V(x) \right ) d\tau 
\; .
\end{align} 
Eq.~(\ref{eq:pathint_qm})
is  the path integral expression for the partition function. It involves a sum over paths $x(\tau)$ that are periodic in imaginary time: $x(0)=x(\tau)$.

The extension  of this formula to  field theory is easy.   
The scalar field theory is specified by the lagrangian ${\cal L}$  
\begin{align} 
{\cal L} 
= 
\frac{1}{2}(\partial_\mu \phi)^2 
-\frac{1}{2}m^2 \phi^2 
-V(\phi) 
\; ,
\end{align} 
where I explicitly extracted the mass term from the
potential $V(\phi)$.
I will very often consider a specific form for $V(\phi)$,
the so-called ``$\phi^4$ field theory'',  where  
\begin{align} 
V(\phi) 
= 
\frac{\lambda}{4!}\phi^4
\; .  
\end{align} 
The notation $(\partial_\mu \phi)^2$
is a shorthand  for 
\begin{align} 
\frac{1}{2}(\partial_\mu \phi)^2 
= 
\frac{1}{2}(\partial_t \phi)^2 
- 
\frac{1}{2}(\bnabla \phi)^2 
\; .
\end{align} 
I now move to imaginary time, and    
change $it$ into $\tau$. 
Then ${\cal L}$ is changed to
\begin{align} 
{\cal L} 
\to  
- 
\left[ 
\frac{1}{2}(\partial_{\tau} \phi)^2 
+ 
\frac{1}{2}(\bnabla \phi)^2 
+ 
\frac{1}{2}m^2 \phi^2 
+ 
V(\phi) 
\right] 
\; .
\end{align} 
You see that the change of $dt \to -i d\tau$ produces the same minus
sign here as in Eq.~(\ref{eq:pathint_qm}),
and I can write the partition function of the scalar field  
as the following expression:   
\begin{align} 
{\cal Z} 
= 
\int \limits_{\phi(\beta,\x) = \phi(0,\x)} {\cal D} \phi(t,\x)
\; e^{-S_E} 
\; .
\label{eq:pathint_phi} 
\end{align}  
Let me remind you that $\phi$ is playing here the role of
the coordinate and is a function of time $t$ and the three-vector $\x$.
I am summing over all field configurations which are periodic in the
imaginary time $\tau$, i.e., $\phi(\beta,\x)=\phi(0,\x)$.    
The Euclidean action $S_E$ is 
an integral over $\tau$ from 0 to $\beta$,
\begin{align} 
S_E 
= 
\int_0^\beta d \tau \int d^3 x 
\left[ 
\frac{1}{2}(\partial_{\tau} \phi)^2 
+ 
\frac{1}{2}(\bnabla \phi)^2 
+ 
\frac{1}{2}m^2 \phi^2 
+ 
V(\phi) 
\right] 
\; .
\end{align} 
This is the basic formula which we are going to use in specific calculations.

\noindent{\bf Remarks}

In the operator formalism we deduced
the periodicity of the Green's functions or the propagators   
from an explicit calculation in Fock space, calculating the 
time-dependence of the operators and observing the periodicity.  
I left it as an exercise to you to prove that this periodicity is in fact quite general
and exists for the full Green's functions (i.e., not only for the Green's functions of the non interacting system).   
In the path integral approach, the periodicity emerges directly from the 
boundary condition in 
Eq.~(\ref{eq:pathint_phi}): The trace in the partition function  involves summing over field configurations that are
periodic in imaginary time. 
   
There are  further remarks which I want to make at this stage.  

\begin{itemize}
\item
The first remark  
concerns  the Euclidean metric, and why the action is called
Euclidean.
By Euclidean metric I mean that all the gradient terms 
come with positive signs. 
This makes the action $S_E$  a positive definite quantity,  
provided of course that the potential $V(\phi)$ is well-behaved.  
For example, if it is $\lambda \phi^4$,  
we require $\lambda>0$.  
Then, one can interpret  
this exponential $e^{-S_E}$  
as a Boltzmann weight.  
That is to say, one can interpret $e^{-S_E}$  
as a probability distribution,
which allows us,   
in particular,  to calculate the path integral  
using Monte-Carlo techniques: One discretizes the field $\phi$, puts it
on the four-dimensional lattice and select the paths using  ``importance sampling'', i.e., 
with a probability distribution proportional to $e^{-S_E}$.  
You would not be able to do that in real time,  
because then, you would have to add 
wildly oscillating factors $e^{iS}$, which  
no computer knows how to do.  

\item
The second remark concerns the periodicity condition in the expression (\ref{eq:pathint_phi}).  
Let's assume for a minute that we forget about it. What do we get? 
We may rewrite the Euclidean action as
\begin{align} 
S_E=\int d^4 x 
\left[ 
\frac{1}{2}(\bnabla \phi)^2 
+ 
\frac{1}{2}m^2 \phi^2 
+ 
V(\phi) 
\right] 
\; ,
\label{eq:four}
\end{align} 
where $\bnabla$ represents the gradient in four dimensions.
This may be interpreted as the energy of a classical field configuration in four dimension, and the whole path integral can be viewed as the partition function for a classical field theory in four dimension (a sum over all classical field configuration weighted by the factor ${\rm e}^{-S_E}$). 
So, 
 if one ignores the temporal periodicity, one is left with a four
dimensional statistical field theory.
 
\begin{description}
\item {\bf Q}:  
In this case,  isn't there a temperature factor in front of the action?  

\item {\bf A}:  
Indeed, strictly speaking, the Boltzmann factor is the exponential of $1/T$ times the energy of the field configuration (here $S_E$ in Eq.~(\ref{eq:four})). However, this factor $1/T$ is just a multiplicative normalization. This is not the main point that I want to emphasize here, which is that  finite temperature effects in a quantum field theory  
could be viewed as finite size effects  
in a problem of classical statistical mechanics in one dimension more. In fact, by abandoning the periodicity condition and letting the integration over imaginary time extend to infinity, one is looking at the zero temperature limit, that is, one is doing quantum field theory for the vacuum. 

\end{description}

\item
The third remark concerns again the integration over the imaginary time in $S_E$.  
I have argued that, if I let $\beta \to \infty$,  
I can treat the time $\tau$ as an ordinary coordinate,   
and  I end up with a four-dimensional classical field theory.  
Now, I want to argue the other way.  
Let's consider the limit where $\beta \to 0$
or 
$T\to \infty$, that is,  the limit of very large temperature,  large
compared to all typical energy scales in the problem.  
For instance, when you have a mass $m$, this limit applies when $T \gg m$.  
If $\beta$ is very small,  and unless  extremely singular field
configurations play a role,
 I can ignore the time dependence of the field $\phi$.  
Then I can rewrite the Euclidean action  as  
\begin{align} 
S_E 
= 
\beta \int d^3 x 
\left[ 
\frac{1}{2}(\bnabla \phi)^2 
+ 
\frac{1}{2}m^2 \phi^2 
+ 
V(\phi) 
\right] 
\; . 
\end{align} 
Because the fields are considered to be independent of $\tau$, the $\tau$-integration can be done trivially and leads to the factor $\beta$ in front. 
Now we get again a classical field theory, this time in three dimensions.  
The reason why physics is becoming classical here is because  
I am ignoring the time dependence. 
The role of the imaginary-time dependence 
is  clear in the operator formalism.  
The time dependence comes because the operators do not commute.  
You can look at the equation of motion 
\begin{align} 
\frac{d O}{d \tau} 
= 
[O,H] 
\; .
\end{align} 
If the hamiltonian commutes with the operator $O$, 
then  $O$ is time independent.  
That is, the imaginary-time dependence of an operator  is  related to  
the non-commutation of the operator in question with the hamiltonian.  
Thus, in  the formula (\ref{eq:Z}),  
the time-ordered exponential is  there  to keep track  of the time-ordering of quantum mechanical operators.  
Now, I would like to remind you of one thing.
Recall the formula for the propagator that I wrote at  the beginning of the lecture  
\begin{align} 
G^0_\p (\tau) 
= 
e^{-\epsilon_\p (\tau_1 - \tau_2)} 
\left[ 
\theta (\tau)(1\pm n_\p) 
\pm 
\theta (-\tau) n_\p 
\right] 
\; , 
\end{align} 
where $\tau=\tau_1-\tau_2$.  
Clearly, in  the limit where $\tau \to 0^-$:
\begin{align} 
G^0_\p(\tau\to 0-) 
=\pm n_\p 
\; . 
\end{align} 
Let me focus on bosons for which 
\begin{align} 
n_\p 
= 
\frac{1}{e^{\epsilon_\p/T} - 1} 
\; . 
\end{align} 
When $\epsilon_\p/T$ is a small number,  
I can expand the exponential to obtain 
\begin{align} \label{classicaln}
n_\p 
\sim  
\frac{T}{\epsilon_\p}  
\; . 
\end{align} 
This is the occupation factor that you get 
in classical  field theory.  

Now remember also that $G^0_\p(\tau)$ can be expanded in a Fourier series  
\begin{align} 
G^0_\p(\tau\to 0^-) 
=\frac{1}{\beta} 
\sum_n  
e^{i\omega_n \tau} \frac{1}{\epsilon_\p - i\omega_n} 
\; .
\end{align} 
The approximate statistical factor in Eq.~(\ref{classicaln}) is obtained by ignoring all Matsubara frequencies
except  the one with $n=0$.  

These are   features that we are going to meet repeatedly.

\end{itemize}

\begin{description}
\item {\bf Q}:  
You are assuming $\hbar=1$ here, but if you put $\hbar$'s back in this formula
one of $\hbar$'s appears in front of $S_E$,
and I guess $\hbar$ is something like temperature...

\item {\bf A}:   
I am going to leave as an exercise
to you to put back the $\hbar$ everywhere
and understand how the classical aspects enter as $\hbar \to 0$.
But the classical physics which we are discussing is actually subtle.  
One sometimes say, as a joke,  that there are ``two'' $\hbar$'s:  there is 
the $\hbar$ of Mr.~Pauli and that of Mr.~Heisenberg.  
These are of course the same $\hbar$, that of Mr. Planck (and we set it equal to one !).  
What I want to emphasize here  is that there are two physics issues.  
We are going to deal with systems of  
particles which are intrinsically quantum,  like the  black body radiation. 
There  the $\hbar$ involved is that of Pauli, that of the ``Pauli principle''.  
And we are going to deal also  with  long wavelength excitations.  
These  behave like classical field oscillations
whose wavelength is large compared  
with the typical distance between the particles.  
That allows us to use the small gradient expansion  
and that expansion is controlled by the $\hbar$ of Heisenberg (that of the ``uncertainty principle'').
  These features are often mixed in  a subtle way.  
You can have long wavelength, classical-like, oscillations of a gas
of  particles which are intrinsically quantum.  
So, taking the limit $\hbar \to 0$ is tricky here.  
\end{description}

\subsection{Thermodynamics of the scalar field}

With all this preparation,  
we can now begin the discussion  of the thermodynamics of the scalar field. 
I shall do that using perturbation theory, and shall  use a mixed formalism,  
just to train you to go 
from one language to the other. 
Let me view the scalar field
in terms of the hamiltonian  to start with. 
This is 
\begin{align}
H =
\int d^3 x 
\left (
 \frac{1}{2}\pi^2 + \frac{1}{2}(\bnabla \phi)^2
+\frac{1}{2}m^2 \phi^2 + \frac{\lambda}{4!}\phi^4
\right )
= H_0 + H_1
\; ,
\end{align}
where $\pi(x)$ is the canonical momentum conjugate to $\phi$, and $H_1 \sim \frac{\lambda}{4!}\int \phi^4$. 
When I said earlier that $H_1$ is small compared to $H_0$, it means in the present context that  the coupling constant $\lambda$ 
is a small number. 
And I want to calculate the thermodynamic potential $\Omega$
as  a power series in $\lambda$, 
\begin{align}
\Omega 
=
-
\frac{1}{\beta}
\log Z
=
\Omega_0
+
\lambda \Omega_1
+
\lambda^2 \Omega_2
+
\cdots
\; .
\end{align}

The reason why I write the hamiltonian first is 
that I don't want to spend time on $\Omega_0$,
which is  associated with the free hamiltonian $H_0$. 
I can write $H_0$ in terms of the normal modes of the
field, that is, as a collection of harmonic oscillators with frequencies $\omega_\k=\sqrt{\k^2 +m^2}$ (corresponding to the wave numbers $\k$):
\begin{align}
H_0
=
\sum_\k \omega_\k \left( a^\dagger_\k a_\k +\frac{1}{2} \right)
\; .
\end{align}
As you probably all know, 
from this expression 
one can easily calculate $\Omega_0$ as 
\begin{align}
\frac{1}{2}
\sum_\k \omega_\k
+
\frac{1}{\beta}
\sum_\k
\log (1-e^{-\beta \omega_\k}), 
\label{eq:Omega0}
\; .
\end{align}
that is, the sum over all the modes of the corresponding oscillator thermodynamic potentials.

\subsubsection{Short wavelength modes}

This expression (\ref{eq:Omega0})
reveals a problem which we are going
to face repeatedly: ultraviolet divergences. 
I want to spend a few minutes on this. 

The sum over $\k$ for the zero point energy is a shorthand for 
\begin{align}
\sum_\k \omega_\k
=
\int
\frac{d^3 \k}{(2\pi)^3}
\sqrt{\k^2 +m^2}
\; .
\end{align}
As it stands, this integral has no meaning, 
because $\sqrt{\k^2 +m^2} \sim |\k|$ for very large $k$ 
and 
if I put a cutoff $\Lambda$ at the upper end of this integral
the result will grow like $\Lambda^4$.
So it is an infinite number when $\Lambda\to \infty$.
On the other hand, 
it does not depend on  the temperature $T$, and can be interpreted as a correction to the vacuum energy. 
Accordingly, we are going to redefine the zero of the energy
and simply subtract it. 
This is a ``poor man's renormalization''
and we will do more sophisticated things soon.
What I am doing here is simply redefining the zero of the energy 
in such a way that the  contribution of the zero point energies (the first term in  Eq.~(\ref{eq:Omega0})) drops out. 
The second term in Eq.~(\ref{eq:Omega0}) 
is finite because when $k \to \infty$ 
the factor $e^{-\beta \omega_\k}$ goes to zero.

\begin{figure}[tb]
\begin{center}
\includegraphics[width=4cm]{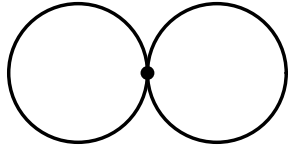}
\caption{First order correction to the thermodynamic potential.}
\label{fig:Omega1}
\end{center}
\end{figure}

Now I want to move on and proceed with the calculation of the first-order
correction, $\Omega_1$. 
It is  given by the diagram in Fig.~\ref{fig:Omega1}. 
Because this is the first order of perturbation theory, this is simply 
$\lambda/4!$ times the expectation value of $\phi^4$
calculated in the non-interacting ground state.
Since the non-interacting energy is quadratic in the fields, the average is a Gaussian integral, and therefore
$\left < \phi^4 \right>_0=
3 \left < \phi^2 \right>_0\left < \phi^2 \right>_0$.
Now, $\left< \phi^2 \right>_0$ is nothing but the propagator
evaluated at $\tau=0$ and $\x=0$. Thus 
\setbox1=%
\hbox to 1.7cm{\resizebox*{1.7cm}{!}
{\includegraphics{./figures2/fig_Omega1.ps}}}
\begin{equation}
\beta \Omega_1/V=
\raise -4mm \box1=
\frac{\lambda}{4!}
\left< \phi^4 \right>_0
=
\frac{\lambda}{4!}
3
\left< \phi^2 \right>_0^2
=
\frac{\lambda}{8}
\left[\Delta (\tau=0, x=0) \right]^2
.
\label{eq:Omega1}
\end{equation}
The boson propagator is, in   a mixed representation, 
\begin{align}
\Delta_\k(\tau)
=
\frac{1}{2 \omega_\k}
\left[
(1+n_\k)\, e^{-\omega_\k |\tau|} + n_\k \, e^{\omega_\k |\tau|}
\right]
\; .
\label{eq:propagator}
\end{align}
You can also verify that
this is a periodic function of $\tau$ 
and it has the Fourier decomposition like the propagators that I was
discussing earlier. The Matsubara representation is
\begin{align}
\Delta_\k(i\omega_n)
=
\frac{1}{\omega_n^2 + \omega_\k^2}
\; .
\end{align}
The propagators I used before were those
for the creation and annihilation operators, 
and they had the singularity at $\tau=0$ 
because it matters whether you order $a^\dagger$ on the left of
$a$ or $a$ on the left of $a^\dagger$.
For the propagator of the scalar field, there is no singularity at $\tau =0$. In particular for $\tau = 0$ and $\x=0$, 
\beq\label{phi20}
\Delta(\tau =0, \x=0)
=
\int
\frac{d^3 \k}{(2\pi)^3}
\frac{1+2n_\k}{2\omega_\k}
\; .
\eeq
This is an important formula that we are going to use again and again. 
This is the expression of the fluctuation of the field, $\left< \phi^2 \right>_0$ in the absence of interactions.

The calculation can be completed now. 
The expression (\ref{eq:Omega1}) is equal to
\begin{align}
\frac{\lambda}{8}
\left[
\sum_\k \frac{1+2n_\k}{2\omega_\k}
\right]^2
=
\frac{\lambda}{8}
\left[
\left(\sum_\k \frac{1}{2 \omega_\k} \right)^2
+
\left(\sum_\k \frac{n_\k}{\omega_\k} \right)^2
+
2
\left(\sum_\k \frac{1}{2 \omega_\k} \right)
\left(\sum_\k \frac{n_\k}{\omega_\k} \right)
\right]
\; ,
\label{eq:Omega1_expanded}
\end{align}
where I rewrite the integrals as sums over $\k$. 
Actually there are various problems in this expression.

Look at the first term: 
\begin{align}
\left(
\sum_\k \frac{1}{2 \omega_\k} 
\right)^2
=
\left(
\frac{1}{2}
\int \frac{d^3 k}{(2\pi)^3}
\frac{1}{\sqrt{\k^2+m^2}}
\right)^2
\; .
\end{align}
This integral is divergent. But since it doesn't depend on temperature, we can drop it, as we did earlier: it represents an 
 infinite correction, of order $O(\lambda)$,  to the zero of the energy. 
The second term is finite, 
 because of the presence of  the statistical factor. 

\begin{figure}[tb]
\begin{center}
\includegraphics[width=3cm]{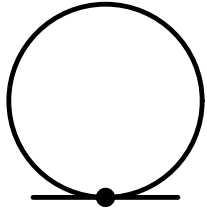}
 \caption{First order  correction  to the mass.
}
\label{fig:tadpole}
\end{center}
\end{figure}

The troublesome term is the third one. 
This is divergent {\it and} temperature dependent. 
We know from general principles that this should not happen. 
What I want to show you  is that it will disappear. 
To that aim, 
it is useful to calculate 
the leading order correction to the mass. 
This is given by the simple diagram 
in Fig.~\ref{fig:tadpole}. 
Let me call that $\Sigma$. 
It is a number which doesn't depend on the momentum of the external lines
and is equal to 
\setbox1=%
\hbox to 1.2cm{\resizebox*{1.2cm}{!}
{\includegraphics{./figures2/fig_tadpole.ps}}}
\begin{equation}
\Sigma
=\raise -4mm \box1=
\frac{\lambda}{2}
\Delta (0,0)
=
\frac{\lambda}{2}
\sum_\k \frac{1+2n_\k}{2\omega_\k}
\; .
\label{eq:sigma}
\end{equation}
The propagator $G_\k$, including the mass correction,  is now given by the Dyson equation
\begin{align}
G_\k^{-1} = \Delta_\k^{-1} + \Sigma
\; ,
\end{align}
where $\Delta_\k$ is given in Eq.~(\ref{eq:propagator}). 

Now, we observe  that,
in $\Sigma$, 
there is a contribution which is independent of temperature, 
so which will exist in the limit of vanishing temperature. 
This contribution is divergent, but  I cannot handle this divergence the way  I did for the energy --
just subtract it -- 
because the correction to the mass  propagates 
 in all the lines of the Feynman diagrams. 
So I need to do a more elaborate treatment. 
I introduce a term,  called a ``counter-term'',
\begin{align}
\frac{1}{2}\delta m^2 \phi^2
\; ,
\end{align}
and add this to the hamiltonian density. 
Then I require that 
this  correction doesn't change the mass at zero
temperature, that is, 
\begin{align}
\frac{\lambda}{2}
\sum_\k \frac{1}{2\omega_\k}
+
\delta m^2
=
0
\; .
\end{align} 
This determines $\delta m^2$:
\begin{align}
\delta m^2
=
-
\frac{\lambda}{2}
\sum_\k
\frac{1}{2 \omega_\k}
\; .
\end{align}

Of course, the term added to $H$ contributes to the thermodynamic potential. It will  in particular generate a contribution to $\Omega_1$.
Since $\delta m^2$ is already of order $\lambda$, in order to obtain this contribution, one just needs to calculate the expectation value of $\phi^2$
in the non-interacting system, which we know already (see Eq.~(\ref{phi20})). Thus:
\setbox1=%
\hbox to 1.0cm{\resizebox*{1.0cm}{!}
{\includegraphics{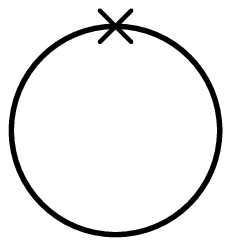}}}
\begin{equation}
\frac{1}{2}
\delta m^2 \Delta(0,0)
=\raise -4mm \box1=
\frac{1}{2}
\left(
-
\frac{\lambda}{2}
\sum_\k
\frac{1}{2 \omega_\k}
\right)
\left(
\sum_\k
\frac{1+2n_\k}{2\omega_\k}
\right)
.
\label{eq:Omega1_renormalize}
\end{equation}
Now if you compare what we have here 
with  Eq.~(\ref{eq:Omega1_expanded}), 
you have $\lambda/4$ $\times$ (divergent sum) 
$\times$ (finite sum) there,
and now $\lambda/4$ with minus sign $\times$ (divergent sum) 
$\times$ (finite sum) here.
So they cancel, as anticipated. This simple example illustrates a general result: there cannot be quantities which are (ultraviolet) divergent 
and which depend on temperature. Such contributions are eliminated by  a proper treatment of the subdivergences of Feynman diagrams, of which I have given you a very elementary example. This property can be verified to all orders. The final result for $\Omega_1$ reads
\beq
\Omega_1=\frac{\lambda}{1152} T^4.
\eeq

\begin{figure}[tb]
\begin{center}
\includegraphics[width=4cm]{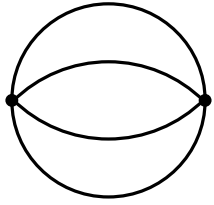}
 \caption{Three-loop contribution to the thermodynamic potential.
}
\label{fig:3loops}
\end{center}
\end{figure}

I would like to show you now the results of 
the three loop calculation, which involves in particular the diagram displayed in Fig.~\ref{fig:3loops}, in order
to illustrate other important 
features of field theoretical calculations at finite temperature. 
When we do such higher loop calculations, 
we need more sophisticated 
techniques for the renormalization procedure than what I've done so far. 
In particular in gauge theory it is essential, 
and  in the scalar case 
it is convenient,  to use dimensional 
regularization. 
The outcome is that, 
in higher orders, quantities like the coupling constant, 
or the mass, start to be dependent on the ``renormalization scale''. 
This dependence is what I would like to discuss now.

Let me just quote the result for $m=0$
($\Omega=-PV$):
\begin{align}
P
=
\frac{\pi^2 T^4}{9}
\left\{
\frac{1}{10}
-
\frac{1}{8}
\frac{\lambda}{16\pi^2}
+
\frac{1}{8}
\left[
3 \log \frac{\mu}{4 \pi T}
+
\frac{31}{15}
+ C
\right]
\left(
\frac{\lambda}{16\pi^2}
\right)^2
\right\}
\; .
\label{e:3loops}
\end{align}
You could have guessed (from last lecture)
the first term $\pi^2 T^4/90$,
which is just the pressure 
of non-interacting massless modes. 
$C$ is a number which has an explicit expression in terms of 
dilogarithms and the Riemann zeta function,  
but this expression doesn't  matter here. 
In this expression,
$\mu$ is the renormalization scale
which can be chosen at will; 
it can be 1 MeV, can be 1 GeV, can be whatever you want, a priori. 

\begin{figure}[tb]
\begin{center}
\includegraphics[width=4cm]{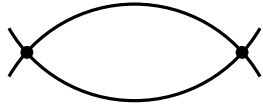}
\caption{One-loop correction to the coupling constant $\lambda$.}
\label{fig:lambda_correction}
\end{center}
\end{figure}

Now, the pressure is {\it a physical quantity}. 
It cannot depend on what you choose for $\mu$,
and therefore $dP/d\mu$ has to be zero: 
\begin{align}
\frac{dP}{d\mu}
=
0
\; .
\end{align}
How is that possible since there is an explicit $\mu$ dependence
which just comes from the calculation of
the diagrams?
The point is that  $\lambda$ depends also on $\mu$:
\begin{align}
\mu 
\frac{d}{d\mu}
\lambda (\mu)
=
\beta (\lambda)
=
3
\frac{\lambda^2}{16 \pi^2}
+
{\cal O}(\lambda^3)
\label{eq:beta}
\; .
\end{align}
(Note that the $\beta$-function is  here positive, in contrast to QCD.)
This dependence on $\mu$ appears essentially 
because $\lambda$ receives contribution 
from diagrams, like that in Fig.~\ref{fig:lambda_correction}, that are logarithmically divergent. 
The factor 3 comes from the three independent channels. 

Now let us calculate $dP/d\mu$. This contains two contributions,  the
$\mu$ dependence implicit in $\lambda$, and the explicit one. We have
\begin{align}
\mu \frac{dP}{d\mu}
=
\frac{\pi^2T^4}{9}
\left\{
-
\frac{1}{8}
\frac{1}{16\pi^2}
\mu
\frac{d\lambda}{d\mu}
+
\frac{3}{8}
\left(
\frac{\lambda}{16\pi^2}
\right)^2
\right\}
+
{\cal O}(\lambda^3)
\; .
\end{align}
Here, in line with the  weak-coupling expansion, I am ignoring the $\mu$-dependence of the $\lambda^2$ term, as this is of order $\lambda^3$. 
 By using Eq.~(\ref{eq:beta}), one verifies that the term within the braces is zero,
In other words, 
if you do a calculation at the order of three loops, 
which is a calculation at the order of $\lambda^2$, 
then the pressure $P$
is independent of the renormalization scale $\mu$
at this order. 
That can also be verified in higher order calculation. 
It is a general result.

But there is more to be said. 
You see that $\mu$ enters in a logarithm. 
This is also a fairly generic situation when doing high order calculations. 
Now, in the logarithm,  $\mu$ 
is divided by some scale, and the typical scale that appears is that in Eq.~(\ref{e:3loops}), namely 
 $\mu$ appears generically as $\mu/(2\pi T)$. 
The  logarithm 
can be large if $\mu$ is very big compared to $2\pi T$, 
or if $\mu$ is very small compared to $2\pi T$. 
You don't want that 
because you are doing an expansion in powers of $\lambda$, and in  order for the successive terms to be as small as possible, you want  these logs to be as small as possible. 
This is why it is natural to choose $\mu$ of the order of $2\pi T$.

This is an important observation, which leads one to expect  the thermodynamics of QCD at high temperature  to  be 
very close to that of a free gas of quarks and gluons. 
The reason is that, in QCD, 
there is a minus sign in the $\beta$-function, so that the coupling decreases as $\mu$ increases.
Since the optimized $\mu$ is related to the temperature as we have just indicated,  
at very large temperature $\mu$ is  large 
and the coupling is  small. 

This is the summary of the arguments 
which deal with short wavelength fluctuations
-- the modes with very large momenta.
There are ultraviolet divergences 
that we can control by the standard process of renormalization. 
Once this is properly done, we  get finite results 
at any finite temperature. 
I have shown to you one particular example 
where
indeed terms, that may appear in intermediate stages of a calculation, which are
divergent and depend on the
temperature, do cancel. 
And finally I've shown  that in higher order, 
you have to worry about running coupling constant effects. The running of the coupling is in particular essential to guarantee that  the physical observables are independent of the renormalization scale $\mu$, at the order at which we calculate. 
I have also indicated that a natural scale may be chosen, $\mu\sim 2\pi T$, in order to optimize the apparent convergence of  perturbation theory. 

\subsubsection{Long wavelength modes}

I would like to
go now into another regime, that of the long wavelength modes. 
I'm going to address another technical issue 
which is related to  infrared divergences. 
This is actually where the major difficulties of perturbation theory 
at finite temperature lie, 
and this will also occupy us next time.

To introduce the subject, let me return to my 
calculation of the correction to the mass,  Eq.~(\ref{eq:sigma}).
I am going to drop systematically the vacuum contribution. 
I shall  work in the limit $m\to 0$ where  I can do explicitly the relevant integral: 
\begin{align}
\frac{\lambda}{2}
\int \frac{d^3 \k}{(2\pi)^3}
\frac{n_\k}{\omega_\k}
=
\frac{\lambda}{2}
\int \frac{d^3 k}{(2\pi)^3}
\frac{1}{e^{k/T}-1}\frac{1}{k}
=
\frac{\lambda}{2}
\frac{T^2}{12}
\; .
\end{align}
Note that  this integral would be quadratically divergent in the absence of the statistical factor. (This factor provides a cut-off at the scale $k\sim T$, and  dimensional analysis shows  that the integral  is proportional to $T^2$. ) Thus it  is dominated by 
``hard'' modes, with $k= {\cal O}(T)$. 
The modes which contribute to the integral 
are those plasma particles which have wavelength of the order $1/T$. 
This mass correction is actually important, and 
I'm going to call it the Debye mass $m_D^2$,
\begin{align}
m_D^2
=
\frac{\lambda}{2}
\frac{T^2}{12}
\; .
\end{align}
The inverse of  $m_D$ plays the same role as  the Debye screening length in an ordinary plasma. 
This is an example of what 
we will meet later and what is called a {\it hard thermal loop}. 
This terminology has its origin 
precisely in the fact that the integral is dominated by  hard modes.
This  correction  is also  sometimes 
called ``thermal mass''.

\begin{figure}[tb]
\begin{center}
\includegraphics[width=2.5cm]{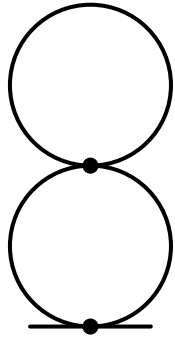}
\caption{A second order correction to the mass.}
\label{fig:mass_2nd}
\end{center}
\end{figure}

Let me now imagine calculating the second-order 
correction to the mass, given by the diagram in Fig.~\ref{fig:mass_2nd}. 
What do I get? 
There is a minus sign because there are two vertices.
Then I have  a sum
over the Matsubara frequency, and an integral over the momentum. 
The insertion is $m_D^2$.
We get (I am assuming again that the particle is massless at zero temperature):
\begin{align}
-
\frac{\lambda}{2}\; T
\sum_n \int
\frac{d^3 \k}{(2\pi)^3}
\frac{m_D^2}{(\omega_n^2+k^2)^2}
\; .
\end{align}
Now look at what happens 
in the  particular term with $n=0$ in this sum, 
\begin{align}
-
\frac{\lambda}{2}\; 
T \int
\frac{d^3 \k}{(2\pi)^3}
\frac{m_D^2}{(k^2)^2}
\; .
\end{align}
That is awful. 
Because $k^2 dk/k^4 = dk/k^2$, this is infrared divergent
as $\int dk/k^2$. 
We have a problem.

\begin{figure}[tb]
\begin{center}
\includegraphics[width=5cm]{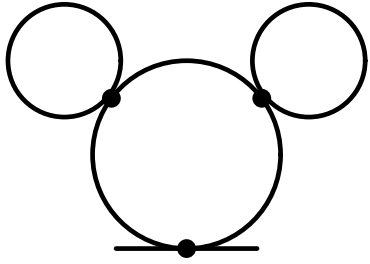}
 \caption{Mickey Mouse diagram for the mass correction.
}
\label{fig:mass_3rd}
\end{center}
\end{figure}

And this problem is going to be worse and worse as we try and calculate more diagrams. 
Let's indeed calculate the ``Mickey Mouse'' diagram in
Fig.~\ref{fig:mass_3rd}. 
This is proportional 
to the integral 
\begin{align}
\int \frac{d^3 \k}{(2\pi)^3}
\frac{(m_D^2)^2}{(\omega_n^2+k^2)^3}
\sim
\int \frac{dk}{(k^2)^2}
\; .
\end{align}
That's indeed  worse! 
The more mass insertions you add, 
the more serious the problem is.

\begin{figure}[tb]
\begin{center}
\hfil
\includegraphics[width=5cm]{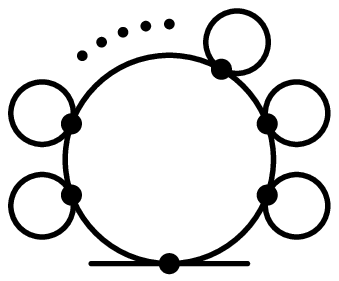}
\hfil
\includegraphics[width=5cm]{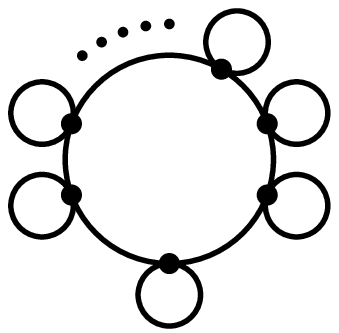}
\hfil
\caption{Ring diagrams for mass correction (left)  and thermodynamic potential (right).
}
\label{fig:ring}
\end{center}
\end{figure}

But now something happens. 
You see that what we are doing here is something 
a little bit stupid, 
once you understand the physics. 
I will have the opportunity to
come back to this later. 
What we have recognized is that, in the thermal bath, 
the particle acquires a mass even if they are massless to start with.
This is a non-perturbative effect,
although the calculation of the mass itself can be done within perturbation theory. 
But the fact that 
the particle acquires a mass is something that you should keep 
in mind when you do a higher order calculation. 
When we do this calculation order-by-order, 
we extract the mass and we treat it as a correction. 
We should not do that. 
If the particle has a mass, 
let's take this properly into account! 
Let me do that by calculating the sum of an infinite 
number of such insertions in Fig.~\ref{fig:ring},
which I will call ring diagram.
You can write the expression for that because
that is the same as Fig.~\ref{fig:mass_2nd} but with a slight
modification:
\begin{align}\label{sigmaring}
\Sigma_{\rm ring}=
\frac{\lambda}{2}\; T
\sum_n \int
\frac{d^3 \k}{(2\pi)^3}
\left(
\frac{1}{\omega_n^2 + \k^2+m^2_D}-\frac{1}{\omega_n^2+\k^2}
\right)
\; , 
\end{align}
where the first  propagator, 
$1/ (\omega_n^2 + \k^2+m^2_D)$, is that of a massive particle with mass $m_D$.  
You can verify that if you expand this propagator 
in powers of $m_D^2$, then you generate back all the diagrams that we have considered before and that are   infrared divergent. The second term in Eq.~(\ref{sigmaring}) subtracts the contribution of the  diagram in Fig.~\ref{fig:tadpole}, 
which is calculated   differently (there is no infrared divergence in this hard thermal loop contribution).
The remaining integrand is then dominated by soft momenta. 
Now, for the particular contribution with $n=0$, I get
\begin{align}
\Sigma_{\rm ring} \to
\frac{\lambda}{2}\; T \int
\frac{d^3 k}{(2\pi)^3}
\left(
\frac{1}{\k^2 + m^2_D}-\frac{1}{\k^2}
\right)
=
\frac{\lambda}{2}\; T
\sum_n \int
\frac{d^3 k}{(2\pi)^3}
\frac{-m^2_D}{\k^2(\k^2 + m_D^2)}
\; .
\end{align}
You see that   the mass $m_D$ in the denominator
provides an infrared cut-off that allows you to  calculate explicitly the integral. Simple dimensional analysis reveals that it is proportional to $ m_D\sim \sqrt{\lambda} T$. An explicit calculation, using the explicit expression for $m_D$ given above, yields
\begin{align}\label{Sigmaring}
\Sigma_{\rm ring} 
=-\frac{\lambda T^2}{8\pi} \left (
\frac{\lambda}{24}\right )^{1/2}
\; .
\end{align}
Likewise, you can calculate 
the ring contribution, $\Omega_{\rm ring}$, 
in Fig.~\ref{fig:ring}
to the thermodynamic potential
and you get
\begin{align}
\Omega_{\rm ring}/V &=
\frac{1}{2}\; T
\sum_n \int
\frac{d^3 \k}{(2\pi)^3}
\left(
\log \left [
1+ \frac{m^2_D}{\omega_n^2 + \k^2}
\right]
-\frac{m_D^2}{\omega_n^2 + \k^2}
\right)
\nonumber \\
&\to
\frac{1}{2}\; T \int
\frac{d^3 \k}{(2\pi)^3}
\left(
\log \left [
1+\frac{m^2_D}{\k^2}
\right]
-\frac{m_D^2}{\k^2}
\right)
\nonumber \\
&=
\frac{T^4}{12\pi}
\left (\frac{\lambda}{24}\right )^{3/2}
\; .
\end{align}
This is an important result. 
It tells you that  if you do an infinite resummation, 
you get rid of the infrared divergences. 
Infrared divergences appear then an artifact 
of the fact that we did not do the calculation in a proper way, i.e., we attempted an expansion that is not valid. 
If we include,  correctly, the thermal mass in the propagator,  we get a finite result. 

The second lesson is that you see something very funny appearing. 
I told you before that we assume a power series in $\lambda$:
$\Omega=\Omega_0 + \lambda \Omega_1+\lambda^2 \Omega_2+\cdots $. 
And this is indeed what you expect if you calculate $\Omega$ 
by just calculating  Feynman diagrams one after the other. 
But what you get here is a non integer power,  $\lambda^{3/2}$!
An unexpected term is sneaking in the series,
which is $\lambda^{3/2} \Omega_{3/2}$. 
That means that the weak coupling expansion of the 
thermodynamic potential is not what you naively expected. 
You get a contribution which is not analytic in the coupling constant. 
And there is no way to get such a contribution by summing a finite 
number of terms. 

There is a lot of material in this lesson. 
We will come back on several points next time. 
I will then try to explain what is going on in very simple terms.

\begin{description}
\item {\bf Q}: 
In this result for the thermal mass, $\lambda$ should be positive, shouldn't it? 

\item {\bf A}: 
The $\lambda$ should be positive in any case, because 
the potential should be bounded from below for the system to be stable. 

\item {\bf Q}: 
Is the instability related to the appearance of the 
$\lambda^{3/2}$ term? 

\item {\bf A}: 
No, but indeed, when $\lambda$ is negative, you
have to pay attention to the 
meaning of $\lambda^{3/2}$. But this is not really related to this 
instability issue that I am referring to.  
If the system is truly unstable, worse things happen. 

\item {\bf Q}: 
If you have an attractive interaction then the mass term 
will be eaten up by the interaction, 
which can give a vanishing mass. 
I guess that is what he is asking. 

\begin{figure}[tb]
\begin{center}
\includegraphics[width=5cm]{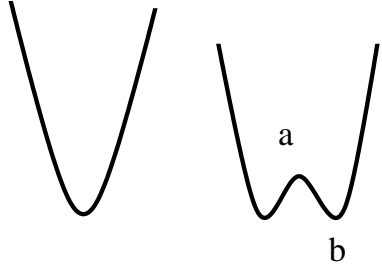}
 \caption{The potential of the $\phi^4$ theory wihtout ($m^2>0$) and with ($m^2<0$) spontaneous symmetry breaking. 
}
\label{fig:qa}
\end{center}
\end{figure}

\item {\bf A}: 
The potential of the $\phi^4$ theory is like this (Fig.~\ref{fig:qa}, right). 
It could also be like this (left of Fig.~\ref{fig:qa}), corresponding to $m^2<0$.
The effect of thermal fluctuations would be typically to restore the symmetry, i.e., 
to transform the potential  
from its initial shape (at $T=0$) on the right hand side of  Fig.~\ref{fig:qa} to the convex shape of the l.h.s of Fig.~\ref{fig:qa}. That is, thermal fluctuations contribute positively to the mass squared, and if $m^2<0$ to start with  (at $T=0$), the thermal fluctuations will eventually turn it positive at sufficiently high temperature.

\item {\bf Q}: Is $m_D^2$ proportional to $\lambda$?

\item {\bf A}: Yes, $m_D^2$ is proportional to $\lambda$ (at least its leading contribution),  and it is positive if $\lambda$ is positive, which we assume to be the case. 

\item {\bf Q}:
It is unless you are on the top of the effective potential. 

\item {\bf A}:
Yes, but if you are here (a), you have to be careful about what you calculate. 
Usually, one wants to expand from here (b), i.e., around the local minimum. 
But I shall  not  discuss symmetry breaking  in these lectures (except a little towards the end).

\item{\bf Q}: You mentioned that you need to 
put $\mu$ equal to $T$ in perturbation theory. 
But I think in the end a physical quantity does not depend on $\mu$...

\item{\bf A}: 
Yes, that is a very good point. 
Indeed, a physical quantity does not depend on $\mu$. 
But remember that if you do the calculation in perturbation theory, 
the physical observables are independent of $\mu$ only up to terms that are of the same order of magnitude as those that are explicitly neglected. 
For example, in the case that  I treated, I considered terms up to, and including, order $\lambda^2$, and the $\mu$ dependence will be of order $\lambda^3$. So at any finite order, there will be a residual $\mu$-dependence. What I have argued is that one can exploit this dependence in order to improve the (apparent) convergence of the perturbative expansion. Such a strategy is sometimes referred to as the ``principle of minimal sensitivity'', or ``principle of fastest apparent convergence''. At finite temperature, the coefficients of the expansion in powers of the coupling constant contain typically logarithms of the ratio $\mu/T$. In order to avoid these logarithms to become too large (and hence spoil the apparent convergence), it is judicious to choose $\mu\sim T$ (and detailed calculations suggest the more specific choice $\mu\simeq 2\pi T$).

\end{description}

\section{Lecture 3}

\subsection*{Summary of last lecture}

Let me start by a brief summary of what we learned
in the last lecture.

What we did last time was to
 consider a scalar field with the hamiltonian density given by
\begin{align}
{\cal H}=
\frac{1}{2}\pi^2 + \frac{1}{2}(\bnabla \phi)^2 
+ \frac{1}{2}m^2 \phi^2+ V(\phi)
\; ,
\end{align}
where we explicitly wrote $V(\phi)=\lambda \phi^4 /4!$.
Then we applied the 
general formalism to calculate thermodynamics,
essentially the partition function,
${\cal Z}={\rm Tr}e^{-\beta H}$. Remember that this could be written as an
integral over field configurations $\phi(x)$, each configuration being weighted by the factor  $e^{-S_E[\phi]}$, with $S_E$  the Euclidean action corresponding to the hamiltonian $H$.
The field configurations included in the integral are periodic 
in imaginary time. 
Here 
the calculation was organized as a power series, and we calculated
the grand potential as 
\begin{align}
\Omega = -(1/\beta) \log Z = 
\Omega_0 + \lambda \Omega_1+\lambda^{3/2}\Omega_{3/2}+ \lambda^2\Omega_2+\cdots
\; .
\end{align}
In doing this calculation,
we encountered two types of difficulties,
namely divergences of some momentum integrals.
Remember that loop integrals in  Feynman diagrams  involve the sum over the Matsubara frequencies $\omega_n$ and the
integral  $d^3 \k$.

We met two types of divergences. 
Ultraviolet divergence -- these are  familiar in field theory
and can be handled by the general procedure called renormalization.
In doing so, 
we introduced the notion of the running coupling constant, with
the $\beta$-function which describes how  $\lambda$
varies with the renormalization scale $\mu$,
\begin{align}
\mu \frac{d \lambda}{d\mu}= \beta(\lambda)=3 \frac{\lambda^2}{16\pi^2}+ \cdots
\end{align}
Physical quantities, like the pressure, should be independent of $\mu$,
that is, 
\begin{align}
\frac{d\Omega}{d\mu}=0
\; . \label{dOmegadmu}
\end{align}
Of course, this holds if $\Omega$ is calculated exactly. If you do an approximation
on $\Omega$, for example 
if you calculate $\Omega$ up to order $\lambda^n$,
Eq.~(\ref{dOmegadmu})  will hold up to corrections of order $\lambda^{n+1}$ 
which are not included in the calculation. In that case some residual dependence on the renormalization scale will subsist in $\Omega$. This residual dependence can often be used as an indication of the accuracy of the calculation. 

Ultra-violet divergences are common 
in all field-theoretical calculations and
they have nothing to do  with the temperature.
The formalism to handle these is well-established.
We encountered also another type of divergences,
called infrared divergences.
These are intimately related to finite temperature effects. At the
end of the last lecture I showed you that, 
in some particular cases,  these infrared divergences
are eliminated by performing an infinite resummation of Feynman diagrams,  which corresponds to taking into account the generation
of a ``thermal mass''. 
We've seen also that this resummation is
 responsible for the fractional power of $\lambda$ that occurs 
 in weak coupling expansion of the pressure, namely, the  term proportional to
$\lambda^{3/2}$. Such a term  is completely unexpected from the point of view of perturbation
theory because 
if you expand naively the partition function in powers of $\lambda$, 
you will only generate  terms 
with integer powers of $\lambda$, not $\lambda^{3/2}$. A term such as $\lambda^{3/2}$ can only be obtained from resumming an infinite number of 
Feynman diagrams. 
Finally, we understood that there was one particular sector which is
important, 
the sector with  the Matsubara frequency $\omega_n=0$. 
Since $\omega_n = 2 \pi n T $, 
as soon as $n$ is equal to 1 or bigger than 1, $\omega_n = 2\pi nT$ provides an infrared cut-off in the propagator $1/(\omega_n^2 + \k^2)$;  therefore there is no infrared divergence
for these  
non vanishing Matsubara frequencies.
This ends  the summary of the main issues that we discussed last time.

\subsection{QCD perturbation theory and its breakdown}

My purpose today is to show you, from a very general perspective,
why we have these specific difficulties at finite temprature
and give you hints about how you can get around these difficulties
by techniques that I shall explain in the next lecture.

Before I do that, let me summarize the results that have been obtained for the pressure as a function of the coupling constant. 
To do so, let me change the notation slightly, and set 
 $g^2=\lambda/24$ (then $\lambda^{3/2}$ becomes $\sim g^3$).
\begin{figure}
\begin{center}
\includegraphics[width=10cm]{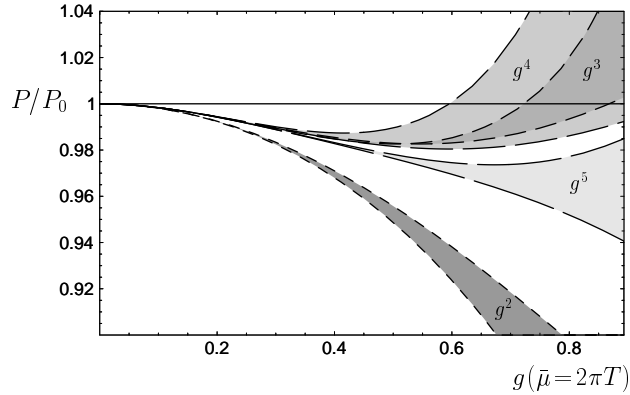}
 \caption{
Weak-coupling expansion for the pressure normalized to that of an
ideal gas as a function of $g(2\pi T)$ in $\phi^4$ theory. Taken from
the review by Blaizot, Iancu and  Rebhan \cite{Blaizot:2003tw}.}
\label{fig:pphi4}
\end{center}
\end{figure}
The pressure $P$ of $\phi^4$ theory
is known up to order $g^6$ (in fact, up to order $g^8\ln g$ \cite{Andersen:2009ct})
\begin{align}
P=P_0 + a_2 g^2 +a_3 g^3 + \cdots + a_6 g^6 + \cdots 
\end{align}
The result is plotted in Fig.~\ref{fig:pphi4}, divided by the free gas pressure $P_0=(\pi^2/90) T^4$ (this has been calculated in the first lecture), and as a function of the running coupling constant defined at the scale $2 \pi T$. 
As I argued last time, 
this is a natural scale to choose in the calculation of the thermodynamical functions. 

As the plot clearly indicates, perturbation theory does not appear to be very predictive unless the coupling constant is very small, $g\lesssim 0.3$. 
This same feature is also met in more complicated theories at finite temperature, such as QCD. In such theories, we know that the coupling constant decreases with the temperature, but unless the temperature is very large, so that the coupling is very small, the same pattern of bad apparent convergence is revealed. 
These calculations of high order contributions are technically demanding, but the result is disappointing. 
One of the questions that  we have to answer is why it is so bad.

\subsubsection{Breakdown of perturbation theory}

In fact, before I turn to general considerations which will shed light on this question, let me show you that in QCD, we can reach a point where perturbation theory completely breaks down, whatever the strength of the coupling. 
Let me show you one particular example
of a class of Feynman diagrams where the problem in question manifests itself. 

\begin{figure}
\begin{center}
\includegraphics[width=8cm]{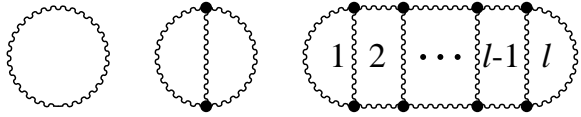}
 \caption{
Gluon loop diagrams contributing to the QCD thermodynamic potential.}
\label{fig:linde}
\end{center}
\end{figure}

I shall consider the particular family of diagrams displayed in  Fig.~\ref{fig:linde}.
The wavy lines represent gluon propagators.
 With the experience that we gained last time, 
we can expect infrared divergences to occur when $\omega_n=0$. Since I am interested in these divergences,  I am going to assume that $\omega_n=0$ in all the propagators.
Let us then examine  the contribution to the pressure, or to the thermodynamics potential, of the  $\ell$-loop diagram. 
For each loop, there is a factor of $T$ that  normally accompanies the sum over the Matsubara frequencies. I shall keep only the term $n=0$, so there is no sum, but the factor  $T$ remains.
And there is an integral $d^3 k/(2\pi)^3$ over the momentum (the factor $(2\pi)^3$ is not  relevant here and I shall drop it).
Now there are also vertices. It  is not hard to count their number: 
If there is one loop, there is no vertex; 
if there are two loops, there are two vertices; 
each time I add one loop, I add two vertices.
Therefore, the power of $g$ is $2 \ell  -2$.
Now, remember that in the first lecture I emphasized the fact 
that the three-gluon vertex carries a momentum.
Therefore, at each vertex is also attached a  momentum $k$.
Then there are propagators, of the form $1/(\omega_n^2 + k^2)$.
I set $\omega_n=0$, but I shall  add a term $m^2$ which is a ficticious gluon mass, so that the propagator is $1/(k^2+m^2)$.
The number of the propagators has to be counted. This is $3(\ell-1)$. You can understand this number in the following way:
Start with two loops (second diagram in Fig.~\ref{fig:linde}), there are three propagators . 
Each time you add a loop, you add two propagators.
I can therefore write, very schematically, and for $\ell>1$) 
\begin{align}
\Omega_{\rm QCD}^{(\ell)}  &\simeq
\left (
T  \int d^3 k\right )^\ell 
\; \; 
\frac{g^{2\ell -2} k^{2\ell-2}}{(\k^2+m^2)^{3(\ell -1)}}.
\end{align}
I am interested in what happens when all momenta are going to zero at the same rate. In this situation, all individual momenta can be combined to form a big momentum, $K$, in a space of
$3\ell$ dimensions. 
Then, I separate the integral into an angular integral and
a radial integral, and focus on the radial integral. In other words, 
in the same way as $d^3 k = k^2 dk d\Omega_3$ with solid angle $d\Omega_3$,
I'm going to write the integral as $\int (d^3 k)^\ell \to \int K^{(3\ell -1)}dK d\Omega_{3\ell}$,
and drop the angular integral because, from the angles, no trouble is going to come.
Let me do that. Then I   get  
\begin{align}
\Omega_{\rm QCD}^{(\ell)}  
&\sim T^\ell g^{2(\ell -1)} 
\int dK K^{3\ell -1}\; \frac{ K^{2\ell -2}}{(K^2 + m^2)^{3(\ell -1)}}
\end{align}
This integral is infrared divergent when $m\to 0$. To see that more clearly, let me replace the denominator $K^2+m^2$ by $K^2$, 
and put $m$ as a lower cutoff in the integral:
\begin{align}
\Omega_{\rm QCD}^{(\ell)}  
\sim T^\ell g^{2(\ell -1)} \int _m dK \frac{1}{K^{\ell -3}}
\; .
\end{align}

If $\ell \leq 3$, this integral is perfectly well-behaved 
because the denominator has a vanishing or negative power of $K$. No problem.
If $\ell =4$, we have a problem -- we have an integral $dK/K$.
There is a logarithmic divergence and I have put a cutoff $m$.
In order to fix the scale there has to be another factor:
the natural scale here is $T$. 
Thus, if I was doing the calculation at four loop, I would expect the
final result to be of the form
\begin{align}
\Omega_{\rm QCD}^{(\ell=4)}  \sim T^4 g^6 \; \log (m/T)
\; .
\end{align}
You see this is a divergent result:  If I let $m$ go to zero, 
which I should do because the gluons don't have any mass,
then I would get an infinity.

\begin{table}
\begin{center}
\begin{tabular}{ll}
\hline
$\ell \leq 3$ \hspace*{1cm}& well-behaved integral \\
$\ell =4$     & $T^4 g^6 \log (T/m)$\\
$\ell > 4$    & $g^6 T^4 \left ( g^2 T/m \right )^{\ell -4}$
\\
\hline
\end{tabular}
\caption{Order of magnitude of the $\ell$-loop contributions to the pressure.
All the $\ell$-loop diagrams with $\ell > 4$
are of the same order of magnitude when  $m={\cal O}(g^2 T)$.}
\end{center}
\end{table}

Let's look at $\ell > 4$. 
This is even worse because,  for $\ell =5$ for instance, 
I get $\int dK/K^2$, which is a power divergence  $\sim 1/m$. 
 More generally, for $\ell>4$, we have
\beq\label{lindeomega}
\Omega_{\rm QCD}^{(\ell)}  \sim
T^\ell g^{2(\ell-1)} \int_m dK/K^{\ell-3}
\; \sim \; 
T^\ell g^{2(\ell-1)} (1/m)^{\ell-4}
\; = \;
g^6 T^4 \left ( g^2 T/m \right )^{\ell -4}
\; .
\eeq
This   situation is  not quite the same, 
but is very  reminiscent of the situation that we met  last time. 
I showed then that there were Feynman diagrams which  where
 infrared divergent, 
the higher the number of loops, the more severe the divergence. 
Now, in the case of the scalar field theory, 
you know that  there is a cut off 
because, as I showed you last time, the modes of the scalar field acquire  a thermal mass of  order of $gT$. 
Remember $m_D^2 \sim \lambda T^2 \sim g^2 T^2$ (with $\lambda\sim g^2$). 
You see, if the mass is of the order of $gT$, there would be no problem 
because $g^2 T/m \sim g \ll 1$ in the weak coupling regime. In other words, the thermal mass cures the potential infrared divergences, and makes the successive loop corrections proportional to $g^\ell$.

In the case of QCD, not all the modes acquire a mass of order $gT$. Only the electric modes do (the thermal mass is then identical to the Debye mass related to electric screening). But there are also modes in QCD which are magnetic in nature, and these modes may develop a mass, so-called  ``magnetic mass''. However, this magnetic mass is  expected to be of order $g^2 T$. 
Then, because $g^2 T /m= {\cal O}(1)$,
all the terms in the expansion (\ref{lindeomega}) end up being of the same order of magnitude!
This is a situation where there is not much you can do with a standard weak coupling expansion: if the mass is of the order $g^2 T$, 
then  all the terms in the 
perturbative expansion are of the same order of the  magnitude. 

In the rest of this lecture, I would like to give you a simple understanding for 
why such nasty things happen.

\begin{description}
\item {\bf Q}: 
Is there any different kind of mass other than magnetic mass in QCD?
\item {\bf A}: 
There are two types of modes in QCD, related to oscillations of the electric field and the magnetic field, respectively.  
The electric field  behaves like a scalar field
and it acquires a mass which is related to the screening phenomenon 
about which I will say more in the next lecture. 
That screening mass is of oder $gT$. So, there is  no problem with the electric modes. 
The difficulty in the QCD plasma comes from the  long wavelength magnetic modes. 
Such modes also exist in electrodynamics (a static magnetic field is not screened). 
But in QCD the magnetic modes interact with themselves, 
and it is this interaction which it is hard to calculate. 

\end{description}

\subsection{Expansion parameter at finite temperature}

\subsubsection{Harmonic oscillator at finite temperature}

The above is the review of the difficulties.
Now, I will try to give you an insight into the physical origin of the difficulty 
and go through an elementary discussion of thermal fluctuations 
in quantum mechanics, 
going back to a very simple system, namely the harmonic oscillator, which you all know for sure.
Basically, what we are doing with field theory 
is playing with an infinite collection of harmonic oscillators. 
You will see, this is a detour, but a quite instructive one.

Let's consider a harmonic oscillator in one dimension whose 
hamiltonian reads
\begin{align}
H=\frac{1}{2} p^2 + \frac{1}{2} \omega^2 x^2
\end{align}
In any state at $T=0$, 
the expectation value of $p^2$ is equal to that of $\omega^2 x^2$ and
given by 
\begin{align}
\left< p^2 \right > = \left< \omega^2 x^2 \right > \sim \hbar \omega
\: .
\end{align}
I shall drop the factor $\hbar$ in what follows (i.e., I shall use natural units where $\hbar=1$). Thus, 
\begin{align}
\left< p^2 \right > \sim \omega, \quad
\left< x^2 \right > \sim 1/\omega
\; .
\end{align}
Now, I need to do a little bit more elaborate things. 
Remember that, for the harmonic oscillator, 
it is useful to introduce  creation and annihilation operators such that 
\begin{align}
[a, a^\dagger ] =1 .
\end{align}
In terms of $a$ and $a^\dagger$, we can write 
\begin{align}
x&=\frac{1}{\sqrt{2\omega}} (a + a^\dagger)
\; , 
\nonumber \\
x^2&= \frac{1}{2\omega}(a + a^\dagger)^2 = 
\frac{1}{2\omega} (a^2 + {a^\dagger}^2+a a^\dagger +a^\dagger a)
\; ,
\end{align}
and the eigenstates of $H$ are of the form $\ket{n_0}=\left(a^\dagger  \right)^{n_0}\ket{0}$. 
When we calculate the expectation value of $x^2$ in a given eigenstate of the harmonic oscillator, e.g. $\bra{n} x^2\ket{n}$, we get contributions only from the last two terms. Using the commutation relation, one easily gets
\begin{align}\label{averagex2}
\left <n_0| x^2 |n_0\right > = \frac{1}{2\omega}(1 + 2n_0).
\end{align}
Here you must recognize something: the factor $(1+2n_\k)/(2 \omega_\k)$ is indeed that same as that in the integral giving the  fluctuation of the scalar field (see for instance Eq.~(\ref{eq:sigma})).

Now, at finite temperature, $T \ne 0$, we can apply the same results. 
On the average, states will be occupied with a probability given  by the Boltzmann factor, 
and the expectation value of $x^2$ at a given temparature $T$ will be given by a formula analogous to Eq.~(\ref{averagex2}), namely
\begin{align}
\left < x^2 \right >_T &\sim \frac{1}{2\omega}(1 + 2n_T)
\; ,
\end{align}
where 
\begin{align}
n_T = \frac{1}{e^{\omega/T}-1}
\; .
\end{align}

There is a limit which is interesting to us.
It is the high temperature limit, $T \gg \omega$. 
In that case, $\omega/T$ is a small number, and we can expand the exponential in the statistical factor $n_T$.
In this regime, $n_T \sim T/\omega\gg 1$, and  we can  ignore the contribution of the zero point fluctuations (the  ``vacuum part'').
What remains is
\begin{align}
\left < x^2 \right >_T &\sim \frac{T}{\omega^2}
\; .
\end{align}
This result is familiar and reflects the equi-partition of the energy, 
$$\omega^2 \left < x^2 \right >  \sim \left < p^2 \right > \sim T ,$$
as expected from  Boltzmann statistics.

Of course, the harmonic oscillator itself
is fine, but 
what I'm really interested in is to study the effect of the interactions. 
So let me add interactions by changing the hamiltonian $H \to H_0 + H_1$. 
What's a good choice for $H_1$? To stay as close as possible to the $\phi^4$ scalar field theory, I take
\begin{eqnarray}
H_1 = \frac {\lambda}{4!} x^4 \;.
\end{eqnarray} 
We may now calculate the energy levels as a function of $\lambda$ 
which is supposed to be a small parameter. 
So, the energy levels will be written in the form 
\begin{align}
E_{n_0}= (n_0+\frac{1}{2})\; \omega + a_1 \lambda + a_2 \lambda ^2 + \cdots .
\end{align}
where the numbers $a_i$  depend on $n_0$. 
At finite temperature,  the  pressure, or the thermodynamical potential $\Omega$, are functions of $\lambda$, 
and we are interested in the expansion of these functions in powers of $\lambda$. 
Then, the question we want to ask is {\it what controls the expansion.} 
We want $H_1$ to be small compared to $H_0$ in some sense. 
Let this be measured by a parameter,  $\gamma$:
\begin{align}
\gamma \sim \frac{H_1}{H_0} .
\end{align}
At this moment this is not very well-defined.
Let me specify this parameter a little bit better by taking some average values:
\begin{align}
\gamma \sim
\frac{\lambda \left< x^4 \right >}{\omega^2\left < x^2 \right >}
\sim
\frac{\lambda \left< x^2 \right >}{\omega^2}
\; ,
\end{align}
where I assume $ \left< x^4 \right > \sim \left < x^2 \right > ^2$. 
This $\gamma$ is my dimensionless expansion parameter. 
As you see, this depends on the strength of the coupling $\lambda$, and obviously
$\lambda$ should be small for the expansion to make sense. 
But it also depends on $\left < x^2 \right >$ and $\omega^2$.

At $T=0$, what is $\gamma$? 
At $T=0$ we have calculated $\left < x^2 \right > \sim 1/\omega$, so that 
\begin{align}
\gamma \sim
\frac{\lambda }{\omega^2} \frac{1}{\omega}=
\frac{\lambda}{\omega^3}
\; .
\end{align}
You can verify that  the dimension of $\lambda$ is the same as that of  $\omega^3$, so that  $\lambda/\omega^3$ is dimensionless, as it should.
If you are doing a perturbative calculation of ground state properties at zero temperature  
the calculation is valid provided that $\lambda/\omega^3$ is  small compared to one.

At $T \ne 0$, things are different.
In the high temperature regime, $T \gg \omega$, 
we have $\left < x^2 \right >_T \sim T /\omega^2$ and
\begin{align}
\gamma \sim
\frac{\lambda }{\omega^2} \left < x^2 \right > \sim
\frac{\lambda }{\omega^2} \frac{T}{\omega^2}=
\frac{\lambda }{\omega^3}\frac{T}{\omega}
\; .
\end{align}
The factor $\lambda /\omega^3$ is the familiar one, but there is
another factor which can be big when the temperature is big.
So, at finite temperature, it's not enough to have $\lambda / \omega^3 \ll 1$.
If the temperature is big enough, $\gamma$ may become of order unity, even if  $\lambda / \omega^3 \ll 1$.

What happens physically at finite temperature is that  the system expands, because states with large quantum numbers become occupied, leading to an increase of $\left< x^2 \right>$. 
The situation here is comparable to that of perturbation theory for the excited states at $T=0$. There the expansion parameter would be $\gamma\sim (\lambda/\omega^3) n_0$, which may become of order unity if $n_0$ is large enough.

\noindent{\bf Remark. }
One way to improve perturbation theory is to define an effective frequency as 
\begin{align}
\omega^2 \to \omega_{\rm eff}^2 
= \omega^2 + \# \lambda \left < x^2\right >
\; ,
\end{align}
that is, absorb part of the interaction into the ``unperturbed hamiltonian'', 
i.e., replace $x^4 \to x^2 \left < x^2 \right >$, 
and adjust $ \left < x^2 \right >$ self-consistently. 
This  leads to an approximation which is similar to the Hartree approximation of many-body physics. 

\subsubsection{Field theory}

Let's move now to field theory.
The field theory is almost identical except for one important feature, namely that a mode carries a given wavelength. 
We are thus led to distinguish between layers of fluctuations at different
wavelengths.

\begin{description}
\item{\bf Q}: I want to interrupt to make sure that 
this difficulty comes from some kind of divergence at small $\omega$. 
\item {\bf A}:
In the harmonic oscillator there 
is no  divergence because we have only one mode. 
The difficulty comes from the fact that, 
if the temperature is big, i.e., $T\gg \omega$, a large number of quanta are excited, and this may make the expansion parameter of order unity even if  the zero temperature expansion parameter $\lambda /\omega^3$ is small. 
If $\lambda/\omega^3 = 10^{-1}$ and if $T/\omega=100$,
you are expanding in powers of 10's --- that is the  problem. 
\item{\bf Q}: This factor comes from the singularity of the
  distribution function?  
\item {\bf A}: It comes indeed from expanding the distribution
$1/(e^{\omega/T}-1)$.
In the regime where the temperature is huge 
compared to the distance between the energy levels, 
we need to take into account many levels, all of which 
contribute  to 
produce a large value of $\left < x^2 \right>$. (A similar difficulty would occur if you were calculating corrections to the energy of an excited state with large $n_0$.)
\item{\bf Q}:
At high temperature the system becomes classical.
Is that the reason why you can use this mean-field approximation 
to treat the higher-order terms?

\item{\bf A}:
Well, this Hartree approximation that you are referring to can be used also at zero temperature.
It turns out that this is a rather good approximation,
even though it's more useful at finite temperature.
But I don't think that this is related to the classical approximation.
Of course, if you are in the classical regime, 
then $\left < x^2 \right >$ is big, and  it is
advantageous to do this.

\item{\bf Q}: I thought that the quantum fluctuations become less and
  less important ...

\item{\bf A}: Yes, that's right. 
In the expectation value, $\left < x^2 \right >  \sim 1+ 2n_T$, the first
term,  which is the vacuum fluctuation can be ignored if $n_T$ is large enough, which is the case at high temperature where $n_T\sim T/\omega\gg 1$.

\end{description}

So let's move now to the field theory. 
We replace $x^2$ by $\phi^2$, and $\langle\phi^2\rangle$ is now given by  an integral \begin{align}
\left < \phi^2 \right > 
= \int \frac{d^3 \k}{(2\pi)^3}
\frac{1+2 n_\k}{2\epsilon_\k}
\sim  \int \frac{d^3 \k}{(2\pi)^3}
\frac{n_\k}{\epsilon_\k}
\; , \label{eq:phi2}
\end{align}
where $n_\k = 1/(e^{\epsilon_\k/T}-1)$. 
I am also going to assume that $m=0$ so that $\epsilon_\k=k$. 
In the last (approximate) equality,  I have just ignored the vacuum fluctuations. 
Compare this to the formula (\ref{averagex2}) that we had before for the harmonic oscillator. 
The main difference is that we have now  a sum all over the modes which are labeled by the
momentum $\k$.

Now, I want to repeat the analysis of the expansion parameter;
I want to compare ``the kinetic energy'' and the ``the potential energy.'' 
The kinetic energy is 
\begin{align}
\left < (\bnabla \phi)^2 \right >
\;, 
\label{eq:kin}
\end{align}
and the potential energy is 
\begin{align}
\frac{\lambda}{4!}\left < \phi^4 \right > 
\sim
g^2 \left < \phi^2 \right > ^2
\;,
\end{align}
where I assumed that $\left<  \phi^4 \right>\sim \left<   \phi^2 \right>^2$,  which is good enough for the qualitative discussion that I want to present.

Now, comes an important remark. 
The integral in Eq.~(\ref{eq:phi2}) is dominated by the large momenta. 
One way to see that is that, in the absence of the statistical factor, it would be  quadratically  divergent.  
What the statistical factor does is to provide a cutoff at $k={\cal O}(T)$. 
Let us then consider the integral with an upper cutoff 
  $\kappa $:
\begin{align}
\left < \phi^2 \right >_\kappa \equiv \int^\kappa \frac{d^3 \k}{(2\pi)^3}
\frac{n_\k}{\epsilon_\k}
\; .
\end{align}
This integral, whatever $\kappa$ is, is always dominated by the
largest possible momenta,
and therefore what will contribute dominantly in this integral are 
the fluctuations which have momenta $\k$ of the order of $\kappa$. 
I'm going to call $\left < \phi^2 \right >_\kappa$ the ``contribution of fluctuations at scale $\kappa$". 
In the same spirit, I'm going to define the typical kinetic energy in (\ref{eq:kin}), 
at the same scale $\kappa$, as 
\begin{align}
\left < (\bnabla \phi)^2 \right >_\kappa 
\sim \kappa ^2 \left < \phi^2 \right >_\kappa 
\; .
\end{align}

Now, I have all the tools to define the expansion parameter $\gamma$,
as we did in the analysis of the harmonic oscillator. The big difference is that  now this 
depends on $\kappa$. I call $\gamma_\kappa$ the ratio the potential energy
$\sim {g^2 \left < \phi^2 \right >_\kappa ^2}$
to the  kinetic energy
${\kappa^2 \left < \phi^2 \right >_\kappa}$, that is, 
\begin{align}
\gamma_\kappa \sim 
\frac{g^2 \left < \phi^2 \right >_\kappa ^2}
     {\kappa^2 \left < \phi^2 \right >_\kappa}
\sim 
\frac{g^2 }{\kappa^2} \left < \phi^2 \right >_\kappa.
\end{align}

The rest of the discussion will be concerned with the analysis  of the 
typical momentum scales that appear  in the ultra-relativistic plasmas in the weak coupling regime. 
\subsection{Interplay of the various wavelengths}

\subsubsection{Self-coupling of hard modes}

There are  ``natural'' scales in the ultra-relativistic plasma; 
one scale, which I call ``hard", is $\kappa $ of the order of $T$. 
This is the scale for the typical plasma particles 
which are the modes of the field that carry  momenta of the order of the 
temperature $T$.   
Remember that the density of such particles goes like $T^3$, 
and the average distance between them is therefore of order $1/T$. Note that 
$1/T$ is also the typical de Broglie wavelength, so that these hard particles are always quantum (as are the photons in the blackbody radiation). 

Then I want to calculate $\left < \phi^2 \right >_\kappa$,
with $\kappa\sim T$.
This is an integral that we have already done. We get
\begin{align}
\gamma_T \sim  \frac{g^2}{T^2} T^2 \sim g^2
\; .
\end{align}
For these  particular modes,
there is therefore no problem:
The expansion parameter is $g^2$, so that  perturbation theory is OK, provided that $g$ is {\it small}..
(In fact, if you refer to a familiar quantum field theory
like QED, you know that the expansion parameter is 
$g^2/4\pi$ rather than $g^2$. )

\begin{table}
\begin{center}
\begin{tabular}{lllll}
\hline
 &$\quad$ hard  & $\quad$ soft  & $\quad$ ultra-soft &$\quad$ hard-soft$^*$  \\
\hline
$\kappa$ & $\quad T$ & $\quad gT$ &  $\quad g^2  T$ & $\quad gT$-$T$  \\
$\left < \phi^2 \right >_\kappa $&
           $\quad T^2$ &  $\quad gT^2$ &  $\quad g^2 T^2$ &  $\quad $--  \\
$\gamma_\kappa $ &
           $\quad  g^2$ & $\quad  g$ & $\quad 1$ & 1 \\
\hline
\end{tabular}
\caption{Expansion parameter for modes with typical scale $\kappa$.
*see text for definitions.}
\end{center}
\end{table}

\subsubsection{Self-coupling of soft modes}

In an ultrarelativistic plasma, there is another scale,
which is called the  {\it  soft scale}, 
 $\kappa \sim gT $. 
You will see why this  particular scale emerges dynamically. We already have an example of that scale with the thermal mass, which is  proportional to $gT$.

When  $\kappa = gT \ll T$,  
\begin{align}
\left < \phi^2 \right >_{\kappa = gT} \sim 
\int ^\kappa \frac{d^3 \k}{(2\pi)^3}\frac{n_\k}{\epsilon_\k}
\sim
\int ^\kappa k^2 d k \; \frac{T}{k}\, \frac{1}{k} 
\sim
T \times \kappa \sim gT^2
\; ,
\end{align}
so that
\begin{align}
\gamma_\kappa \sim \frac{g^2}{(gT)^2} \left < \phi^2 \right >_{\kappa=gT}
\sim
\frac{g^2}{(gT)^2}\,  gT^2 
\sim g 
\; .
\end{align}
You see an interesting phenomenon here. 
You see that $\gamma_{\kappa}$ is still a good expansion parameter, 
because it proportional to the coupling constant, which is a small number.
But you also see that, while we had $\gamma_T=g^2$ before,
now $\gamma_{gT}=g$.  
It means that the perturbation theory for the self-interaction 
of the soft modes is still valid,
but the expansion is not in powers of $g^2$ but
in powers of $g$, and so is less precise.  
This new expansion parameter, $g$ rather than $g^2$, where $g$ is the strength of the self-interaction between the soft modes,  is the origin of the odd powers of $g$ that we have encountered in the expansion of the pressure.

\subsubsection{Coupling between the soft and hard modes}

But there is something more, which is very important. 
The parameter $\gamma_{gT}\sim g$ compares the strength 
of the self interactions of the soft modes with their kinetic energy. 
Now, you may ask another question which is as relevant as this one.
You may ask what about the possible coupling 
between the soft fluctuations and the hard fluctuations. 

Kinetic energies of the soft modes are of order $(gT)^2$, 
while  the interaction energy between the soft and hard modes
is 
$\sim g^2 \left< \phi^2 \right >_T \sim g^2 T^2$. 
So the motion of a soft  mode, 
with wavelength of order $1/(gT)$, 
  is  non-perburbatively
renormalized by its coupling to the hard degrees of freedom. 
Here you cannot expand because the two effects are comparable.
So, this is a non-perturbative correction, which goes under the name of
{\it Hard Thermal Loops. }
The thermal mass which I have introduced the last time, and also
discussed today,  
is one particular example of this correction.

So you see, at the soft scale the landscape  complicates a bit.
We know that we can treat the soft modes perturbatively, 
although the perturbation theory is not as accurate as for the hard modes. 
But the propagation of the soft modes itself is  affected 
by their coupling to the hard modes in a non-perturbative fashion. 
In order to treat this  phenomenon, 
perturbation theory is not enough. 
But this is well under  control. 
The mass resummation that we have discussed  last time 
is a simple example of what needs to be done to handle this problem.

\begin{figure}
\begin{center}
\hfil
\includegraphics[width=5cm]{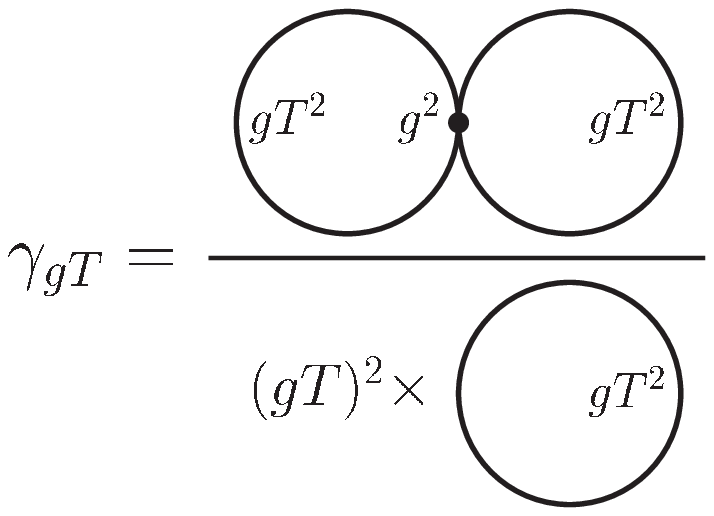}
\hfil
\includegraphics[width=5cm]{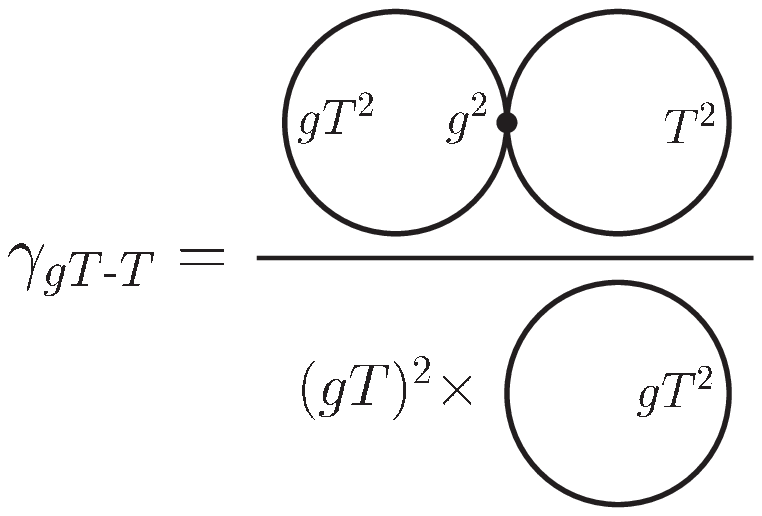}
\hfil
 \caption{
Schematic representation of the expansion parameter  $\gamma_{gT}$ 
for the soft modes $\kappa=gT$
and the coupling $\gamma_{gT-T}$ 
of the soft modes to the hard modes.}
\label{fig:gamma}
\end{center}
\end{figure}


\begin{description}
\item{\bf Q}: I thought that you defined the $\gamma_\kappa$ 
as the ratio of the kinetic energy to the potential energy of each mode
at the scale $\kappa$, right?

\item{\bf A}: Yes, this is what I've done before.
But here I'm defining {\it another} $\gamma$.
I look at a mode with wavelength $1/gT$, much bigger than the 
wavelength $1/T$ of the typical plasma particles.
When such a mode propagates, 
it can  interact with the hard modes through  loop corrections (see Fig.~\ref{fig:gamma}).
The present expansion parameter is coming from the comparison between the kinetic energy
of the soft modes, which is $(gT)^2$,  and the potential energy 
coming from the hard loop, $\sim T^2$, times the coupling $g^2$.
I'm arguing that they are of the same order of magnitude.
\end{description}

\subsubsection{Ultra-soft modes}
Now let me move on to the another  natural scale, and
confront a real catastrophy. 
This is the {\it ultra-soft  scale}, $\kappa \sim g^2 T$.
The reason why this scale occurs is that
it is at this particular momentum
that you  have the complete matching between   kinetic and potential energies.

At the scale $\kappa\sim g^2 T$, we have indeed
\begin{align}
\left< \phi^2 \right>_{g^2 T} \sim T \times \kappa = g^2 T^2, 
\end{align}
so that
\begin{eqnarray}
\left < (\bnabla \phi)^2 \right >_{g^2 T}
&\sim& (g^2 T)^2 \left< \phi^2 \right>_{g^2 T} 
\sim g^6 T^4
\; .
\end{eqnarray}
Then you see that the kinetic energy and the potential energy are of
the same order:   
\begin{align}
\gamma_{g^2 T} \sim 
\frac{g^2 \langle \phi^2 \rangle_{g^2 T}}
     {\kappa^2} \sim 1
\; .
\end{align}
{\it No expansion is possible!} 
In other words, the ultra-soft modes remain strongly coupled for
arbitrary small coupling. 
Even if the coupling is $10^{-23}$,
you cannot expand in the powers of $10^{-23}$ because the kinetic
energy and the potential energy are always  of the same order. 
Note  that the  contribution of the ultra-soft modes to the thermodynamic potential is
\begin{align}
g^2 \left < \phi ^2 \right > ^2_{g^2 T} \sim g^2 ( g^2 T^2 )^2 = g^6
T^4
\; .
\end{align}
This is the order $g^6 T^4$ at which QCD perturbation theory breaks down, as we have discussed earlier. 

You see that the difficulties with QCD perturbation theory at finite temperature can be understood from rather general considerations. 
The key point  here is that 
 ultra-relativistic plasma should be viewed as multi-scale systems. 
These are difficult to treat because, typically, approximations 
 devised for one particular scale
do not work for  other scales.

Perhaps you  have heard about 
``the strongly coupled quark-gluon plasma''
and about the AdS/CFT correspondence which allows us to calculate 
at infinite coupling. From the perspective that I have just outlined, 
the problem with finite temperature filed theory is not so much connected with the absolute strength of the coupling. The picture provided by the AdS/CFT correspondence is one in which all modes interact with infinite strength. What I have tried to argue is that this picture ignores the important fact that the effects of the interaction in the QCD plasma depend on the wavelength of the modes that one considers.

\begin{description}
\item {\bf Q}:
This discussion reminds me very much of the critical phenomena. 
Near the critical phenomena, we have long wavelength massless modes. 
They interacts very strongly and there are many scales intermingled. 
Is there any insight to be gained 
from the experience of the critical phenomena?

\item{\bf A}: Well, in fact, 
in the last two lectures, we will be dealing
 with one particular critical phenomenon, 
which is Bose-Einstein condensation. 
This very same problem will appear.
The techniques to solve it will be borrowed from what 
I will shall discuss next time.
Here, as I said before,  
the ultra-soft scale does not occur for the ordinary scalar field, 
because a thermal mass is generated. 
But if you are in the vicinity of a second order phase transition, 
you can adjust a parameter in such a way that
the effective mass at some temperature vanishes.
Then we are exactly in the same situation.

\item{\bf Q}:
I think the origin of the classical, long wavelength modes is very
similar.

\item{\bf A}:
You will see that the structure of the infrared divergences in the analysis that  I
will do quickly for the Bose-Einstein condensation is indeed very 
similar.  That is why I put the 
two topics together in these lectures. In the last lecture, I hope
to have some time to 
introduce you to some of the modern techniques of the renormalization group
to handle explicitly such multi-scale phenomena.

\item{\bf Q}:
Can the breakdown of perturbation theory at order $g^6$ be overcome by using other weak coupling techniques?
For example, 
the mass-screened perturbation theory,
hard thermal loop resummation, optimized perturbation theory,
or others?

\item{\bf A}:
 I have discussed here the origin of the difficulty.
I have not discussed at all what we should do in order to
overcome it.
We know that if you use perturbation theory
in a regime where soft momenta are integrated over
you should not be surprised to meet  infrared divergences
because the expansion that  you are trying to use makes no sense.
That we know {\it a priori} without doing any calculation.
So we know {\it a priori}  that we should do better.
 
Now what to do depends very much on what you want to calculate.
I will briefly comment next time 
on the application to the thermodynamics of QCD.
For the thermodynamics, hard degrees of freedom
are dominant.
So you expect that perturbation theory, plus some corrections,
will work, and indeed it does.
So, you can use screened perturbation theory,
you can use the  2PI formalism, you can do hard thermal loop resummation,
all these techniques will essentially deal with
this part of the problem.

If you want to address the situation described here, 
and if you want to calculate 
explicitly the contribution of the very long wavelength modes (the modes that I called ultra-soft),
then there is no other way, that I know of today, than to do 
lattice calculations.

But for the thermodynamics, observe that 
the long wavelength, small momentum,  modes have very small phase space,
so their contribution to the energy density and the pressure is presumably
 small. 
But  if you are asking  about  correlations
at long distances in the plasma, or perhaps transport phenomena,
you may need to worry about these modes.
Their analytical treatment remains an open issue. 

\item{\bf Q}:
You said that  the coupling between the soft and hard modes
becomes in a sense dangerous because the ratio becomes essentially one.
If you consider the coupling of the ultra-soft modes to the other modes,
what happens? 

\item{\bf A}:
This is an interesting question. I let you meditate about it. I shall just point out that the  scale $gT$  is uniquely determined by the requirement  that $\kappa^2\langle \phi^2\rangle_\kappa\sim g^2\langle \phi^2\rangle_\kappa\langle \phi^2\rangle_T$.

\end{description}
Next time I will show you the technique of the effective field theory. This will be directly 
relevant to what I will do for the Bose-Einstein condensation.
And I will also show you how the hard thermal loop emerges
in a dynamical context, i.e., from kinetic theory. This is an interesting perspective, which may be also of special interest to those of you who are  are working on kinetic equations.

\section{Lecture 4} \label{lecture4}
\subsection*{Summary of lecture 3}

One of the things that I have
emphasized in the previous lecture is that the quark-gluon plasma,
or more generally an ultra-relativistic plasma, can be viewed as a  multi-scale system: there is only one scale to start with, namely the temperature, but at weak coupling, other scales are generated dynamically. Such a plasma contains modes  with various wavelengths, in fact there is a
continuum of wavelengths. And the important point is that the effect of the interaction between
these modes depends very much on their wavelengths. In order to characterize
this interaction I have introduced the quantity  $\langle \phi^2 \rangle_\kappa$ 
which I referred to as the fluctuations at  scale $\kappa$: 
\begin{equation}
\label{Eq:def_of_phi}
	\langle 
		\phi^2 
	\rangle_\kappa 
\equiv 
	\int^\kappa 
		\frac{\mathrm{d}^3 k}{(2 \pi)^3} \frac{n_k}{k}
, \quad 
	n_k = \frac{1}{e^{k/T} - 1}
.
\end{equation}
(Remember that the dominant contribution to
this integral is determined by momenta which are of the order of the upper
momentum cutoff $\kappa$.) This allowed me to define  
an expansion parameter which controls perturbation theory, that  I called $\gamma_\kappa$:
\begin{equation}
\label{Eq:gamma_kappa}
	\gamma_\kappa 
= 
	\frac{g^2 \langle \phi^2 \rangle_\kappa}{\kappa^2}\sim \frac{g^2 T}{\kappa}
.
\end{equation} 

The discussion of the last lecture can be summarized in the table below. 
We have considered three particular scales, $T$, $gT$,
and $g^2 T$. I'm assuming here that $g$ is a small number, $g\ll 1$. (This is a condition which in fact does not need to be that strict because 
there are factors $2 \pi$, for instance, that  have been left out. )\begin{center}
\begin{tabular}{|l|c|c|c|} \hline
	 & 
	$\langle \phi^2 \rangle_\kappa$ & 
	$g^2 \langle \phi^2 \rangle_\kappa$ & 
	$\gamma_\kappa$ 
	\\ \hline
	$\kappa = T$    & $T^2$    & $g^2 T^2$ & $g^2$ \\ \hline
	$\kappa = gT$   & $gT^2$   & $g^3 T^2$ & $g$   \\ \hline
	$\kappa = g^2T$ & $g^2T^2$ & $g^4 T^2$ & $1$   \\ \hline
\end{tabular}
\end{center}
The fluctuations at scale $T$ have interactions controlled by 
$g^2$. For these, the perturbative expansion behaves as at zero temperature (and is really an expansion in powers of $g^2/4\pi$). If we move down to the scale
$gT$ then we have again an expansion parameter which is small. But it is 
of order $g$ instead of order $g^2$. That means that perturbation theory
for these particular set of wavelengths will be less precise. The expansion will not be in powers of $g^2$ but will contain odd 
powers of $g$. The contributions of order  $g^3$ will be at the center of our  discussion  
today. And finally at the scale $g^2 T$ the expansion parameter is of order one,
which means that the very long wavelength modes will remain strongly coupled, however  small $g$ may be.

I also made another remark concerning the scale $gT$, which will introduce today's discussion. The remark is that if  I compare the kinetic energy of a mode with momentum $\kappa\sim gT$, with the contribution of its interaction with the  fluctuations  at scale $T$, I find that they are of the same order of magnitude, namely:
\begin{equation}
\label{Eq_lec4:kappa_square}
	\kappa^2 \sim g^2 T^2 
\sim
	g^2 \langle \phi^2 \rangle_T
\end{equation}
This indicates that even though the self-interactions of the modes with momentum  $gT$ are of order $g$, the motion of these modes 
is  strongly modified by their coupling with the fluctuations at the scale $T$.
This feature  is at the 
heart of what is known in the literature as the ``hard thermal loops''. And I will 
tell you a good deal about these hard thermal loops today. 

What I would like to 
do is to explain how we can handle this 
coupling between soft and 
hard degrees of freedom. This will be done by introducing an important 
construct which is that of effective field theories.
I shall do that both in Euclidean, or imaginary time formalism, and  also in real time, where the effective theory takes the form of 
a kinetic theory. This program would require a whole set of lectures in itself, which of 
course I cannot do in one afternoon, so I shall have to skip some details, especially in the second part of the lecture. 

\subsection{Effective theory}

Let me now discuss effective theory.
What we have to deal with is a situation 
where  degrees of freedom with different wavelengths  are coupled together and 
interact differently depending on their momenta. I'm going to 
approach the problem within the path integral formalism. 
Let me remind you the formula we have for the
partition function
\begin{equation}
\label{Eq_lec4:partition_function}
Z = 
	\mathrm{Tr} \, e^{- \beta H} 
= 
	e^{- \beta \Omega} 
= 
	\int_{\phi (\beta , \mathbf{x}) = \phi (0 , \mathbf{x})} D \phi \,
		\, 
		e^{- S_\mathrm{E} }
,
\end{equation}
where
\begin{equation}
\label{Eq_lec4:Euclidian action}
S_\mathrm{E} = 
	\int^\beta_0 \mathrm{d} \tau 
	\int         \mathrm{d}^3 x
		\,
		\mathcal{L} (\phi , \partial \phi )
.
\end{equation}
What I am going to do is a more elaborate version of something that I have already
introduced. Remember two lectures ago. We discussed the situation where $\beta$
goes to $0$. If $\beta$ goes to $0$, and if nothing singular happens then,
we can ignore the time dependence of the field.
Then $S_E$ simplifies since the integration over time gives just a factor $\beta$:
\begin{equation}
\label{Eq_lec4:classical Euclidian action}
S_\mathrm{E} 
\to 
	\beta 
	\int \mathrm{d}^3 x 
		\left( 
			\frac{1}{2} (\nabla \phi)^2 
			+ 
			\frac{m^2}{2} \phi^2 
			+ 
			\frac{\lambda}{4!}\phi^4
		\right)
.
\end{equation}
Since $\beta$ is the inverse of the 
temperature, this corresponds to the high temperature limit. In this limit, the quantum 
filed theory reduces to a classical three dimensional field theory.

What I want to do now is essentially an elaboration of this remark.
I will often refer to this high temperature limit as the ``classical field 
approximation''. As I have also argued earlier, this is the approximation that is obtained by 
ignoring the non vanishing Matsubara frequencies when one expands the field 
in Fourier space. That is, this is the approximation where one keeps only the vanishing  Matsubara 
frequency component of the field. 

So let me  consider the Fourier 
expansion of the field:
\begin{align}
	\phi (\tau , \mathrm{x} )
&=
	\frac{1}{\beta}
	\sum_n
		e^{-i \omega_n \tau} \phi_n (\mathrm{x})
\notag
\\
&=
	T \phi_0 (\mathrm{x})
	+
	T \sum_{n \neq 0}
		e^{-i \omega_n \tau} \phi_n (\mathrm{x}).
\label{Eq_lec4:Fourier_expansion_of_field}
\end{align}
What is done in Eq.~(\ref{Eq_lec4:classical Euclidian action}) is just 
keeping $\phi_0$ and ignoring all the rest.
What I would like to do now is to show you how we can take into account the rest. 
I do this in the framework of a general $\phi^4$ scalar field theory. (The procedure generalizes to more complicated theories, in particular to QCD, but we shall not have time to discuss this in detail.)  This approximation will play a crucial role in  the solution of the problem that I shall discuss in the next two lectures,  namely Bose-Einstein condensation.

How are we going to take into account the effects of the components with non vanishing $\omega_n$? 
Note that I can make a change of variables, going from a path integral over 
field configurations in spatial coordinates $\times$ time,  towards configurations 
in spacial coordinates $\times$ frequency.  In other words I can write the measure of the path integral 
as a product $D \phi_0\times \prod_{n\ne 0}D \phi_n$. (I am concerned here 
just with the  time dependence, and I do not write explicitly the dependence of the field on the spatial coordinates.)
Then I can rewrite the path integral as follows
\begin{align}
{\cal Z} &=
	\int 
		D \phi_0  
			\prod_{n \neq 0} D \phi_n 
		e^{
			-S_\mathrm{E} \left[ \phi_0 , \phi_n \right]
		}
\notag \\
&=
	\int D \phi_0
		\,
		e^{
			-S_\mathrm{eff} \left[ \phi_0 \right]
		}\; ,
\label{Eq_lec4:path_integral_of_Seff}
\end{align}
where
\begin{equation}
\label{Eq_lec4:Seff}
	e^{
		-S_\mathrm{eff} \left[ \phi_0 \right]
	}
\equiv
	\int
			\prod_{n \neq 0} D \phi_n 
		e^{
			-S_\mathrm{E} \left[ \phi_0 , \phi_n \right]
		}\;.
\end{equation}
How can we calculate $S_\mathrm{eff}[\phi_0]$? In fact, you know enough to 
be able to calculate $S_\mathrm{eff}$ from what I have told you already. 
Look indeed at Eq.~(\ref{Eq_lec4:partition_function}), and compare it with Eq.~(\ref{Eq_lec4:path_integral_of_Seff}). 
You see that $\exp\left( -S_\mathrm{eff} \right)$ is a partition function, for a system 
in which $\phi_0$ appears as  a frozen, given parameter. So we know how to calculate that. 
We know that the thermodynamic potential $\Omega$ in Eq.~(\ref{Eq_lec4:partition_function}) is given by the set of all connected Feynman diagrams. Similarly, 
\begin{equation}
\label{Eq_lec4:how_to_calculate_Seff}
	S_\mathrm{eff} \left[ \phi_0 \right]
\equiv
	\textrm{set of all connected Feynman diagram with external lines}
	\;
	``\phi_0 "
\end{equation}

We know the leading order contribution to $S_{eff}[\phi_0]$: this is the ``tree-level'' action, obtained by evaluating the original classical action with the field $\phi(\tau, \mathbf{x})=T\phi_0(\mathbf{x})$, and it is identical to Eq.~(\ref{Eq_lec4:classical Euclidian action}). In fact it is convenient to rescale $\phi_0$  by a factor $\sqrt {T}$, that is to define  $\phi_0$  as
\begin{equation}
\label{Eq_lec4:rescale_of_filed}
	\phi_0(\mathbf{x})
=
	\sqrt{T}
	\int_0^\beta \mathrm{d} \tau\,
		\phi (\tau, \mathbf{x} )
\end{equation}
With this definition, you see that a term such as $\beta \phi^2$ in Eq.~(\ref{Eq_lec4:classical Euclidian action}) would lead a term $T\phi_0^2$ with the  definition of Eq.~(\ref{Eq_lec4:Fourier_expansion_of_field}). 
The new definition absorbs the  factor $T$ into $\phi_0^2$, leaving the simpler expression for the leading order effective action:
\beq\label{Seffphi0}
S_{eff}^{(0)}[\phi_0]= \int \mathrm{d}^3 x 
		\left( 
			\frac{1}{2} (\nabla \phi_0)^2 
			+ 
			\frac{m^2}{2} \phi_0^2 
			+ 
			\frac{\lambda T}{4!}\phi_0^4
		\right).
		\eeq

\begin{figure}
	\centering
	\includegraphics[width=6cm,clip]{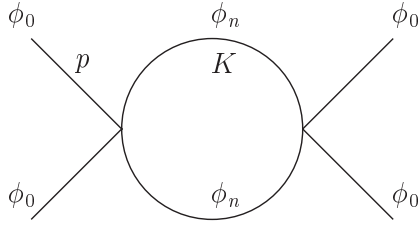} \\
	\caption{One loop diagram with external lines $\phi_0$, and  internal lines
			  $\phi_n$.}
	\label{Fig_lec4:effective_phi4}
\end{figure}

Corrections to the leading order are generated, for instance, from a loop expansion in the theory for the hard modes,  whose  Euclidean action is given by
\beq\label{Eq_lec4:Fourier_expansion_of_field2}
S_E[\phi_0,\phi_n]&=&\sum_{n\ne 0} \left\{ \frac{1}{2} \nabla \phi_{n}\nabla\phi_{-n}  +\frac{1}{2}\left (m^2+\lambda T\phi_0^2+\omega_n^2   \right)\phi_n\phi_{-n}\right\}\nonumber \\
&+&\frac{\lambda T\phi_0}{3!} \sum_{nmk} ^\prime\phi_n\phi_m\phi_k+\frac{\lambda T}{4!} \sum_{nmkl} ^\prime\phi_n\phi_m\phi_k\phi_l,
\eeq
where the fields $\phi_n$ have been rescaled by the same  factor $\sqrt {T}$ as $\phi_0$. For instance a one loop contribution is easily obtained from the part of $S_E[\phi_0,\phi_n]$ that is written in the first line of Eq.~(\ref{Eq_lec4:Fourier_expansion_of_field2}), and which is quadratic in $\phi_n$. The one loop diagrams produce a correction to the mass (to be discussed later), a correction to the four-point function (see the Fig.~\ref{Fig_lec4:effective_phi4}), a contribution to the six point function, etc.  There will be other corrections generated by the terms in the second line of Eq.~(\ref{Eq_lec4:Fourier_expansion_of_field}), but these will be of higher order in $\lambda$. For instance the first correction to the mass generated by the term with three $\phi_n$ will be of order $\lambda^2$ at least. 

At this point I need to specify more carefully the  separation between what is meant by ``soft''  and ``hard''. Naively 
we may attempt to call soft the mode with $n=0$, 
and hard the modes with $n \neq 0$. This is what we have done so far, but this is not enough. 
If we just leave things at this stage, we are going to generate 
 an effective action which will contain arbitrary powers of $\phi_0$,
but which will be mostly non local. For instance, consider the diagram in Fig.~\ref{Fig_lec4:effective_phi4}. As it stands, it is a complicated function of the momenta carried by the various lines labelled $\phi_0$. We would like however to continue working with a local effective action, that is, we would like to be able to consider this diagram as a correction to the coupling constant.  This can be achieved by 
introducing a separating scale $\Lambda$, chosen so that (here I rename $\lambda\to  g^2$)
\begin{equation}
\label{Eq_lec4:constraint_of_separation_scale}
gT \ll \Lambda \ll 2 \pi T,
\end{equation}
because I'm now assuming ``soft'' will involve some energy scale of order 
$gT$, while ``hard'' will involve some energy scale of order $2 \pi T$.
Then I redefine the separation between soft and hard in the following way. In the soft sector I consider the mode $n=0$, but I   also  
assume that the momenta are limited by $\Lambda$. So, when I calculate 
an integral in the effective theory,  I do all the 
momentum integrations up to the scale $\Lambda$. On the other hand, 
the hard sector contains all the modes with non vanishing Matsubara frequencies, plus 
the sector with the large momentum components ($k \ge \Lambda$) of the mode $n=0$.  We have now a cleaner separation. In all 
cases you see that, with  $K^2 = \omega_n^2 + k^2$, the separation which I 
have introduced guarantees that the hard momenta satisfy $K \ge \Lambda$ 
and the soft momenta satisfy $K \le \Lambda$.

\begin{center}
\begin{tabular}{|l|l|} \hline
	hard & $n \neq 0 \quad$ + $\quad n=0$ and $|k| \gtrsim \Lambda$ \\ \hline
	soft & $n=0, |k| \lesssim \Lambda$ \\ \hline
\end{tabular}
\end{center}
Thus defined, the effective theory for soft external momenta is valid only when the  momentum $p$ carried by the field $\phi_0$ is smaller 
than $\Lambda$.  On the other hand, the loop integrals are  dominated by hard momenta.

Now, we have gained something with respect to the 
argument of locality  that I alluded to earlier. We know that the momenta which are inside the 
loop are all going to be large compared to the momenta outside. The locality 
will result from the fact that one can expand in $p/K$ where $K$ is a typical 
momentum inside the loop, and $p$ an external momentum. The expansion in powers of $p/K$ means the expansion 
in the field and its derivatives. Therefore that means that the effective action 
is going to be given by a  series of monomials built from the field $\phi_0$ and its derivatives:
\begin{equation}
\label{Eq_lec4:effective_Lagrangian}
	\mathcal{L}
=
	a \phi_0^2
	+
	b \left( \nabla \phi_0 \right)^2
	+
	c \phi_0^4
	+
	d \phi_0^6
	+
	e \left( \phi_0 \nabla \phi_0 \right)^2
\end{equation}
Note that this expansion preserves the symmetry of the original lagrangian 
under $\phi_0 \to - \phi_0$. In principle we have 
an infinite collection of terms, and we need a guiding principle to 
truncate this expansion, otherwise we cannot do calculations. We shall use here a weak coupling approximation,  where the successive terms in the effective action can be calculated using perturbation theory. 

Let me give you the form of the effective action with more standard notation:
\begin{equation}
\label{Eq_lec4:expansion_of_Seff}
	S_\mathrm{eff}
=
	f_\Lambda
	+
	\int \mathrm{d}^3 x
		\left(
			\frac{1}{2}
			\left(
				\nabla \phi_0
			\right)^2
			+
			\frac{1}{2} M^2_\Lambda \phi_0^2
			+
			\frac{g_3^2}{4!} \phi_0^4
			+
			\frac{h}{6!} \phi_0^6
			+
			\cdots
		\right), 
\end{equation}
where $f_\Lambda$ represents the contribution of the hard mode to the thermodynamical potential. Note that the dimensions of the various terms are characteristic of a three dimensional field theory: 
$\phi_0$ have mass dimension $1/2$, $\left[ \phi_0 \right] = M^{\frac{1}{2}}$,  $g_3^2$ has a dimension of mass, $\left[ g_3^2  \right] = M$,  and $h$ is dimensionless, $\left[ h      \right] = M^0$.

The partition function of the system can  be written as
\begin{equation}
\label{Eq4_lec4:partition_function_in_terms_of_Seff}
{\cal Z} =
	\int D \phi_0\,
		\,
		e^{
			-S_\mathrm{eff} [ \phi_0 ]
		}
.
\end{equation}
If $S_{eff}$ is calculated exactly, this is identical to Eq.~(\ref{Eq_lec4:partition_function}). The strategy now is to obtain ${\cal Z}$ from an approximate $S_\mathrm{eff} [ \phi_0 ]$. 
\begin{figure}
	\centering
	\includegraphics[width=5cm,clip]{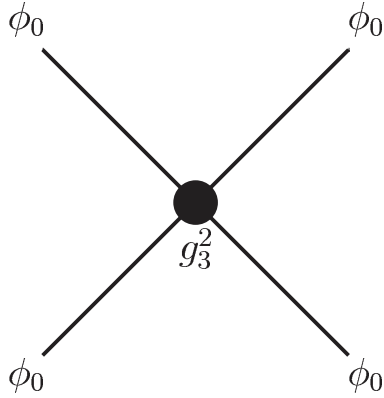} \\
	\caption{Tree level diagram contributing to $g_3^2$.}
	\label{Fig_lec4:g3}
\end{figure}%
The coefficients in $S_{eff}$ can be calculated using perturbation theory. 
Some are easy to get, because they exist already at tree level. This is the case for instance of $g_3^2$ (the corresponding tree level diagram is given by Fig.~\ref{Fig_lec4:g3}, whose value can be  red off Eq.~(\ref{Seffphi0}):
\begin{equation}
\label{Eq4_lec4:value_of_g_3^2}
g_3^2 = g^2 T
\end{equation}
The next vertex, 
$h$, has no contribution at the tree level because there is no term like 
$\phi^6$ in the original lagrangian. The term of order 6 in $\phi_0$ 
is  induced, at leading order,  by the one loop diagram of Fig.~\ref{Fig_lec4:h}.
\begin{figure}
	\centering
	\includegraphics[width=6cm,clip]{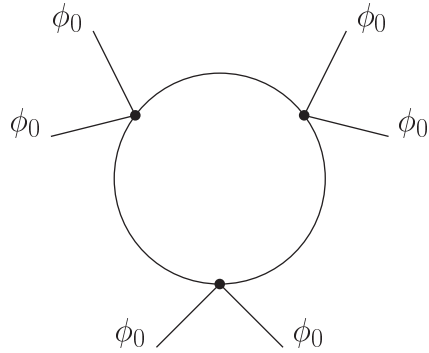} \\
	\caption{One loop diagram contributing to the $\phi_0^6$ vertex of the effective theory.}
	\label{Fig_lec4:h}
\end{figure}%
This diagram is proportional
to $g^6$, and is therefore subleading if  $g$ is a small number. 

Everything is fine so far with the general strategy. We have a very 
systematic way of calculating the contributions to the effective action. 
One can do that in perturbation theory, i.e., one can calculate the coefficients of the effective action from Feynman diagrams. Note that there is no contradiction here with the fact that perturbation theory cannot be used to calculate within the effective theory. I am using perturbation theory here in 
the sector where it is safe, because  in calculating the coefficients of the effective theory, I am only integrating over the hard modes. To calculate with the effective theory, I have to do something more sophisticated.

\subsection{Calculation of the thermal mass $M^2$}

What I shall do now is perform a simple calculation using the effective theory, in order to illustrate how things work. I shall calculate the thermal mass. The calculation will proceed in two steps. First, I shall calculate 
explicitly the  one-loop correction to the coefficient of $\phi_0^2$ in the effective action (\ref{Eq_lec4:expansion_of_Seff}). Then, I shall use the effective theory to calculate the correction due to the soft modes.  
One issue that I want to address is that of the 
dependence of the results on the arbitrary scale $\Lambda$ which comes in as soon as we consider loop corrections. I shall verify that when the calculation of a physical observable is correctly performed, this dependence disappears from the final result. 

\subsubsection{Contribution of the hard modes}

\begin{figure}
	\centering
	\includegraphics[width=4cm,clip]{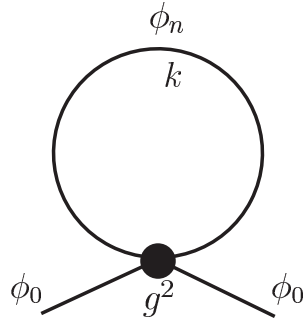} \\
	\caption{Tadpole diagram that contributes to the coefficient of $\phi_0^2$ in the effective theory}
	\label{Fig_lec4:M2_leading}
\end{figure}

The coefficient of $\phi_0^2$ in the effective action is  given by all the Feynman diagrams which have two external $\phi_0$  lines. At leading order, there is a single diagram, often called the  ``tadpole'' diagram, displayed in Fig.~\ref{Fig_lec4:M2_leading}. 
This is easy to calculate. We have already done so a number of times. But now, we have to pay attention to the fact that the loop integral runs over hard momenta only. Thus, we have\begin{equation}
\label{Eq4_lec4:calculation_of_M^2_A}
	M^2 (\Lambda)
=
	\frac{g^2}{2} T
	\sum_n
		\int \frac{\mathrm{d}^3 k}{(2 \pi)^3}
			\frac{
				1 - \delta_{n0} + \delta_{n0} \theta (k - \Lambda)
			}{
				\omega_n^2 + k^2,
			}
\end{equation}
where I have taken into account  that 
the hard degrees of freedom are all the modes with $n \neq 0$ (hence the contribution proportional to  $1 - \delta_{n0}$),  as well as the contribution from the mode $n = 0$ with momenta bigger 
than the dividing scale $\Lambda$ (hence the term  $\delta_{n0} \theta (k-\Lambda)$).
Now, use the following relation between $\theta$ functions:
\begin{align}
	\theta ( k - \Lambda ) + \theta ( \Lambda - k )
&=
	1,
\notag \\
	1 - \delta_{n0} 
	+ 
	\delta_{n0} 
	\left(
		1 - \theta ( \Lambda - k )
	\right)
&=
	1 - \delta_{n0} \theta ( \Lambda - k ),
\notag
\end{align}
and rewrite the integral in (\ref{Eq4_lec4:calculation_of_M^2_A}) as
\begin{align}
	M^2 (\Lambda)
&=
	\frac{g^2}{2} T
	\sum_n
		\int \frac{\mathrm{d}^3 k}{(2 \pi)^3}
			\frac{
				1 - \delta_{n0} \theta (\Lambda - k)
			}{
				\omega_n^2 + k^2
			}
\notag \\
&= 
	\frac{g^2}{2} T
	\sum_n
		\int \frac{\mathrm{d}^3 k}{(2 \pi)^3}
			\frac{1}{\omega_n^2 + k^2}
	-
	\frac{g^2}{2} T
	\int^\Lambda \frac{\mathrm{d}^3 k}{(2 \pi)^3}
		\frac{1}{k^2}.
\label{Eq4_lec4:calculation_of_M^2_B}
\end{align}
The interpretation of the first line of this equation is simple: the term $ \delta_{n0} \theta (\Lambda - k)$ explicitly removes from the loop integral the contribution of the soft momenta, which needs to be calculated more accurately. This will be done by using the effective theory. 
By performing  the sum over the Matsubara frequencies, one obtains
\begin{equation}
	M^2 (\Lambda)
=
	\frac{g^2}{2} T
	\int \frac{\mathrm{d}^3 k}{(2 \pi)^3}
		\frac{1 + 2 n_k}{2k}
	-
	\frac{g^2}{2} T
	\int^\Lambda \frac{\mathrm{d}^3 k}{(2 \pi)^3}
		\frac{1}{k^2}.
\label{Eq4_lec4:calculation_of_M^2_C}
\end{equation}
As we have done earlier, we shall drop the term which is ultraviolet divergent and independent of the temperature (this goes away with a standard ultraviolet renormalization of the  mass at zero temperature). 
The result can be then written as follows 
\begin{equation}
	M^2 (\Lambda)
=
	\frac{g^2 T^2}{24}
	\left(
		1 - \frac{\Lambda}{T} \frac{6}{\pi^2}
	\right),
\qquad
	\Lambda \ll T.
\label{Eq4_lec4:calculation_of_M^2_D}
\end{equation}
Since  $\Lambda$ is much smaller than $T$, 
the second term of (\ref{Eq4_lec4:calculation_of_M^2_D}) can be viewed as a correction,
but which depends on $\Lambda$. As we shall see, this  $\Lambda$ dependence will cancel against an analogous contribution from the effective theory calculation.

\subsubsection{Contribution of the soft modes}
\begin{figure}
	\centering
	\includegraphics[width=5cm,clip]{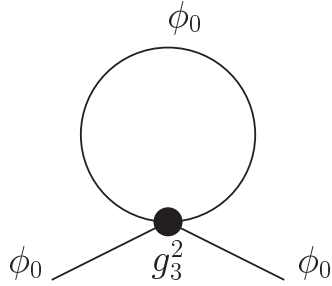} \\
	\caption{Tadpole diagram in the effective theory. The loop involve the propagator of the field  $\phi_0$}
	\label{Fig_lec4:M2_correction}
\end{figure}

What I have done so far is to calculate the coefficient $M_\Lambda^2$ in
the effective action. What I want to do next is to calculate the 
correction to the thermal mass in the effective theory. What is the correction to the mass? This is the correction that results from the self interaction of the field $\phi_0$. The correction is given by the same diagram as in Fig.~\ref{Fig_lec4:M2_leading}, but redrawn in Fig.~\ref{Fig_lec4:M2_correction} in order to emphasize the elements of the calculation. The diagram in Fig.~\ref{Fig_lec4:M2_correction} is a 
 diagram in the effective theory. So the vertex is $g_3^2$. 
And the loop integral involves  soft modes only, that is, the  momenta are limited to $\Lambda$. As for the propagator, it's inverse  can be read off  the 
effective action itself in Eq.~(\ref{Eq_lec4:expansion_of_Seff}): $k^2 + M_\Lambda^2$. The calculation then proceeds as follows
\beq
	\delta M^2
&=
	\frac{g_3^2}{2}
	\int^\Lambda \frac{\mathrm{d}^3 k}{(2 \pi)^3}
		\frac{1}{k^2 + M_\Lambda^2}
\notag \\
&=
	\frac{g_3^2}{4 \pi^2}
	\int^\Lambda_0 \mathrm{d} k
		\,
		\frac{k^2}{k^2 + M_\Lambda^2}
\notag \\
&=
	\frac{g_3^2}{4 \pi^2} \Lambda
	\left[
		1 - \frac{M_\Lambda} {\Lambda} \tan^{-1} \frac{\Lambda}{M_\Lambda}
	\right].
\label{Eq4_lec4:correction_of_the_mass_A}
\eeq
Remember that $M_\Lambda$ is of order $g T$ (see Eq.~
(\ref{Eq4_lec4:calculation_of_M^2_D})), so that $
\Lambda \gg M_\Lambda \sim g T
$.
By expanding the second term of (\ref{Eq4_lec4:correction_of_the_mass_A})
for large  $\Lambda/M_\Lambda$, and keeping only the leading order correction in $M_\Lambda$, that is, $M^2_\Lambda\approx M_\Lambda^2=g^2T^2/24$ ($g_3^2 = g^2 T$),  one obtains
\begin{equation}
\label{Eq4_lec4:correction_of_the_mass_B}
	\delta M^2
\simeq 
	\frac{g^2 T \Lambda}{4 \pi^2}
	-
	\frac{g^2}{8 \pi} MT. 
\end{equation}
It follows that the physical mass  is given by 
\begin{align}
	M^2 \to M^2 ( \Lambda ) + \delta M^2
&=
	\frac{g^2 T^2}{24}
	\left(
		1 - \frac{\Lambda}{T} \frac{6}{\pi^2}
	\right)
	+
	\frac{g^2 T \Lambda}{4 \pi^2}
	-
	\frac{g^2}{8 \pi} MT
\notag \\
&=
	\frac{g^2 T^2}{24}
	-
	\frac{g^2 T}{8 \pi}
	\frac{g T}{ \sqrt{24} }.
\label{Eq4_lec4:mass_of_effective_theory}
\end{align}
You see that, as anticipated,  the $\Lambda$ dependent term  coming from the parameter 
of the effective theory, $M_\Lambda$, and that coming from the cutoff 
in the loop integral within the effective theory, precisely 
cancel. The correction, of order $g^3 T^2$, agrees with the calculation done in  the
second lecture (see Eq.~(\ref{Sigmaring})).

I have presented a very simple example of the construction of the effective theory at 
the scale $gT$.  In the  particular context of QCD, such an effective theory (also called dimensional reduction), has been pushed to a 
 high degree of accuracy. When I mentioned earlier in the lecture 
the calculation up to $g^6$ or $g^6 \log g$ for the QCD pressure, these 
were obtained by relying on such techniques (see for instance \cite{Kajantie:2002wa} and references therein). Of course in QCD, these techniques are more 
elaborate. In a gauge theory, you cannot simply put a cut off on integrals, as we did. You have to use more sophisticated regulators. The 
technicalities are more difficult to master. But the basic 
concepts can be understood from  the simple scalar field theory discussed in this lecture.

\subsection{Real time Hard Thermal Loops}

In this second part of the lecture, I would like to give you another perspective on hard thermal loops, using kinetic theory.  
As we shall see, kinetic theory emerges as the effective theory that allows us to efficiently handle the coupling between hard and soft degrees of freedom in ultrarelativistic plasmas.  I shall not proceed through a systematic derivation, which would require more lectures, but shall try to indicate the main steps in such a derivation and emphasize the main physical aspects. 

\subsubsection{Real time propagators}

As a preliminary, I would like to comment  about connections between the imaginary time  and the real time formalisms, and  in particular remind you of some relations based on the analyticity of the propagators. 
In real time, we define
\begin{eqnarray}
&&G(t_1,t_2)=\langle T\phi(t_1)\phi(t_2)\rangle =\frac{1}{Z}{\rm Tr}\bigl[e^{-\beta H}T(\phi(t_1)\phi(t_2))\bigl]\\
&&\langle T\phi(t_1)\phi(t_2)\rangle =\theta (t_1-t_2)\langle\phi(t_1)\phi(t_2)\rangle + \theta (t_2-t_1)\langle\phi(t_2)\phi(t_1)\rangle
\end{eqnarray}
where I have omited the spatial coordinates (which play no role in the discussion) in order to alleviate the notation. 
We also define  $G^{>}(t_1,t_2)=\langle\phi(t_1)\phi(t_2)\rangle$ and $G^{<}(t_1,t_2)=\langle\phi(t_2)\phi(t_1)\rangle$. \\

\begin{figure}
\centering
\psfrag {0}{$0$}
\psfrag {-beta}{$-\beta$}
\psfrag {Ret}{Re $t$}
\psfrag {Imt}{Im $t$}
\includegraphics[width=8cm,clip]{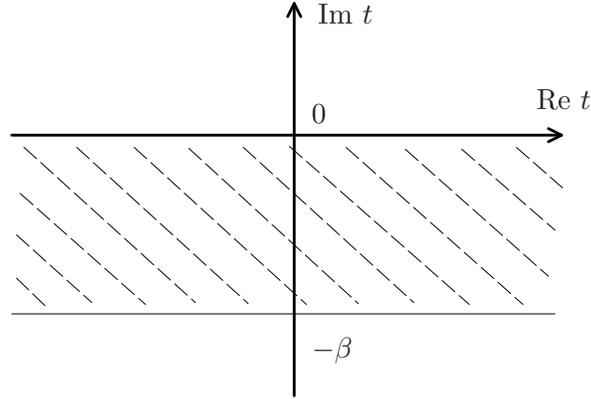}
\caption{Analyticity domain  of $G^{>}(t)$}
\label{akamatsu4-1}
\end{figure}

Let us focus on $G^{>}(t_1,t_2)$. By making explicit the time dependence, and expanding on a complete set of eigenstates of the hamiltonian, we get
\begin{eqnarray}
G^{>}(t_1,t_2)&=&\frac{1}{Z}{\rm Tr} \bigl(
e^{-\beta H}e^{iHt_1}\phi\, e^{-iHt_1}e^{iHt_2}\phi \,e^{-iHt_2}\bigr)\nonumber\\
&=&\frac{1}{Z}\sum_{n,m}e^{-\beta E_n}e^{iE_n t_1}\langle n|\phi|m\rangle e^{-iE_m t_1}e^{iE_m t_2}\langle m|\phi |n\rangle e^{-iE_n t_2}\nonumber\\
&=&\frac{1}{Z}\sum_{n,m}e^{-\beta E_n}e^{i(t_1-t_2)E_n}e^{-i(t_1-t_2)E_m}\bigl|\langle n|\phi |m\rangle\bigr|^2, 
\end{eqnarray}
or, setting $t_1-t_2=t$, 
\begin{eqnarray}
G^{>}(t)=\frac{1}{Z}\sum_{n,m}\bigl\{
e^{-\beta E_n}e^{iE_nt}e^{-iE_mt}\bigl|\langle n|\phi|m\rangle\bigr|^2
\bigr\}.
\end{eqnarray}
In this expression, we can, as we have already done several times in another context, set $it=\tau$ or $t=-i\tau$.
We then get
\begin{eqnarray}\label{Ganaltau}
G^{>}(-i\tau )=\frac{1}{Z}\sum_{n,m}\bigl\{e^{-\beta E_n}e^{\tau E_n}e^{-\tau E_m}\bigl| \langle n|\phi |m\rangle\bigr|^2\bigr\}.
\end{eqnarray}
Let us see under which conditions this substitution is legitimate, or more generally, under which conditions the time can be given an imaginary part. 
In most cases, the convergence of the sum will be controlled by the exponential factors.
If $\tau>\beta$, $e^{\tau E_n}$ will win compared to $e^{-\beta E_n}$ and the sum will explodes.
However if $\tau <\beta$, $e^{-\beta E_n}$ dominates and this will kill the other terms as $E_n$ gets large.
Therefore the sum over states in Eq.~(\ref{Ganaltau}) that allows the calculation of 
$G^{>}(t=-i\tau)$ is  finite if $\tau<\beta$. One concludes that $G^{>}(t)$ is an analytic function of $t$   in the strip $-i\beta<{\rm Im} t<0$, as indicated in Fig.~\ref{akamatsu4-1}.

\begin{figure}
\centering
\psfrag{Imomega}{Im $\omega$}
\psfrag{Reomega}{Re $\omega$}
\psfrag{iomegan}{$i\omega_n$}
\includegraphics[width=8cm, height=6cm, clip]{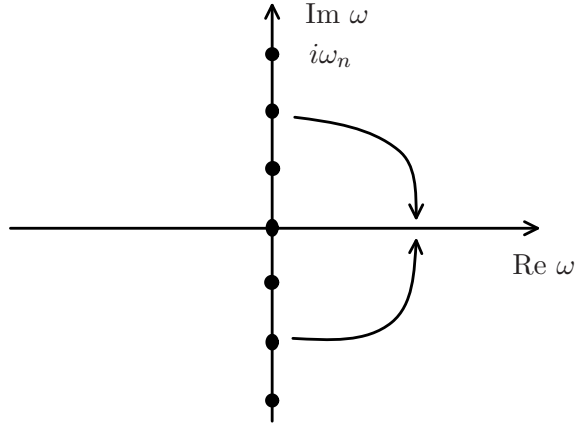}
\caption{Analytic continuation in the complex $\omega$ Plane}
\label{akamatsu4-2}
\end{figure}

After Fourier transform, the propagator enjoys also analyticity properties in the frequency complex plane. Recall the form of the propagators in terms of Matsubara frequencies 
\begin{eqnarray}
\frac{1}{\omega_n^2+k^2+m^2}=\frac{1}{-(i\omega_n)^2+k^2+m^2}.
\end{eqnarray}
By changing $i\omega_n$ to $\omega$, I transform this into the familiar propagator of a relativistic particle 
\begin{eqnarray}
\frac{1}{-(i\omega_n)^2+k^2+m^2}\rightarrow\frac{1}{-\omega^2+k^2+m^2}.
\end{eqnarray}
Once you have continued the propagator from discrete imaginary frequency to an arbitrary complex frequency $\omega$ (see Fig.~\ref{akamatsu4-2}), you make apparent the pole at a real frequency $\omega_k$ corresponding to the excitation energy of the system with plus or minus one particle,  $\omega_k=\pm\sqrt{k^2+m^2}$. Thus, the analytic continuation of the Matsubara propagator allows us to get information on the excitation energies of the system.

However, it is sometimes difficult to do the analytic continuation explicitly. For instance you may know the propagator only  numerically, at all values, or only at a given subset of values,  of the Matsubara frequencies. Then,  performing the analytic continuation in order to extract the physical singularities may be an (almost) impossible task. Therefore, it may be advantageous to be able to perform calculations directly in real time. This is what we shall do in this lecture. 
\\
\subsubsection{An example of hard thermal loop}

After this reminder about analyticity property, I would like to show you one particular Feynman diagram calculation of a hard thermal loop.
This is actually how hard thermal loops were discovered,  by explicitly calculating a series of Feynman diagrams, and making the appropriate kinematical simplifications.
I shall do a calculation of a one-loop  self-energy in a scalar theory with a  $\phi^3$ interaction, in order to be able to compare with corresponding results in electrodynamics that I shall consider next. 
The scalar $\phi^3$ field theory is not completely stable, but this difficulty is not relevant for the present discussion. 

\begin{figure}
\centering
\psfrag{p}{$\vec p$}
\psfrag{k}{$\vec k$}
\psfrag{p+k}{$\vec p+\vec k$}
\psfrag{tau1}{$\tau_1$}
\psfrag{tau2}{$\tau_2$}
\includegraphics[width=8cm, height=3cm, clip]{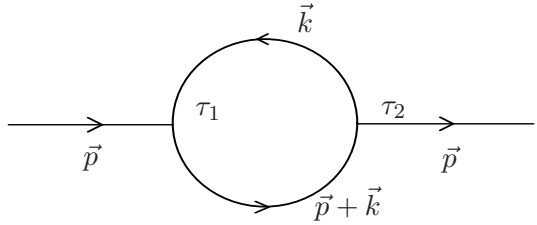}
\caption{Self-energy diagram in $\phi^3$ scalar field theory}
\label{akamatsu4-3}
\end{figure}

The diagram that  I want to calculate is displayed in Fig.~\ref{akamatsu4-3}.
 I am going to use the mixed representation of the propagator,  $D_\k(\tau)$, that I introduced in previous lectures.
I shall call the self-energy $\Pi(\tau,\p)$, and I shall focus on the regime where  $p$ is a soft momentum while the loop integral is dominated by  hard momenta $k\sim T$.
I get first
\begin{eqnarray}
\Pi(\tau,\p)&=&g^2\int\frac{d^Dk}{(2\pi)^D}D_{\p+\k}(\tau)D_{\k}(-\tau) \qquad
(\tau_2-\tau_1\equiv\tau)\\
D_{\k}(\tau)&=&\frac{1}{\omega_\k}\bigl[(1+n_\k)e^{-\omega_\k|\tau|}+n_\k e^{\omega_\k|\tau|}\bigr],
\end{eqnarray}
so that,  for $\tau>0$
\begin{eqnarray}
\Pi(\tau,\p)&=&g^2\int\frac{d^D\k}{(2\pi)^D}\frac{1}{2\omega_{\p+\k}}\frac{1}{2\omega_{\k}}\bigl[(1+n_{\k+\p})e^{-\omega_{\k+\p}\tau}+n_{\k+\p}e^{\omega_{\k+\p}\tau}\bigr]\nonumber \\
& & \ \ \ \ \ \ \ \ \ \ \ \ \ \ \ \ \ \ \ \ \ \ \ \ \ \times
\bigl[(1+n_\k)e^{-\omega_\k\tau}+n_\k e^{\omega_\k\tau}\bigr].
\end{eqnarray}
Now, I take the Fourier transform by integrating over the imaginary time from 0 to $\beta$
\begin{eqnarray}\label{Piphi3}
\Pi(i\omega_n,\p)&=&\int^{\beta}_{0}d\tau e^{i\omega_n\tau}\Pi(\tau,\p)\nonumber \\
&=&g^2\int\frac{d^D\k}{(2\pi)^D}\frac{1}{2\omega_{\p+\k}}\frac{1}{2\omega_{\k}}
\Bigl\{
\frac{1+n_{\p+\k}+n_{\k}}{\omega_{\p+\k}+\omega_{\k}-i\omega_n}
+\frac{1+n_{\p+\k}+n_{\k}}{\omega_{\p+\k}+\omega_{\k}+i\omega_n}\nonumber \\
& & \ \ \ \ \ \ \ \ \ \ \ \ \ \ \ \ \ \ \ \ \ \ \ \ \ \ \ \ \ \
+\frac{n_{\k}-n_{\p+\k}}{\omega_{\p+\k}-\omega_{\k}+i\omega_n}
+\frac{n_{\p+\k}-n_{\k}}{\omega_{\k}-\omega_{\p+\k}+i\omega_n}
\Bigr\}.
\end{eqnarray}
At this point, I can perform the analytic continuation, $i\omega_n\to\omega$.
In doing so I may run into trouble because the denominators may vanish.
When the denominators vanish, a priori the integral blows up, but this singularity is associated with well understood physics, that I am going to discuss. Look at the first term in the integrand 
\begin{eqnarray}
\frac{1+n_{\p+\k}+n_{\k}}{\omega_{\p+\k}+\omega_{\k}-i\omega_n}
\rightarrow
\frac{1+n_{\p+\k}+n_{\k}}{\omega_{\p+\k}+\omega_{\k}-(\omega\pm i\eta)}.
\end{eqnarray}
When the denominator vanishes, it is a signal that there is a process which is allowed. Here it is the process by which an excitation carrying momentum $\p$ (and energy $\omega$)  decays into a set of two excitations carrying momentum $\k$ and momentum $\p + \k$.
This  translates into an imaginary part, which is proportional to the rate of such decay.
To get this imaginary part, we add a little imaginary part to $\omega$.
In other words, in the continuation, we start from Matsubara frequency and we  continue up to the real axis but stop a little bit below or above (depending on which propagator we want to consider, retarded, advanced, etc).
We just do not touch the real axis (see Fig.~\ref{akamatsu4-2}).

\begin{figure}
\centering
\psfrag{p}{$\vec p$}
\psfrag{k}{$\vec k$}
\psfrag{p+k}{$\vec p+\vec k$}
\psfrag{-}{$-$}
\psfrag{decay}{$(1+n_k)(1+n_{p+k})$}
\psfrag{production}{$n_kn_{p+k}$}
\includegraphics[width=7cm, clip]{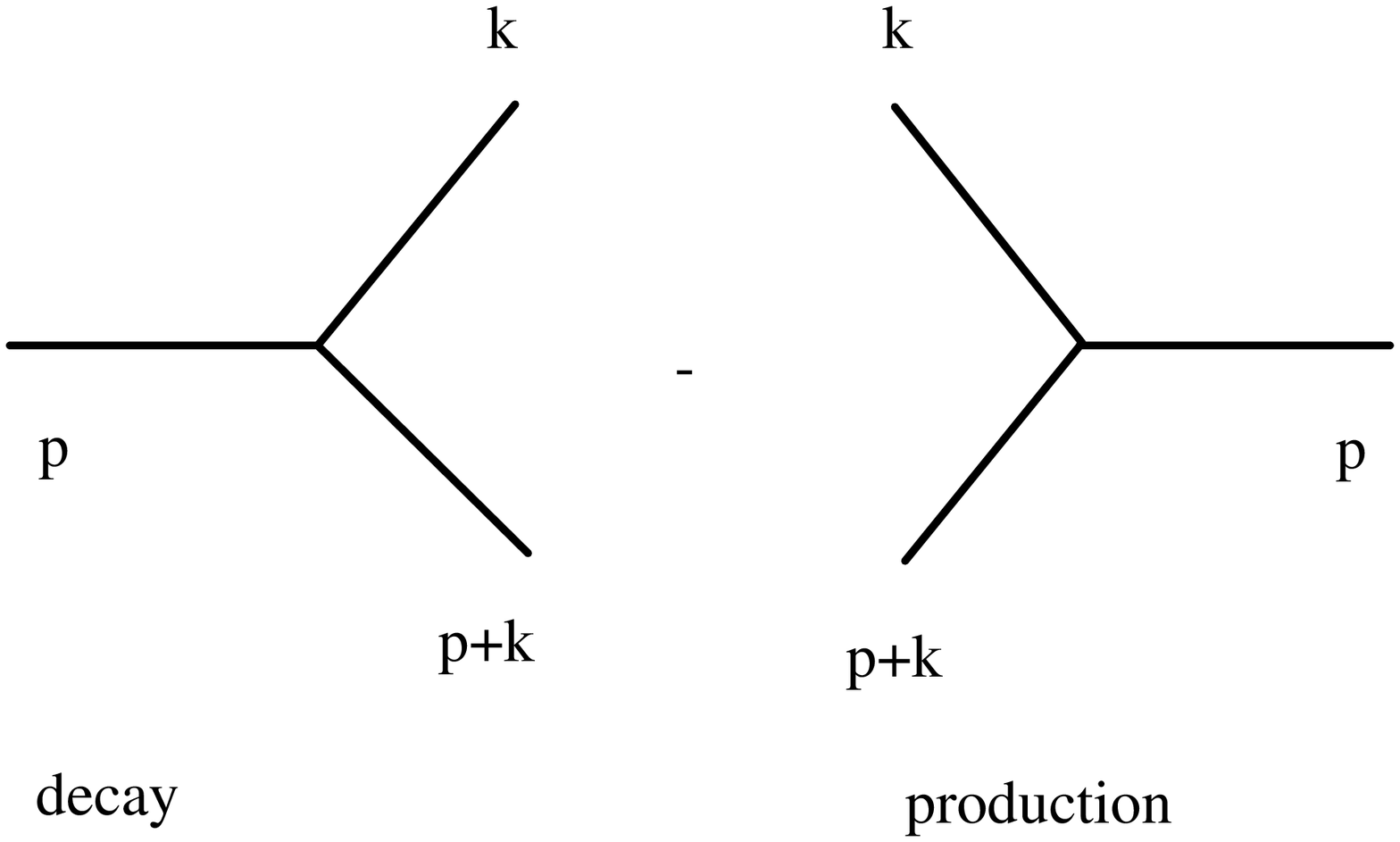}\\
\psfrag{p}{$\vec p$}
\psfrag{k}{$\vec k$}
\psfrag{p+k}{$\vec p+\vec k$}
\psfrag{-}{$-$}
\psfrag{decay}{$n_k(1+n_{p+k})$}
\psfrag{production}{$n_{p+k}(1+n_k)$}
\includegraphics[width=7cm, clip]{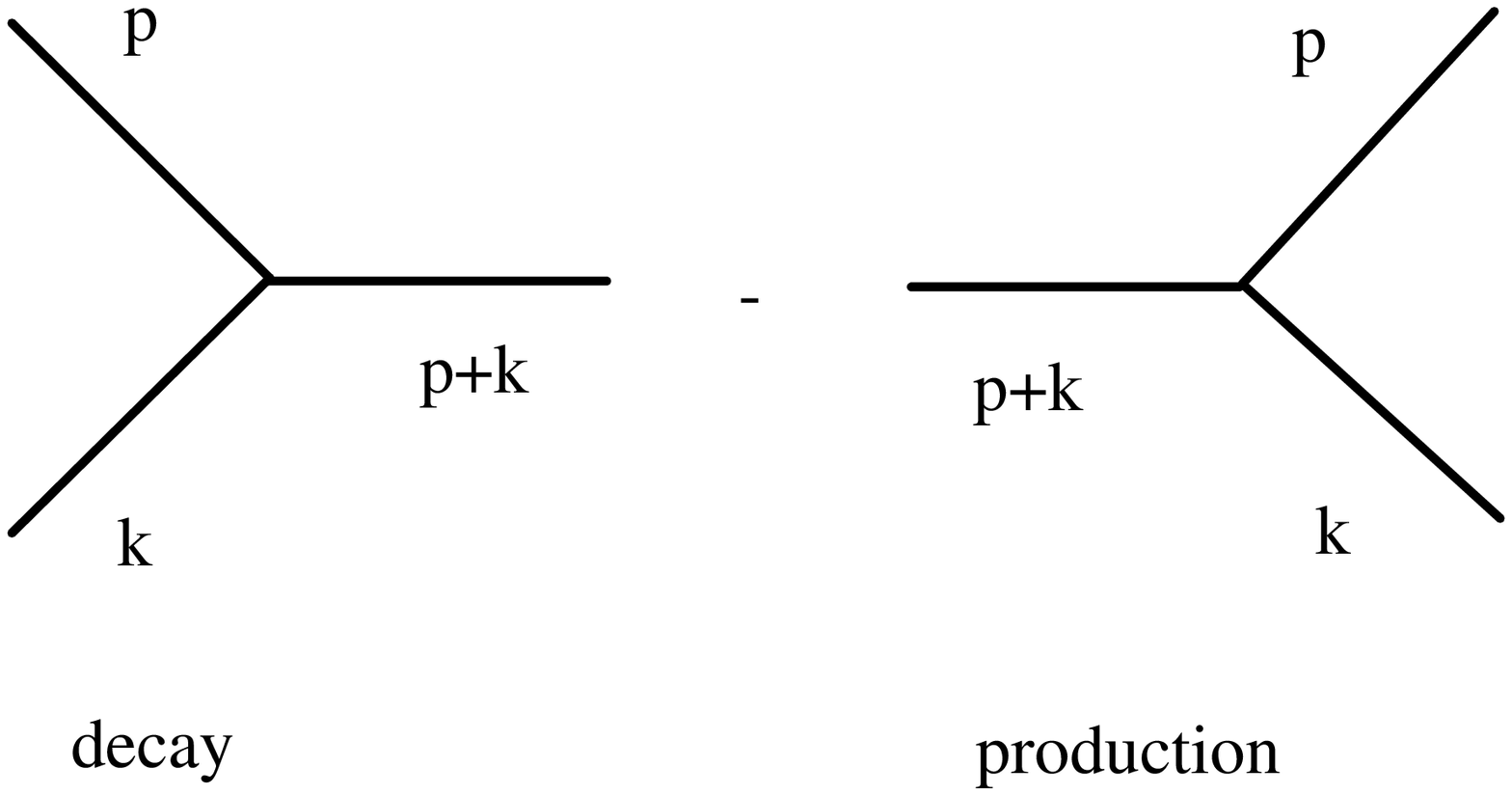}
\caption{The various processes contributing to  Eq.~(\ref{Piphi3}).}
\label{akamatsu4-4}
\end{figure}

Now let me turn to the numerators, and focus again on the first process in Fig.~\ref{akamatsu4-4}.
If you view the imaginary part as  part of a rate calculation (the denominator providing the delta-function that expresses energy conservation), the numerator accounts for the statistical factors that accompany  the direct process, by which the mode with momentum $\p$ decays into two other modes, as well as the reverse process by which two excitations recombine to form the initial excitation (the second term in the first line of Fig.~\ref{akamatsu4-4}).
The direct process is enhanced by the factor  $(1+n_\k)(1+n_{\p+\k})$, while the reverse process will be simply proportional to $n_\k n_{\p+\k}$, the probability that the modes $\k$ and $\k+\p$ are occupied. 
Subtracting the two yields
\begin{eqnarray}
(1+n_\k)(1+n_{\p+\k})-n_\k n_{\p+\k}=1+n_\k+n_{\p+\k},
\end{eqnarray}
which is indeed the numerator in the first term of Eq.~(\ref{Piphi3}). You may verify that all the other terms can be obtained from this simple reasoning.

Let us consider for instance the third  term of Eq.~(\ref{Piphi3}). 
This represents a process by which the mode with momentum $\p$ can absorb a mode with momentum $\k$ and go into a state with momentum $\p+\k$ (see Fig.~\ref{akamatsu4-4}).
The reverse process is a mode with $\p+\k$ going to the modes with $\k$ and $\p$.
The statistical factors here are  $n_\k$ for the incoming line ($n_\p$ does not count because it corresponds to the particle which I am looking at), and the  induced emission factor  $1+n_{\p+\k}$ on the outgoing line, giving a factor $n_\k(1+n_{\p+\k})$ for this process.
There is a factor $n_{\p+\k}(1+n_{\k})$ for the inverse process.
You see again that the products $n_\k n_{\p+\k}$ cancel out, leaving a term linear in $n$:
\begin{eqnarray}
n_\k(1+n_{\p+\k})-n_{\p+\k}(1+n_\k)=n_\k-n_{\p+\k}.
\end{eqnarray}
Note that this process is a genuine finite temperature effect, and it disappears at $T=0$.  This is in contrast to the the one considered previously, which exists also in the vacuum (the 1 in the numerators of Eq.~(\ref{Piphi3})), and which represents a decay process. The processes that we consider now are scattering processes involving particles of the heat bath.  
As you will see in a moment, these are the dominant contributions at high temperature.

Now comes the relations with the rest of the lectures.
What we have done so far is  an exact one loop calculation. Let us pursue a little bit the analysis of the diagram here.
The loop integral,  for the same reason as I discussed already several times, is dominated by the largest momenta, that is by momenta of the order of the temperature.
Observe the first two terms in the self energy, which contain the vacuum contribution, and which represent decay processes:
\begin{eqnarray}
\frac{1+n_{\p+\k}+n_{\k}}{\omega_{\p+\k}+\omega_{\k}-i\omega_n}
+\frac{1+n_{\p+\k}+n_{\k}}{\omega_{\p+\k}+\omega_{\k}+i\omega_n}
\end{eqnarray}
The denominators are of the order of $k\sim T$ ($p\sim gT$), which is big.
In contrast, in the last two terms, \begin{eqnarray}
\frac{n_{\k}-n_{\p+\k}}{\omega_{\p+\k}-\omega_{\k}+i\omega_n}
+\frac{n_{\p+\k}-n_{\k}}{\omega_{\k}-\omega_{\p+\k}+i\omega_n},
\end{eqnarray}
 you have difference between two large energies $\omega_{\p+\k}$ and $\omega_{\k}$,  and 
\begin{eqnarray}
\omega_{\p+\k} -\omega_\k\approx  \p\cdot\frac{\partial \omega_\k}{\partial\k}=\p\cdot\v_\k,
\end{eqnarray}
where $\v_\k$ is the velocity, whose modulus is the  speed of light (assuming massless particles). Thus the denominator, if  $\omega$ is of order $gT$, is a soft energy denominator.
And indeed, at high temperature, the dominant contribution is obtained form these last two terms
\begin{eqnarray}\label{htlPiphi3}
\Pi_{\rm HTL}(\omega,\p)&\approx& g^2\int\frac{d^D\k}{(2\pi)^D}\frac{1}{(2\omega_\k)^2}\Bigl\{
\frac{n_\k-n_{\p+\k}}{\omega+\v\cdot\p}+\frac{n_{\p+\k}-n_\k}{-\v\cdot\p+\omega}
\Bigr\}\nonumber\\
&\approx&g^2\int\frac{d^D\k}{(2\pi)^D}\frac{1}{(2\omega_\k)^2}\v\cdot\p\, \Bigl\{
\frac{1}{\omega-\v\cdot\p}-\frac{1}{\omega+\v\cdot\p}
\Bigr\}
\frac{\partial n_\k}{\partial\omega_\k},
\end{eqnarray}
where I have used the relation
\begin{eqnarray}
n_{\p+\k}-n_\k\approx \p\cdot\frac{\partial}{\partial\k}n_\k=\p\cdot\v_k\, \frac{\partial n_\k}{\partial \omega_\k}.
\end{eqnarray}
The energy denominators reflect the well-known phenomenon  of Landau damping, that takes place when the phase velocity of the soft mode, $\omega/p$, equals the velocity of the hard particle, $\v_\k$,  in the direction of the propagation of the soft mode.

What we will do now in the rest of the lecture is to recover similar expressions  for QED,  starting from kinetic theory.
My way of showing you the connection with the kinetic theory is not a formal way.
There is a formal route to deduce things but I shall only be able here to give you hints of how things work and are tied together.

%
%
%
%
%
%
%
%
%

\subsection{Calculation of $\Pi _\text{QED}(\omega ,\v)$ using  kinetic theory}

What I shall do now is to do a similar calculation, but for electrodynamics, and using  kinetic theory. I shall show you how to get the self-energy, 
$\Pi _\text{QED}(\omega ,\p)$ by solving a simple kinetic equation. Doing so, we shall in fact get immediately the hard thermal loop approximation for $\Pi_\text{QED}$. 
What is kinetic theory? 
It is a theory which describes the evolution of distribution functions, $f_q (\p,X)$,  which are the phase space densities of particles at space time point $X=(t,\x)$, carrying momentum $\p$, and  electric charge $qe$ (with $q=\pm 1$). 

\subsubsection{Linearized Vlasov equation}

The Vlasov equation
is  the simplest  of kinetic equations.
It describes the evolution of particles under the action of a force, and reads
\begin{equation}
\frac{\partial f_q}{\partial t} +\v \cdot \del_{\x} f_q +\F_{\!orce} \cdot \del_{\p} f_q =0. 
\label{Vlasov}
\end{equation}
The force, in the case of QED, is
\begin{equation}
\F_{\!orce} = q(\E +\v \wedge \B),
\end{equation}
where $\E$ is an electric field and $\B$ is a magnetic field and 
$\v =\frac{\partial \epsilon _\p}{\partial \p}$.

I assume that the system is  initially  in thermal equilibrium,  with a distribution function $f^0$ which is independent of $X$: $\partial _X f^0 =0$, and function only of the energy of the particle.
At time $t=0$, the system is weakly perturbed away from its equilibrium state, and the distribution function becomes \begin{equation}
f^0 \longrightarrow f^0 +\delta f \label{df}.
\end{equation}
I'm going to assume that the perturbation is small so that I can treat $\delta f$ as a small quantity 
and linearize the Vlasov equation in order to determine $\delta f$:
\begin{equation}\label{linearVlasov}
v\cdot \partial_X \delta f_q (\p,X) = -q\v \cdot \E \frac{d f^0}{d \epsilon _\p}, 
\end{equation}
where a covariant notation 
\begin{equation}
v\cdot \partial_X = v^\mu \frac{\partial}{\partial X^\mu } 
 = \frac{\partial}{\partial t} +\v \cdot \frac{\partial}{\partial \x} 
\end{equation}
is used. 
To obtain Eq.~(\ref{linearVlasov}), 
I have just replaced in Eq.~\eqref{Vlasov} $f$ by $f_0 +\delta f$ and took advantage of the fact 
that $f_0$ is an equilibrium distribution. In particular,  I have used the fact that  $f_0$ is isotropic to eliminate the contribution 
from the magnetic field. 
\\
\subsubsection{Induced current}

Let me now introduce a new function $W(X,\v)$ 
\begin{equation}
\delta f_q (\v,X) = -qW(X,\v) \frac{d f^0}{d \epsilon _p}. \label{defW}
\end{equation}
If you compare \eqref{df} and this expression \eqref{defW}, you can write the following 
\begin{equation}
f_q (\v,X) =f_q ^0 (\epsilon_p ) -qW(X,\v) \frac{d f^0}{d \epsilon _p}. \label{fW1}
\end{equation}
You see this is just the beginning of Taylor expansion of the quantity $f_q ^0 (\epsilon_p -qW(X,\v))$:
\begin{equation}
f_q (\v,X) = f_q ^0 \left( \epsilon_p -qW(X,\v)\right). \label{fW2}
\end{equation}
In other words, the distribution function in the presence of the linear perturbation is 
just the thermal distribution function for an energy which is shifted by an amount $qW$ which depends on both 
the velocity of the particle and the coordinates and time. 

The equation in terms of $W$ reads simply
\begin{equation}
v\cdot \partial_X W(X,\v) = \v \cdot \E.
\end{equation}
This  is a first order partial differential equation, which can be solved by the method of characteristics. One gets
\begin{equation}
W(X,\v) = \int _{-\infty} ^0 dt^\prime \,\v \cdot \E(\x -\v(t-t^\prime ),t^\prime ). \label{Wsol}
\end{equation}
You can verify by a direct calculation that this satisfies the equation. 
The interpretation is simple. 
The characteristic line is a straight line 
which is represented by $\x =\v(t-t^\prime )$.
Along the characteristic line the electric field does the work $\v \cdot \E$ in time $dt$. 
The work of the electric field adds up to make the quantity $W$ in Eq.~(\ref{Wsol}). 

Once we know $W(X,\v)$,  we can calculate the induced current $j_{ind}^\mu(X)$, which is the current generated by the perturbation (in equilibrium the current vanishes). This is given by 
\begin{equation}
j_\text{ind}^\mu(X) = e\int \frac{d^3 \p}{(2\pi )^3} v^\mu \left[ f_+ (\p,X)-f_- (\p,X)\right],
\end{equation}
where the  positive charges going in one direction contribute as the  negative charges going in the  opposite direction. 
By using Eq.~(\ref{Wsol}), we get
\begin{align}
j_\text{ind}^\mu = -2e^2 \int \frac{d^3 p}{(2\pi )^3} v^\mu \frac{d f^0}{d \epsilon _p}
 \int _0 ^\infty d\tau \v \cdot \E(X-v\tau ), \label{jsol}
\end{align}
where  $\tau= t-t^\prime$ (not the imaginary time !). 

Now, there is a simple relation between the (retarded) polarization tensor and the induced current:
\begin{align}
j_\text{ind}^\mu = \int d^4 y\, \Pi_{\text{Ret }}^{\mu \nu} (x-y) A_\nu (y).
\end{align}
Since 
\begin{align}
\E= -\nabla A_0-\frac{\partial \A}{\partial t}, 
\end{align}
$\Pi_{\text{Ret }}^{\mu \nu} $ can be easily obtained from Eq.~(\ref{jsol}).
It is in fact convenient to perform first a Fourier transform (F.T.), using 
\begin{align}
\text{F.T.} \int _0^\infty e^{-\eta \tau} f(X-v\tau ) = \frac{i\tilde{f} (Q)}{v\cdot Q+i\eta }, \label{FTformula}
\end{align}
where $\tilde{f} (Q)$ is the Fourier transform of $f(X)$. 
Then one gets
\begin{align}
\Pi ^\text{Ret } _{\mu \nu} (\omega ,\q ) = m_\text{D} ^2 \left\{ -\delta _{\mu 0} \delta _{\nu 0} 
 +\omega \int \frac{d\Omega _v }{4\pi} \frac{v_\mu v_\nu}{\omega -\v \cdot \q +i\eta} \right\}, \label{pol}
\end{align}
where
\begin{align}\label{mDebyeQED}
m_\text{D} ^2 = -\frac{2e^2}{\pi ^2} \int _0^\infty dp \ p^2 \frac{df^0}{d\epsilon _p}. 
\end{align}

The structure of this equation is very similar to that of Eq.~(\ref{htlPiphi3}). The energy denominators in particular are identical and reflect the Landau damping processes, while the integration over the modulus of the hard momentum (which factorizes) involves the derivative of the equilibrium distribution function. The Debye mass (\ref{mDebyeQED}) is the analog of the thermal mass of the scalar field.

\begin{figure}
 \begin{tabular}{cc}
  \begin{minipage}{0.5\textwidth}
   \begin{center}
    \includegraphics[scale=0.6,clip]{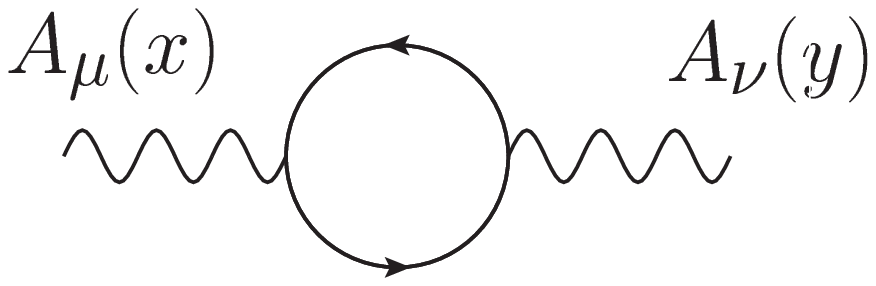}
   \end{center}
  \end{minipage} & 
  \begin{minipage}{0.5\textwidth}
   \begin{center}
    \includegraphics[scale=0.6,clip]{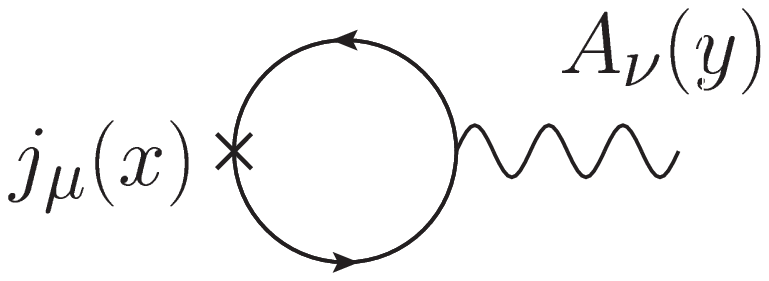}
   \end{center}
  \end{minipage} \\[+20pt]
  (a) & (b)
 \end{tabular}
 \caption{Diagrammatic representation of the (one-loop) polarization tensor (left) and induced current (right). }
 \label{fig:loop}
\end{figure}

Let me  summarize what I have done.  I have obtained an approximate expression for the QED polarization tensor which could have been obtained also by calculating the one-loop diagram in 
Fig.~\ref{fig:loop}(a), and doing the appropriate kinematical approximation valid when the external momentum is soft. Instead of calculating approximately a diagram, I have used a kinetic equation to calculate the induce current, from which the polarization tensor was obtained. In the kinetic theory, the hard particles are described by a distribution function $f(\p,X)$ whose slow variations in space time describe the soft, collective, excitations. 
You see how kinetic theory  manages to separate the hard and soft degrees of freedom: 
the hard degrees of freedom are those which govern the momentum dependence of $f(\p,X)$. In $f(\p,X)$, $\p$ is a hard momentum.  The slow degrees of freedom  are described by the slow variations of the distribution function in space and time (i.e., the dependence of $f(\p,X)$ on $X$). 
  In a way, the kinetic theory based on the Vlasov equation can be viewed as an effective theory for describing real time phenomena, somewhat analogous to the (Euclidean) effective theory used earlier.

As I said before, it is possible to establish the correspondence between the kinetic theory and the diagrammatic hard thermal loop calculation. Such an  approach has been generalized to QCD, where  it provides a microscopic effective theory for the quark-gluon plasma in the regime  
where the separation between the hard and soft degrees of freedom can be meaningfully realized (see \cite{Blaizot:2001nr} for more details).


\section{LECTURE V}
\subsection*{Introduction}

Today we start exploring a new system, the dilute Bose gas. The specific phenomenon that  I want to discuss concerns  the effect of  weak repulsive interactions on the critical temperature.
As you know, in a Bose gas at high density or low temperature, there is a phase transition, called the Bose-Einstein condensation. This phase transition occurs in the absence of any interaction among the atoms.
The question I shall address is what happens to this phase transition when the atoms repel each other very weakly.
This problem is interesting in many respects. In the particular context of these lectures it is interesting because, as you will see, the techniques employed to calculate the shift of the critical temperature are very similar to the techniques that I explained to you in the previous lectures when we dealt  with general aspects of quantum fields at finite temperature. In particular the technique of dimensional reduction, and of effective field theory will play an important role.

Let me be now  more specific.
The  interaction between the atoms will be characterized by a scattering length. I am going to assume that atoms interact only in the s-wave partial wave, and I will denote the corresponding scattering length by $a$. This has the dimension of a length.
There is another parameter with the dimension of a length, namely $n^{-1/3}$, where $n$ is the density of particles.
I am assuming that  the gas is dilute, which means that somehow $n$ is small. To be more precise, let me remark that together with $a$ and $n$, I can form a dimensionless parameter, $n^{1/3}a$. Then the diluteness condition reads $n^{1/3}a\ll 1$.

Two issues arise in addressing the question that I was mentioning a minute ago. The first issue is  whether the phase transition which is observed in the absence of any interaction survives in the interacting system.
I am simply going to assume that this is the case. The second issue,   assuming that the phase transition indeed takes place, is connected to the evaluation in the change of the critical temperature, and this is  what I want to calculate, that is $$\frac{\Delta T_{\rm c}}{T_{\rm c}}=\frac{(T_{\rm c}-T_{\rm c}^{0})}{T_{\rm c}^{0}},$$ where $T_{\rm c}$ is the critical temperature in the presence of the interaction, while $T_{\rm c}^0$ is the corresponding critical temperature (i.e., at the same density) in the absence of interaction.
What I will show you is that \beq \label{shiftTc}\frac{\Delta T_{\rm c} }{ T_{\rm c}} = c(an^{1/3}),\eeq where $c$ is a dimensionless positive number. 

This is a non trivial result. The shift in $T_{\rm c}$ is a quantity which is small if the scattering length is small, which is the case when the interactions are weak. Now, given that $\Delta T_{\rm c}\rightarrow 0$ as $a\rightarrow 0$, you could think naively that the change of the critical temperature can be calculated  by perturbation theory.
However perturbation theory is useless -- well, I mean, not completely-- but strict perturbation theory is useless because if you start calculating the Feynman diagrams order by order in an expansion in powers of $a$, you will meet infrared divergences.
That is a situation we have already met. It is an indication that we are doing something wrong.
What I shall  do in the next couple of lectures is to show you the origin of the difficulty, and present the techniques that can be used to overcome it,  namely the techniques based on effective field theory. The exact renormalization group sheds a more complete light on this problem, but I shall not have time to discuss it.

\subsection{Bose-Einstein Condensation}

\subsubsection{Non-interacting uniform systems}

Let me now remind you of a few basic facts about Bose-Einstein condensation. This is  textbook material, so I shall skip many details. 
I consider a collection of non-relativistic atoms. These atoms are (spinless) bosons, and  at finite temperature the average occupation of the single particle level of momentum $p$ is given by\begin{eqnarray}
n_{\bf p}=\frac{1}{e^{(\epsilon_{\bf p}-\mu)/T}-1}, \ \ \ \epsilon_{\bf p}=\frac{p^2}{2m},
\end{eqnarray}
where $\mu$ is the chemical potential.

\begin{figure}
\centering
\includegraphics[width=8cm, clip]{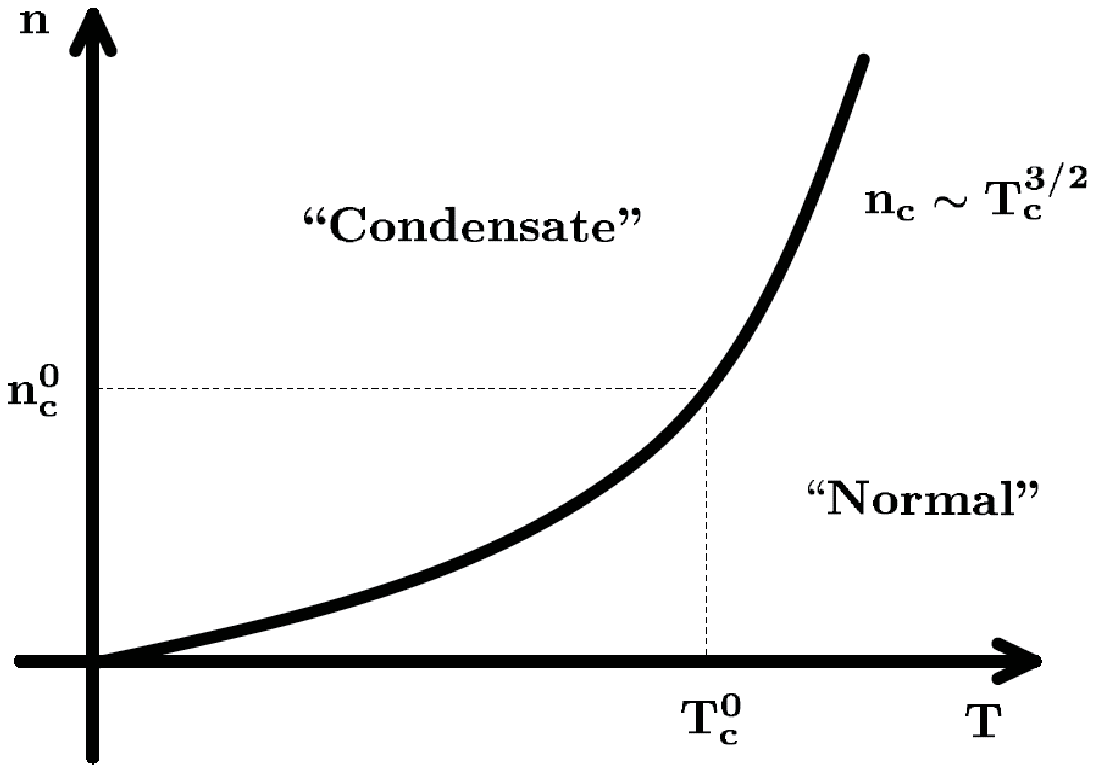}
\caption{Phase Diagram of Ideal Bose Gas}
\label{akamatsu5-1}
\end{figure}

Let us start by considering  a dilute gas at  high temperature.  Then the chemical potential is negative and large, so that the factor $e^{(\epsilon_{\bf p}-\mu)/T}$ is large, and I can  approximate $n_{\bf p}\approx e^{-(\epsilon_{\bf p}-\mu)/T}$. In this regime, the gas is essentially classical, the effects of quantum statistics (the $-1$ in the denominator)  can be ignored. 
The density is easily obtained 
\begin{eqnarray}\label{lowdensity}
n=\int \frac{d^3 {\bf p}}{(2\pi)^3} \,n_{\bf p}\approx e^{\mu/T}\lambda^{-3},
\eeq
where 
\beq
\lambda\equiv\sqrt{\frac{2\pi}{mT}}
\end{eqnarray}
 is the thermal wavelength, an important length scale in the problem.
(As in most of these lectures, I am using the natural units with $\hbar=1$.)
It is convenient to rewrite Eq.~(\ref{lowdensity}) as a formula for $\mu/T$ (valid when $n\lambda^3\ll 1$):
\begin{eqnarray}
\frac{|\mu|}{T}=-\ln (n\lambda^3).
\end{eqnarray}
Let us now examine what happens when one  decreases the temperature, keeping the density fixed. The  formula above gives you the trend (when the gas is very dilute so that $n\lambda^3\ll1$): the chemical potential decreases in absolute value. As the temperature continues to decrease, it eventually reaches the value $\mu=0$.
When $\mu=0$,  the number of particles is given by (quantum statistic can no longer be ignored then)
\begin{eqnarray}\label{nc0}
n^0_{\rm c}=\int\frac{d^3{\bf p}}{(2\pi)^3}\frac{1}{e^{\epsilon_{\bf p}/T}-1}=\frac{\zeta (3/2)}{\lambda^3} , \ \ \ \zeta(3/2)=2.612.
\end{eqnarray}
At that point, something happens.
The chemical potential can no longer increase: if it would,  the statistical factor would become singular (with in particular negative values for $\epsilon_{\bf p}<\mu$).  What happens is that particles start to accumulate in the state with  vanishing momentum $ p=0$. Thus,  the total number of particles is split into two contributions: one contribution, $n_0(T)$,  from particles  in the state  $ p=0$,   another contribution from the particles populating all other momentum states. That is 
\beq
n=n_0(T)+\frac{\zeta(3/2)}{\lambda^3}. 
\eeq
Equation (\ref{nc0}) can be viewed as the condensation condition relating the critical temperature $T_c^0$ to the density $n_c^0$. 
This leads to the  phase diagram drawn in Fig.~\ref{akamatsu5-1}, where the phase boundary is the curve that relates  the critical density $n^0_{\rm c}$ to the critical temperature $T_c^0$: because $\lambda\sim T^{-1/2}$, we get from Eq.~(\ref{nc0}) $n_{\rm c}^0\propto (T^0_{\rm c})^{3/2}$.  Above this phase boundary, we have the condensed phase and below we have the normal phase.
It is not difficult to verify that the density of particles in the condensate is given by 
\begin{eqnarray}
n_0(T)=n\Bigl[1-\Bigl(\frac{T}{T^0_{\rm c}}\Bigr)^{3/2}\Bigr], \ \ \ (T<T^0_{\rm c}),
\end{eqnarray}
where $n$ is the total density. The condensate density  vanishes, as it should, at the critical point. At zero temperature all particles are in the condensate. 

At this point I want to empasize that the phase transition that we have just described, and which occurs in the absence of interaction, exhibits several unphysical features,  which will be cured by  interactions, however small these may be. 
Consider for instance the compressibility (at fixed temperature). This is given by 
\begin{eqnarray}
\chi=-\left.\frac{1}{V}\frac{\del V}{\del P}\right|_T=\frac{1}{n^2}\frac{dn}{d\mu}=\frac{1}{n^2T}\int\frac{d^3{\bf p}}{(2\pi)^3}n_{\bf p}(1+n_{\bf p}).
\end{eqnarray}
This integral is dominated by  the low momentum region, where 
I can replace $n_{\bf p}$ in the vicinity of $\mu=0$ by
\begin{eqnarray}
p\rightarrow 0, \ \ \ n_p\sim\frac{2mT_{\rm c}^0}{p^2} \ \ \ (\mu=0),
\end{eqnarray}
an approximation which I have already used in previous lectures. 
When $p\rightarrow 0$, the integral is infrared divergent.
So the compressibility diverges at the transition, which reflects  the existence of anomalously large fluctuations of the density.
As we shall see, this pathological behavior will disappear in the interacting system. 

\subsubsection{A first look at the effect of interactions}

As I already mentioned, I am working in a system where the interaction is dominated by s-wave scattering.
I am going to assume that this interaction can be described by an effective two body potential, function only of the distance between the two atoms, and of the form $V(\r_1-\r_2)=g\delta^{(3)}(\r_1-\r_2)$. The relation between the coupling strength $g$ and the scattering length $a$ can be obtained by solving the scattering problem and is  $g=4\pi a/m$. (In fact, there are ultraviolet divergences in the scattering calculation, so an ultraviolet cutoff needs to be introduced. The coupling strength should be considered as a function of this cut-off so that the relation to the scattering length remains valid for any choice of the cut-off. I shall not discuss this in detail here since this play no role in our main discussion. I just want to alert you about this subtlety.) The hamiltonian density  is composed of the kinetic energy term and the interaction term
\begin{eqnarray}\label{hamiltonianbosons}
H(\r)=-\frac{1}{2m}\nabla\psi^{\dagger}(\r)\cdot\nabla\psi(\r)+\frac{g}{2}\psi^{\dagger}(\r)\psi^{\dagger}(\r)\psi(\r)\psi(\r),\qquad  g=\frac{4\pi a}{m},
\end{eqnarray}
where $\psi(\r)$ and $\psi^{\dagger}(\r)$ are quantum fields which describe the atoms and obey the usual commutation relation $[\psi(\r),\psi^{\dagger}(\r')]=\delta^{(3)}(\r-\r')$.

\begin{figure}
\centering
\includegraphics[height=3cm, width=6cm, clip]{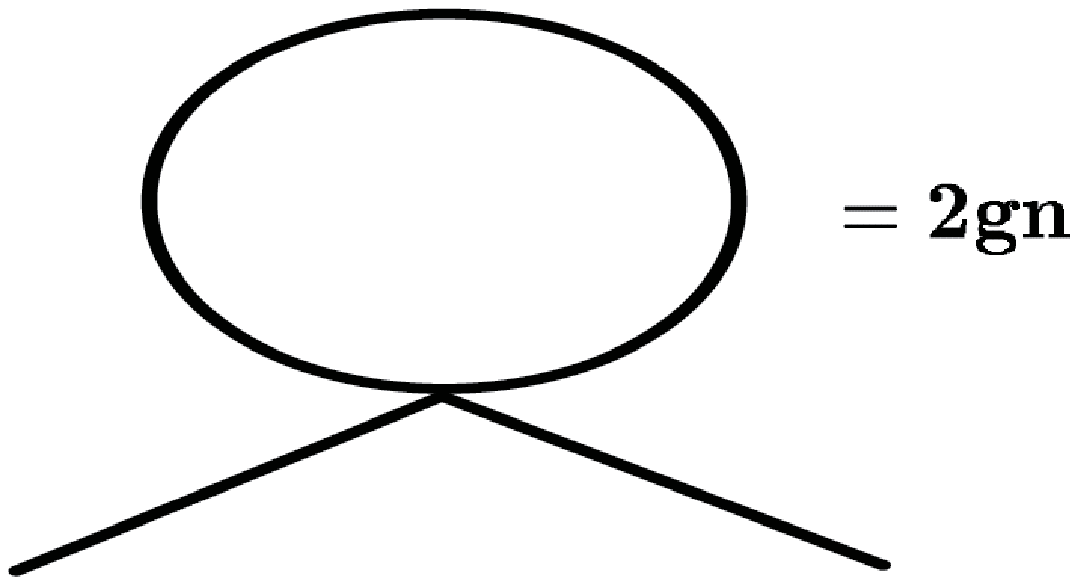}
\caption{Self-energy diagram of leading order in $a$}
\label{akamatsu5-2}
\end{figure}

As I indicated at the beginning of the lecture, we expect the shift in the critical temperature to be proportional to the strength of the interaction, i.e., proportional to the scattering length $a$.
It is therefore natural to try and estimate it using perturbation theory at leading order in $a$.
Let us then calculate the correction to the single particle energy $p^2/2m$ due to the interaction.
This  is obtained from the simple Feynman diagram shown in Fig.\ref{akamatsu5-2}.
Let me just give you the result: $2gn$ (it is proportional to $g$, and the  loop integral is proportional to the density, so the only hard work is the determination of  the factor  2).
In the presence of the interaction, the single particle energy becomes $\epsilon_\p =\epsilon_\p^0+2gn$, where $\epsilon^0_\p=p^2/2m$.
The main feature of this correction is that it is a constant  shift, by the quantity $2gn$.
What is the effect of this shift on the transition temperature?
Does it move the transition temperature up or down when the interaction is repulsive, i.e.,  $g>0$?
That is, is the shift  positive, negative, or zero?
The right answer is zero!
Let me explain to you why this is zero.
Remember that the criterion for condensation is $n(\mu=0,T)=2.612$, where
\begin{eqnarray}
n(\mu,T)=\int\frac{d^3\p}{(2\pi)^3}\frac{1}{e^{(\epsilon_\p^0-\mu)/T}-1}.
\end{eqnarray}
This is for $a=0$.
When $a\neq0$, the density is given by
\begin{eqnarray}
n=\int\frac{d^3\p}{(2\pi)^3}\frac{1}{e^{(\epsilon_\p^0-(\mu-2gn))/T}-1} = n(\mu-\Delta\mu,T), \ \ \ \Delta \mu=2gn,
\end{eqnarray}
where  $\mu-\Delta\mu=\mu-2gn$ can be viewed as a modified chemical potential since $2gn$ is a constant.
The condensation condition that we had for $a=0$,  $n(0,T)=2.612$, is  unaffected; it just occurs at a different chemical potential ($\mu=2gn$), but the  relation between the critical temperature  and the critical density is independent of $g$.
So the first-order perturbation theory or, if you wish, the mean-field calculation, does not produce any shift in the critical temperature.
This result,  simple to establish, shows immediately that the linear relation between the  shift in $T_c$ and  $a$ is not going to be obtained by a trivial procedure.\\

\noindent
\begin{description}
\item {\bf Q: } If you have temperature dependence in the mean-field, somehow...
\item {\bf A: } Somehow? What do you mean by ``somehow"?
\item {\bf Q: } If you calculate only  this diagram, of course you do not get a temperature dependence.
\item {\bf A: } You can think of a more complicated calculation. But at this level,  the message is fairly robust, although it has in fact  been overlooked by quite a number of people.
Perhaps I should say that I am approaching the transition temperature from above where the gas is classical, and  there is no...
Well let me not say that now.  I was about to say ``symmetry breaking'',  but I have not mentioned that concept.
I shall do that shortly. 
\end{description}

Although in leading order the transition temperature is not affected by interaction, the interaction has nevertheless a profound effect on the properties of the system.
Let me show that in the case of the compressibility.
 I leave it  as an exercise to you to  recalculate $\frac{1}{n^2}\frac{dn}{d\mu}$. Let me just give you the result,
\begin{eqnarray}\label{compressibility2}
\frac{1}{n^2}\frac{dn}{d\mu}=\frac{\frac{1}{T}\int_\p n_\p(1+n_\p)}{1+\frac{2g}{T}\int_\p n_\p(1+n_\p)}\approx \frac{T}{2g}.
\end{eqnarray}
(You see why you get the denominator: when you  take the derivative with respect to $\mu$ of the density
\begin{eqnarray}
n=\int\frac{d^3\p}{(2\pi)^3}\frac{1}{e^{(\epsilon_\p^0-(\mu-2gn))/T}-1} ,
\end{eqnarray}
you have an explicit derivative, but you have also to take the derivative of the factor $n$ inside the integral, and that brings another factor $dn/d\mu$ (proportional to $2g/T$).)
The momentum integral in Eq.~(\ref{compressibility2}) are infrared divergent,  but  now the divergences  actually cancel out since the same divergent integral appears in the numerator and in the denominator. 
Of course, if  $g$ is very small, the compressibility is very big but even for infinitesimal $g$ the compressibility becomes finite.
So there is a deep modification of the properties of the system.
What is being done here, in terms of  Feynman diagrams, is actually a resummation of a chain of bubble diagrams which contribute to the screening of the long wavelength density fluctuations.

\subsubsection{Symmetry breaking}
There is another aspect which is qualitatively new in the presence of the interaction:  we can discuss the phase transition in terms of symmetry breaking.
I am going to consider very low, in fact zero, temperature and show you that  Bose-Einstein condensation is what one may call  a quantum phase transition seen here as a change in the properties of the Bose gas as one tunes the chemical potential. To carry out the discussion in simple terms, let me introduce a quantum state, which I call $|N_0 \rangle$,
\begin{eqnarray}
|N_0\rangle\sim {\rm exp}(\sqrt{N_0}a_0^{\dagger})|0\rangle.
\end{eqnarray}
This is a coherent state, with  $a_0^{\dagger}$  the zero momentum component of the field creation operator, $[a_0,a_0^{\dagger}]=1$.
The expectation value $\langle N_0 |H-\mu a^\dagger_0a_0| N_0\rangle$ contains no contribution from the kinetic energy because $a^{\dagger}_0$ carries no momentum. The result is 
\begin{eqnarray}
\langle N_0|H-\mu a^\dagger_0 a_0|N_0\rangle=-\mu N_0+\frac{g}{2}N_0^2.
\end{eqnarray}
\begin{figure}
\centering
\includegraphics[height=5cm, width=6cm, clip]{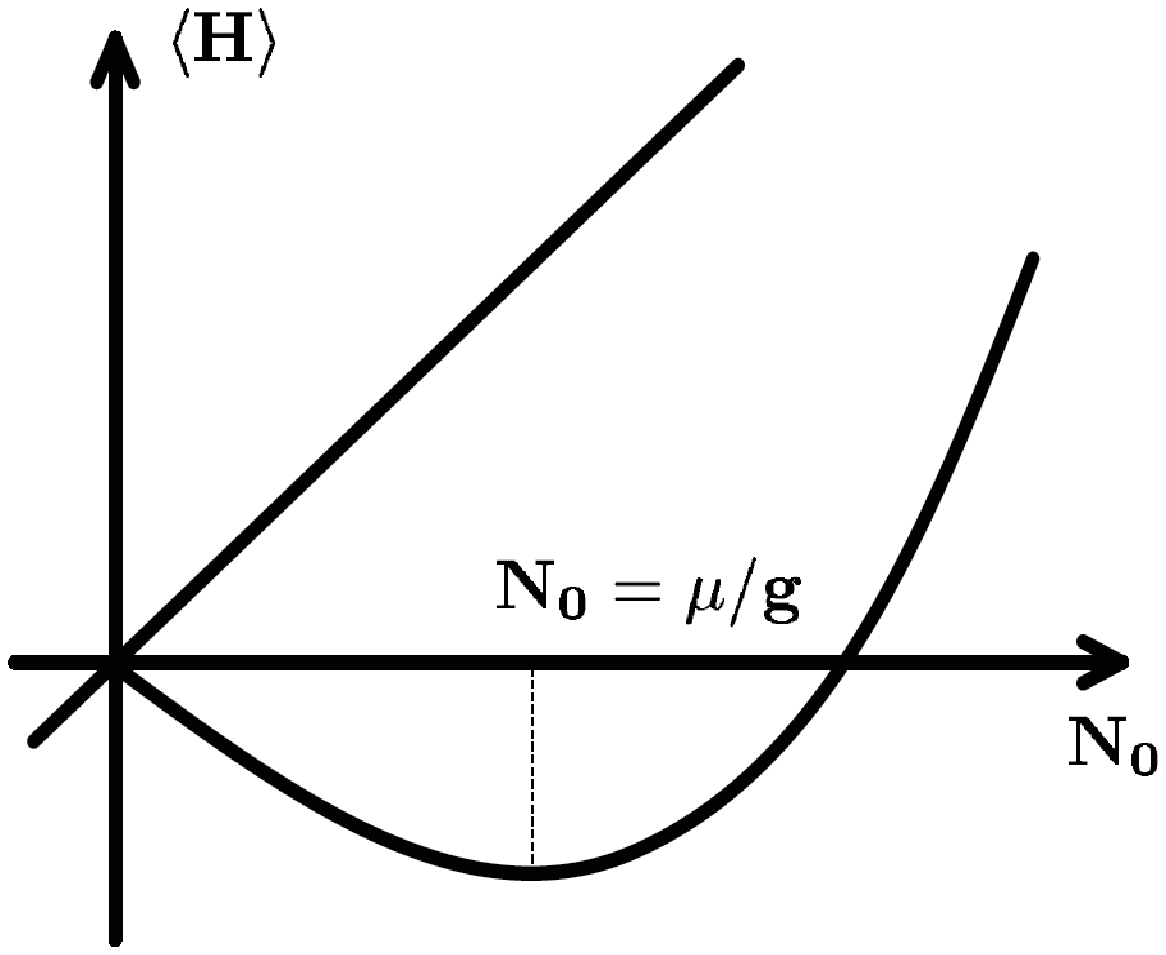}
\caption{Energy of the coherent state $\ket{N_0}$ as a function of $N_0$ for $\mu<0$ and $g=0$ (straight line) and for $\mu>0$, in which case a minimum occurs for $N_0=\mu/g$.}
\label{akamatsu5-3}
\end{figure}
If $g=0$ and $\mu<0$,  the ground state in Fock space is $N_0=0$: this is the vacuum state with no particle.
On the other hand if $\mu=0$ the system is completely degenerate: we have an arbitrary number of particles in the ground state because it cost no energy to add one. This degeneracy is the source of the large density fluctuations in the non interacting system at the transition. 
If $\mu>0$, the system is completely unstable:  you can decrease its energy by an arbitrary amount by adding more and more particles.
The presence of an interaction controls this phenomenon and cures the instability. 
When $\mu>0$, the quadratic term coming from the interaction generates a minimum at the value $N_0=\mu/g$ (see Fig.~\ref{akamatsu5-3}).
As a function of $\mu$,  the phase diagram looks like Fig.~\ref{akamatsu5-4}. \begin{figure}
\centering
\includegraphics[height=3cm, width=8cm, clip]{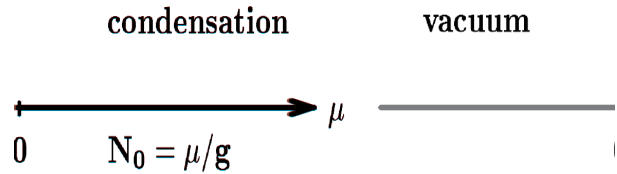}
\caption{Phase diagram of the weakly Interacting Bose Gas at zero-Temperature}
\label{akamatsu5-4}
\end{figure}
This is a pattern which allows us to view this Bose-Einstein condensation using 
techniques which are  familiar in quantum field theory, namely borrow all what we know about symmetry breaking and associated Goldstone modes and related phenomena.
The symmetry which is broken here is a $U(1)$ symmetry:
Remember that the hamiltonian of the system does not depend on the (global) phase of the bosonic field. However,   the field acquires an expectation value (that is explicitly taken into account here in our description with a coherent state), the $U(1)$ symmetry is broken.
\noindent
\begin{description}
\item {\bf Q}:  Excuse me, did you include $-\mu N$ to hamiltonian?
\item {\bf A}: Yes. 
Whenever I talk about the hamiltonian $H$, it is the true hamiltonian with kinetic energy, potential energy and $-\mu N$.
When I talk about degenerate states, I am referring to eigenstates  of $H-\mu N$.
\end{description}
 This is a very rapid digression, but it indicates to you that indeed even an infinitesimal interaction  changes the properties of the system in a qualitative way (you can extend this discussion to finite temperature). In the presence of interactions the ground state is indeed different and we can use languages which we are familiar with in field theory to understand what is going on.
This actually leads us to deep consequences; when there is a symmetry breaking there are Goldstone modes associated with a rotation of the phase of the order parameter and those Goldstone modes have a strong impact on the physics of Bose-Einstein condensates.
But I shall not discuss this too much here. 

There is another digression that  I want to make to illustrate the role of mean-field effects on $T_{\rm c}$: these are indeed very different in uniform and finite sytems.

\subsubsection{Atoms in a trap}
I would like to discuss briefly what happens for the atoms in a trap. This for two reasons. The first is that the physics of cold atoms in traps is what  has triggered the renewal of interest in Bose-Einstein condensation and 
much of the works that I am discussing. 
As you know, one is able now to cool atoms in a trap to
a sufficiently low temperature to observe Bose-Einstein condensation, 
in a system where the interaction strength is very small. Till then, the prototype of systems in which one could observe a phase transition akin to Bose-Einstein condensation was  liquid helium. But in liquid helium  the interaction between the atoms is very strong and the fraction of the particles that are sitting in the condensate  is never bigger than about  10 \%, even at very low 
temperatures. Nowdays, one is able to prepare small condensates of
few tens of  thousands of  atoms in a trap,  tune their mutual interaction to be as small as desired, and observe genuine Bose-Einstein condensation. However, and this is 
the second motivation for this digression,  the presence of the trap inhibits some of the effects of the interactions that occur in uniform system. So the study of atoms in a trap gives, so to speak by contrast,  an interesting perspective on some aspects of the effects of the interactions, related to long wavelength phenomena that are characteristics of uniform systems. These are these long wavelength phenomena that, in my view,  make Bose-Einstein condensation so interesting.

I am going to consider the following situation: 
the trapping potential is described by a harmonic oscillator, with a typical level spacing  $\hbar\omega$.
I  assume that the  temperature $T$ is high, 
such that $k_{\text{B}}T\gg \hbar\omega$.
Under this condition, I can use a semiclassical approximation 
to describe the atoms. 
What does it amount to?
The semiclassical approximation essentially states that
one can consider the gas of the atoms locally as a piece of uniform matter whose  density is equal to the  local density in the trap.
In other words, one assumes that  the 
energy of an atom can be written as
\begin{eqnarray}
\epsilon(\p,\r)
=\frac{\p^2}{2m}+\frac{m\omega^2}{2}\r^2,
\end{eqnarray}
 ignoring  the fact that $\p^2$ and $\r^2$ do not commute.
The density  is, as usual, obtained by integrating the distribution function over the momentum of the particles
\begin{eqnarray}
n(\r,\mu,T)=\int\frac{d^3\p}{(2\pi)^3}
\frac{1}{e^{(\epsilon(\p,\r)-\mu)/T}-1}.
\end{eqnarray}
This is  now a function of $\r$. The total number of particles is
\begin{eqnarray}
N=\int d^3r\, n(\r,\mu),
\end{eqnarray}
and is kept fixed.
The number density at the center of the trap, where it is  is the biggest, is given by
\begin{eqnarray}
 n(0,\mu,T)
=\int\frac{d^3\p}{(2\pi)^3}
\frac{1}{e^{(\epsilon(\p,
0)-\mu)/T}-1}.
\end{eqnarray}
When \begin{eqnarray}
n(0,\mu=0,T)=2.612,
\end{eqnarray}
the atoms at the center of the trap will undergo Bose-Einstein condensation. That is, 
condensation occurs 
whenever the density at the center of the trap satisfies the same relation with temperature as in a uniform system. 
The density at the center of the harmonic trap
can be calculated easily.
You find then, that in the absence of interactions,  
\begin{eqnarray}
\frac{k_{\text{B}}T^0_{\rm c}}{\hbar\omega}=\left(\frac{N}{\zeta(3)}\right)^{1/3}.
\end{eqnarray}
So you see that if $N$ is big enough (in typical experiments, $N\sim 10^5\div 10^7$), the condition for
the validity of the semiclassical approximation, $k_{\text{B}}T\gg \hbar\omega$  is well fulfilled.

Now we can discuss easily the effect of the
interaction on the transition temperature of the trapped gas.
I can ask again my question on the shift in $T_{\rm c}$:
Will it  be positive, negative or zero?
Let me tell you that this is not going to be the same result as before, namely,
this is not going to be zero.
That leaves two possibilities: 
 positive or negative.
You should be able to answer this question because that requires
no calculation at all.
What will the interaction do if you put particles in the trap? Let us imagine putting particles in a trap, and keep adding particles (at fixed temperature)  until the density at the center of the trap is high enough for condensation to occur. When 
the particles in the middle of the trap just begin to condense, let us switch on the interaction.
What happens? Because the particles repel each other,  the gas will expand and the density in the middle of the trap will decrease, destroying the condensation. 
How can one recover 
the condensation? 
By decreasing the temperature: this will indeed decrease the kinetic energy of the atoms making them more sensitive to the effects of the trapping potential (that pushes them towards the center of the trap). This discussion shows that  the effect of the repulsion between the atoms is to shift the transition temperature downwards. This effect has been observed in experiments. It's magnitude has been estimated 
\begin{eqnarray}
\frac{\Delta T_{\rm c}}{T_{\rm c}}=-1.32\frac{a}{a_{\text{ho}}}N^{1/6},
\end{eqnarray}
where $a_{\text{ho}}=\sqrt{\hbar/m\omega}$.

This  negative shift of $T_{\rm c}$
 is purely a mean-field effect, and its physical origin is transparent, as we have seen. 

\subsection{Towards the calculation of $\Delta T_{\rm c}$ in an uniform system}

In a uniform system, the effect of the interaction leads to a positive  shift in the critical temperature, which is therefore opposite to the effect that we just discussed for atoms in a trap. 
The physics responsible for a change in $T_c$ in uniform systems  has in fact  nothing to do with what happens in a trap where mean field effects dominate: as we have seen, in uniform systems mean field corrections do not produce any shift in $T_c$.

\begin{figure}
 \begin{center}
  \includegraphics[width=0.5\linewidth]{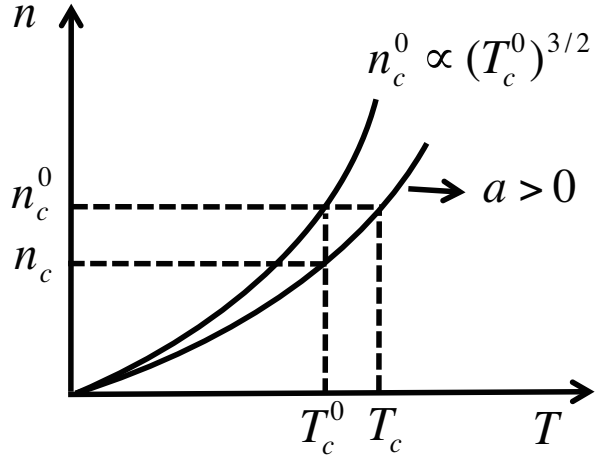}
  \caption{The critical lines $n_c(T_c)$ for the non interacting ($n_c^0$) and the interacting ($a>0$ ) systems. }
  \label{fig1}
 \end{center}
\end{figure}

Let me first of all indicate a useful relation, which is a
purely geometrical relation, valid in leading order in the interaction strength. 
It is a relation between 
the shift in the critical density and that in the critical temperature.
Figure~\ref{fig1} represents the phase diagram in the density-temperature plane
in the cases where $a=0$ and $a>0$.
I am assuming here that
in the presence of the interaction, there is a still a phase transition
similar to Bose-Einstein condensation, and that
 if $a$ is small the critical line is only slightly displaced  from where it is when $a=0$. The relation that I am referring to is 
\begin{eqnarray}
\frac{\Delta T_{\rm c}}{T_{\rm c}}=-\frac{2}{3}\frac{\Delta n_{\rm c}}{n_{\rm c}},
\end{eqnarray}
which just follows from the fact  that the two curves are very close to each other, and the fact that
I am looking at the leading order in $a$. 
The factor  $2/3$ has its origin in the relation $n_c^0\sim (T_c^0)^{3/2}$, while the  minus sign is obvious from the figure.
Why is that relation important? It is important because $\Delta n_{\rm c}$ is much easier to calculate than $\Delta T_{\rm c}$. This is so because it is easier to calculate 
at a fixed temperature than to calculate at a fixed density (in the latter case, you need to adjust the chemical potential to keep the density fixed as you change $a$). 

Let us go back to the interaction. The hamiltonian that we
consider is of the form
\begin{eqnarray}
H=\int d^3r \left\{\psi^{\dagger}(\r)
\left(-\frac{\Delta}{2m}\right)
\psi(\r)+\frac{g}{2}
\psi^{\dagger}(\r)\psi^{\dagger}
(\r)
\psi(\r)\psi(\r)
\right\},\qquad g=\frac{4\pi a}{m},
\end{eqnarray}
which is the hamiltonian (\ref{hamiltonianbosons}), and we have recalled the relation between the coupling constant $g$ and the scattering length $a$. 
When using this  hamiltonian we assume that we can ignore a lot of details of the atomic physics. The potential between the atoms is replaced by  a contact potential, which is of course meaningful only if the atoms are on the average far from each other. Only then can we ignore the details of the atom-atom interaction, its dependence on specific electronic levels, etc. To be more precise, this hamiltonian will be used to describe modes of the Bosonic fields whose wavelengths $1/k$ are large compared to the range $r_0$ of the potential, 
\begin{eqnarray}
kr_0\ll 1.
\end{eqnarray}
We have also mentioned another condition, that involves the scattering length $a$, 
\begin{eqnarray}
na^{1/3}\ll 1.
\end{eqnarray}
This is the statement that the distance between the atoms is large compared to $a$. Recall finally that in the vicinity of the transition, $n\lambda^3\sim 1$, where $\lambda$ is the thermal wavelength. The condition on the scattering length translates then into the relation $a\ll \lambda$. 

In order to calculate $\Delta n_{\rm c}$, it is useful to express 
the density  in terms of  the
propagator (in the imaginary time formalism):
\begin{eqnarray}
n=\lim_{\tau\to 0^-}\langle T\psi(\tau,\r)
\psi^{\dagger}(0,\r) \rangle
=\lim_{\tau\to 0^-} G(\tau,\r).
\end{eqnarray}
The Fourier transform of the propagator obeys the Dyson equation:
\begin{eqnarray}
G^{-1}(i\omega_n,\p)=G_0^{-1}
(i\omega_n,\p)+
\Sigma (i\omega_n,\p),
\end{eqnarray}
 where $G_0$ is the free propagator, $\Sigma$  the self-energy, and $\omega_n=2n\pi T$  is
a Matsubara frequency.
Therefore, the expression for the density of particles can be
written as
\begin{eqnarray}
n=\lim_{\tau\to 0^-}T\sum_n\int \frac{d^3\p}{(2\pi)^3}
\frac{e^{-i\omega_n T}}{\epsilon_\p-i\omega_n
+\Sigma(i\omega_n,\p)},
\end{eqnarray}
where $\epsilon_\p=\epsilon_\p^0-\mu$ and $G_0^{-1}(i\omega_n,\p)=\epsilon_\p^0-\mu-i\omega_n$.
This is an exact relationship: It allows me to calculate $n$, provided I know how  to calculate the self-energy.

Now I need to say a few words about the condensation condition.
I shall give you the result and try to motivate it.
I am going to assume that condensation takes place when
\begin{eqnarray}
G^{-1}(\omega=0,\p=0)=0.
\end{eqnarray}
Note first that, in the non interacting case, the condition $G^{-1}_0(0,0)=0$ yields $\mu=0$, which is indeed the condensation condition that we have already met. 
Turning to 
the interacting system, one may  recognize that $G^{-1}(i\omega_n=0,\p=0)$   
is just the second derivative of the ``effective potential'' (the free energy expressed in terms of the expectation values of the field $\psi$ and $\psi^{\dagger}$). Assuming that
the condensation is a second-order phase transition, the second derivative of this effective potential vanishes at the transition. This is just the statement that $G^{-1}(0,0)=0$.

By using the explicit expression $G^{-1}(i\omega_n,\p)=\epsilon_\p^0-\mu-i\omega_n+\Sigma(i\omega_n,\p)$, 
one sees that the condition  $G^{-1}=0$ reduces to 
\begin{eqnarray}
\mu=\Sigma(0,0).
\end{eqnarray}
Now I can calculate the shift $\Delta n_{\rm c}$ in the critical density. I get
\begin{eqnarray}
\Delta n_{\rm c}=\lim_{\tau\to 0^-}T\sum_ne^{-i\omega_nT}
\int\frac{d^3p}{(2\pi)^3}
\left\{\frac{1}{\epsilon_p^0+\Sigma(i\omega,\p)
-\Sigma(0,0)-i\omega_n}-
\frac{1}{\epsilon_p^0-i\omega_n}
\right\}.
\end{eqnarray}
The first term is the critical density of the interacting system, for which $\mu=\Sigma(0,0)$, the second term is the critical density of the non interacting system at the same temperature, for which $\mu=0$. 
Note that the only place where the interaction enters 
is the self-energy. So we have to evaluate the self-energy.
You see here that when the self-energy is frequency and momentum
independent, there is no correction, because if the self-energy does not
depend on $\omega$ or $\p$
it is the same as for $\omega=p=0$, and then the two terms cancel and $\Delta
n_{\rm c}=0$.

\begin{description}
\item{\bf Q}: This is precisely what happens in the mean-field approximation?
\item{\bf A}: Yes. As I said, the mean field approximation 
leads to a self-energy which is independent of frequency and momentum. That produces no shift of $T_{\rm c}$.  To get a non-trivial effect you need frequency
and momentum dependences in the self-energy.
\end{description}

\begin{figure}
 \begin{center}
  \includegraphics[width=0.5\linewidth]{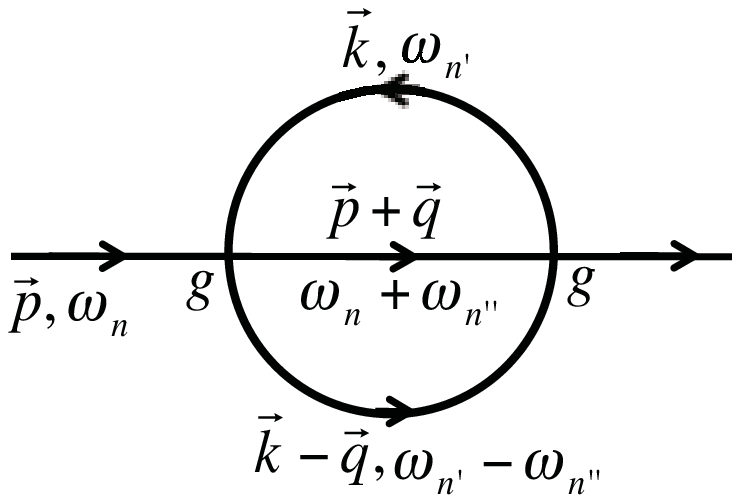}
  \caption{Second order diagram contributing tho the self-energy.}
  \label{fig3}
 \end{center}
\end{figure}

Let me start the calculation of the correction.
We have already done the first-order calculation and shown this to be zero.
Therefore, we have to go to
the second-order.  I shall come back to it next time.
The second-order diagram is shown in Fig.~\ref{fig3}.
The corresponding contribution to the self-energy reads
\begin{eqnarray}
\Sigma (i\omega_n,\p)
=-2g^2T^2\sum_{n^{'},n^{''}}\int\frac{d^3\k}{(2\pi)^3}
\int\frac{d^3\q}{(2\pi)^3}
\frac{1}{\epsilon_{\k-\q}-i(\omega_{n'}-\omega_{n^{''}})}
\frac{1}{\epsilon_{\k}-i\omega_{n'}}
\frac{1}{\epsilon_{\p+\q}-i(\omega_{n}+\omega_{n^{''}})},
\label{second}
\nonumber\\
\end{eqnarray}
where the factor $T^2$ comes from the  double sum over the Matsubara frequencies.
I am going to focus on the contributions with $n^{'}=n^{''}=0$.
Remember that the infrared divergences of the scalar field theory
are coming from this particular sector where all the Matsubara
frequencies vanish.
I am going to look at this particular contribution, and follow the same strategy as a  few lectures ago when I discussed
the divergences in high temperature QCD and I exhibited 
special classes of  Feynman diagrams  which are infrared divergent.
I used a power-counting argument and replaced the multiple
integrations over  the momenta by an integration over a big momentum vector in a 
 large space and I ignored the angular integral.
Let me  proceed in the  same way  quickly today (I  shall come back to this result
next time) because I would like to show you that there is difficulty.
Think therefore of the momentum integrals over $\q$ and $\k$  as an integral of a big vector in 6 dimensions. Then the integral in 
 Eq. (\ref{second}) behaves as \begin{eqnarray}
\sim\int\frac{k^5dk}{k^6}=\int\frac{dk}{k}.
\end{eqnarray}
This is a logarithmically divergent integral.

\begin{description}
\item{\bf Q}: How about the external momentum $p$? It does not provide any cutoff?

\item{\bf A}: Well, it does, but in a subtle way that we shall discuss more precisely next time. For today, let me just observe that what we need to calculate is $\Sigma(0,\p)-\Sigma(0,0)$, and there is no $\p$ in $\Sigma(0,0)$.
\end{description}

What will come after is very much the same thing as what we met when we discussed  QCD at finite temperature.
Namely, when I calculate the third-order, the fourth-order, etc., I find increasingly divergent contributions. 
The pattern of divergences becomes worse and worse as the order increases.
What I want to argue next time is
that this situation is very much similar to what we have met already. You will see that  this pattern of divergences is indeed that of  3-dimensional field theory.
Having recognized that, we will know what to do, namely,
how to construct an effective theory for the sector where the divergences occur. This will allow us to establish the linear relation between $\Delta T_c$ and $a$, and to obtain  an explicit formula for
the coefficient $c$ in the formula (\ref{shiftTc}).
I will then explain to  you why it is hard to  calculate explicitly this coefficient.
\begin{description}
\item{\bf Q}:  In the last integral, I think you used a large momentum approximation?

\item{\bf A}:      Not really. But, of course, the logarithmic  divergent integral involves a ratio of momenta, and it is large when one of the momenta is large (or small) compared to the other. 
But I am really concerned here by the low momentum sector.

\item{\bf Q}: This approximation is on the low momentum?

\item{\bf A}:  Yes.  I am assuming that all the loop momenta are going to zero at the same rate. 
Note that, for this second order case, I can calculate the integral explicitly and
I shall do that next time. I hope then that this question will be completely clarified.
\end{description}


\section {LECTURE V\hspace{-.1em}I} \label{lecture6}

\begin{figure}
	\centering
	\includegraphics[width=6cm,clip]{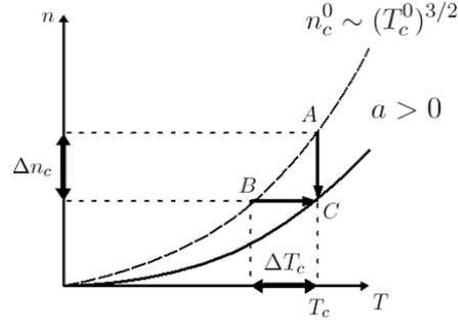} \\
	\caption{The critical lines $n_c(T_c)$ for the non interacting ($n_c^0$) and the interacting ($a>0$ ) systems. }
	\label{Fig_lec6:nT}
\end{figure}

Let me remind you that we want to establish the following formula for the shift in the critical temperature of the Bose-Einstein condensation caused by weak repulsive interactions:
\begin{equation}
	\frac{\Delta T_c}{T_c} = c (a n^{1/3}), \qquad a n^{1/3} \ll 1.
\label{Eq_lec6:shift of Tc}
\end{equation}
We want to establish this formula to leading order in the strength of the interaction, 
measured by the scattering length $a$. 
The plot in Fig.~\ref{Fig_lec6:nT} displays the critical line for the 
non-interacting gas. It  goes like $n_c\sim T_c^{3/2}$. Also drawn is the critical line for 
the interacting system. We assume that it differs very little from that of the non-interacting system when $a$ is small, and the  curve is drawn here  for a positive scattering length. 

I'm interested in the critical temperature $T_c$ at point $C$ on Fig.~\ref{Fig_lec6:nT}: $T_c$ is the transitiion temperature of the interacting system for some small positive value of $a$.  Starting from the non interacting system, $a=0$,  I can approach this point $C$ either by moving the density
downward at constant temperature (going from $A \to C$) or moving upward the temperature at constant 
density (going from $B \to C$).
I argued  last time, based solely on the   
geometrical propeties of Fig.~\ref{Fig_lec6:nT}, that $\Delta T_c$ and $\Delta n_c$ are related by the
simple equation:
\begin{equation}
\label{Eq_lec6:delta_Tc/Tc_result}
	\frac{\Delta T_c}{T_c}
=
	-\frac{2}{3} \frac{\Delta n_c}{n_c}
\end{equation}
I also argued last time that  it is much easier to 
calculate $\Delta n_c$, i.e., to work at fixed temperature. 

The last thing I want to remind you is  the condensation condition, expressed as the vanishing of the inverse propagator for vanishing momentum and Matsubara frequency. This translate into a condition on the self-energy:
\begin{equation}
\label{Eq_lec6:condensation_condition}
	G^{-1} (\omega = 0, \mathbf{p} = 0 ) = 0 \quad
\to \quad
\Sigma (\omega = 0, \mathbf{p} = 0 ) = \mu 
\end{equation}
With this,  we can write the formula for the shift $\Delta n_c$:
\begin{align}
	\Delta n_c
=
	\lim_{\tau \to 0^-}
		T \sum_n
&
			e^{-i \omega_n \tau}
			\int \frac{ \mathrm{d}^3 p }{ (2\pi)^3 }
\notag \\
& \times
				\left\{
					\frac{1}
						 {\epsilon_p^0 + \Sigma (i\omega_n, \mathbf{p}) - \Sigma (0, 0) - i\omega_n}
					-
					\frac{1}
						 {\epsilon_p^0 - i\omega_n}
				\right\}
\label{Eq_lec6:delta_nc}
\end{align}
The first term of the integral is the density of the interacting system 
at criticality. The  subtracted term is the density of the same system at the 
same temperature, but without interaction (we used here the fact that at criticality the chemical potential in the non-interacting system is zero).
From this formula you may observe that if the self-energy is a constant, 
independent of frequency and momentum then 
$\Sigma (i\omega_n, \mathbf{p}) - \Sigma (0, 0)$ vanishes,  and so does $\Delta n_c$. This is what happens for instance in the mean field (one-loop) calculation that I did last time.

Now perhaps I should add some word of caution here. Sometimes you may be led to 
consider a mean field calculation where the dominant effect of the interaction 
leads to a momentum dependent potential, that is, $\Sigma$ doesn't depend on frequency but 
depends on momentum (note that for this to occur, you need to go beyond the approximation that consists in replacing the atom-atom interaction by a contact potential). That dependence on momentum can often be 
described by an effective mass. Such an effect would produce a shift, just 
because the effective mass enters for instance the thermal wavelength. But this is not the kind of effect that I am interested in here. 

\subsection{Power counting and infrared divergences}

\begin{figure}
	\centering
	\includegraphics[width=6cm,clip]{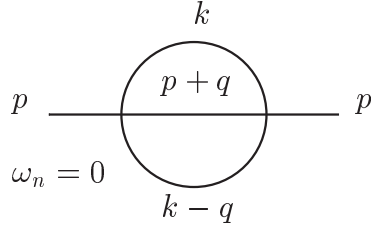} \\
	\caption{Second order diagram contributing to the self-energy.}
	\label{Fig_lec6:second order diagram}
\end{figure}

We also started discussing the interaction. I remind you that the typical 
hamiltonian we use is given as
\beq
	H =
		\int \mathrm{d}^3 \mathbf{r} \,
			\left\{
				\psi^\dagger (\mathbf{r}) 
				\left( -\frac{\nabla^2}{2m} \right)
				\psi         (\mathbf{r})
				+
				\frac{g}{2} \psi^\dagger (\mathbf{r})\psi^\dagger(\mathbf{r}) \psi(\mathbf{r}) \psi (\mathbf{r})
			\right\}
	, \quad
		g =\frac{4 \pi a}{m}.
\label{Eq_lec6:hamiltonian}
\eeq
We argued  that the  leading 
order correction to the  single particle energies, of order $g$, produces no shift in $T_c$, and we started considering  the 
calculation of the second order diagram (see Fig.~\ref{Fig_lec6:second order diagram}). 
Then I argued that one could expect infrared 
divergences when all Matsubara 
frequencies are equal to zero. So, I'm going to calculate this 
diagram only in this particular case. Let me  give you the complete expression, near criticality:
\begin{align}
		\Sigma (i\omega_n=0 , \mathbf{p}) - \Sigma (0, 0)
	&=
		-2 g^2 T^2
		\int \frac{ \mathrm{d}^3 k }{ ( 2\pi )^3 }
		\int \frac{ \mathrm{d}^3 q }{ ( 2\pi )^3 }
\notag \\
	& \times
			\frac{1}{ 
				( \epsilon^0_{ \mathbf{k}-\mathbf{q} } - \mu^{\prime} )
				( \epsilon^0_{ \mathbf{k}            } - \mu^{\prime} ) 
			}
			\left(
				\frac{1}{ \epsilon^0_{ \mathbf{p}+\mathbf{q} } - \mu^{\prime} }
				-
				\frac{1}{ \epsilon^0_{            \mathbf{q} } - \mu^{\prime} },
			\right)
\label{Eq_lec6:second order diagram}
\end{align}
where $\mu^{\prime}$ is given as
\begin{equation}
	\mu^{\prime} \equiv \mu - 2gn = - \frac{\kappa_c^2}{2m}
\label{Eq_lec6:mu prime}
\end{equation}
I'm doing a calculation in second order perturbation theory. But I'm 
incorporating in the chemical potential the shift due to the first order 
diagram. At the mean field level, the
condensation  takes place in exactly the same way as in the non-interacting system:
$\mu^{\prime}$ is negative until one reaches condensation where it vanishes. The notation with $\kappa_c$ is convenient, as it allows me to express the  
energy denominators  as follows:
\begin{equation}
		\epsilon^0_{\mathbf{k}} - \mu^{\prime}
	=
		\frac{1}{2m} ( k^2 + \kappa_c^2 ),
\label{Eq_lec6:energy denominator}
\end{equation}
and $\kappa_c$ will play the role of an infrared regulator 
in the integrals that I'm going to calculate. Within the present perturbative setting, it measures the deviation from criticality. 

Let me give you the value of the integral in Eq.~(\ref{Eq_lec6:second order diagram}). This  can be 
calculated analytically for finite $\kappa_c$:
\begin{align}
		\Sigma(0,\p)-\Sigma(0,0)
	=
		\frac{128 \pi^2}{2m} \left( \frac{a}{\lambda^2} \right)^2
		\left\{
			\frac{3 \kappa_c}{p} \tan^{-1} \frac{p}{\kappa_c}
			+
			\frac{1}{2} \ln \left( 
				1 + \left( \frac{p}{3 \kappa_c} \right)^2
			\right)
			- 
			1
		\right\}
.
\label{Eq_lec6:result of second order diagram}
\end{align}
\begin{figure}
	\centering
	\includegraphics[width=6cm,clip]{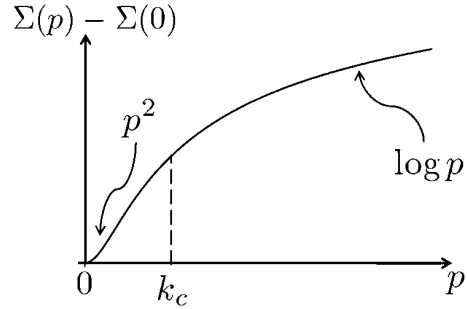} \\
	\caption{$\Sigma(p) - \Sigma(0)$ given by Eq.~(\ref{Eq_lec6:result of second order diagram}) for small $p$ and large $p$}
	\label{Fig_lec6:Sigma}
\end{figure}%
This result  allows me to do a couple of comments. The quantity  $\Sigma (p) - \Sigma (0)$ is plotted schematically  in  Fig.~\ref{Fig_lec6:Sigma}. For 
small $p$, $p\ll \kappa_c$,  it goes like $p^2$, while for large $p$, $p\gg\kappa_c$,  it goes like 
$\ln p$.  When $p$ 
is very large compared to $\kappa_c$, $\Sigma$ has a logarithmic behavior. And when $p$ is small compared to $\kappa_c$, $\Sigma$ is a smooth 
function which has a regular momentum expansion. Of course $\kappa_c$ is 
a quantity that  we would like to let go to $0$, because we want to approach the 
condensation. But when $\kappa_c$ goes to $0$ we get  an infrared divergence. That's the   infrared divergence that I mentioned at  the end of the last lecture. 

Now I will show you that this peculiar infrared behavior is actually not 
limited to the second order diagram. It will show up in higher orders. We 
are going to rediscover the pattern that we have already identified in the case 
of hot QCD, or in the case of the thermodynamics of the scalar field that we 
discussed a few lectures ago. Let me repeat the  analysis that I did last time for the 
second order diagram (Fig.~\ref{Fig_lec6:second order diagram}). Focusing on the region where  all the momenta are going to zero at the same rate, we regroup the two loop-momenta into a single 6-dimensional vector $\bf K$, and write
\begin{align}
		g^2 T^2 \int 
			\frac{ \mathrm{d}^6 K  }
				 { (K^2 + \kappa_c^2)^3 }
			(2m)^3
	& \sim
		\frac{a^2}{m^2} T^2 m^3
		\int_{\kappa_c}
			\frac{ K^5 \mathrm{d} K }{ K^6 }
\notag \\
	& \sim
		\frac{1}{m} \left( \frac{a}{\lambda^2} \right)^2
		\int_{\kappa_c}
			\frac{ \mathrm{d} K }{ K }
\label{Eq_lec6:two loop integral}
\end{align}
This is a two loop diagram, so there are two momentum  integrations, hence the $\mathrm{d}^6 K$. Then are three propagators 
of the form $1/(K^2 + \kappa_c^2)^2$. And there are a number of coefficients: there is a 
factor $g^2$ coming from the two vertices, and a factor $T^2$ coming from the two summations over the Matsubara frequencies ( I keep only the zero Matsubara frequencies, but the 
factor $T^2$ that accompanies the double  sum remains). The  factor $(2m)^3$ comes from the $p^2/2m $ in the three propagators. The last line is obtained by remembering that $g\sim a/m$, and that  $mT\sim1/\lambda^2$. Also, to make it easier to see the divergence of the integral  when $\kappa_c\to 0$, I have  removed $\kappa_c$ from the denominators, and put it at 
the lower end of the integration. Then I ignored the angular integration, which would produce just a numerical factor.
You see now that in the limit where $\kappa_c$ goes to 0 this integral is
logarithmically divergent. Also, it  is proportional to $(a/\lambda^2)^2$. 

\begin{figure}
	\centering
	\includegraphics[width=5cm,clip]{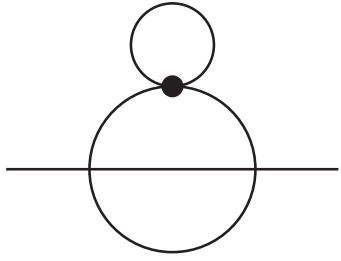} \\
	\caption{2+1 loop diagram}
	\label{Fig_lec6:2+1 loop diagram}
\end{figure}

Let us now  generalize this calculation. I go from this 2-loop diagram to an $l$-loop diagram. 
Let's see what happens by  adding  a simple loop as shown in Fig.~\ref{Fig_lec6:2+1 loop diagram}. 
When I add  a loop I add  one vertex and  two propagators. One can easily verify that this is generic. 
The calculation of the $l$-loop diagram is then going to take the following form:
\begin{equation}
		\frac{1}{m} \left( \frac{a}{\lambda^2} \right)^2
		\int
			\frac{ \mathrm{d}^6 K  }
				 { (K^2 + \kappa_c^2)^3 }
			g^{l-2} T^{l-2}
		\int
			\frac{ ( \mathrm{d}^3 K )^{l-2} }
				 { (K^2 + \kappa_c^2)^{ 2(l-2) } }
			(2m)^{ 2(l-2) }.
\label{Eq_lec6:add a loop}
\end{equation}
In this calculation I have added $l-2$ loops to the 2-loop digram of Fig.~\ref{Fig_lec6:second order diagram},  in order to 
get an $l$-loop diagram.
By doing so, I have added  $l-2$ loops,  $l-2$ vertices, hence the factor $g^{l-2} T^{l-2}$.
Then I have an integration $(\mathrm{d}^3 K)^{l-2}$ and $2(l-2)$ propagators 
$(K^2 + \kappa_c^2)^{-2(l-2)}$,  and a factor $(2m)^{2(l-2)}$. You are still with me?
 I can rewrite these factors as 
\begin{equation}
		(g)^{l-2} (Tm)^{l-2} m^{l-2}
	\sim
		\left( \frac{a}{m} \right)^{l-2}
		\left( \frac{1}{\lambda^2} \right)^{l-2}
		m^{l-2}
	=
		\left( \frac{a}{\lambda^2} \right)^{l-2}
	.
\label{Eq_lec6:rewrite factor}
\end{equation}
In the integral, I remove $\kappa_c$ in the denominator, and put it as a lower 
bound. After all this, I can rewrite the expression (\ref{Eq_lec6:add a loop}) as 
\begin{equation}
		\frac{1}{m} \left( \frac{a}{\lambda^2} \right)^l
		\int_{\kappa_c}
			\frac{ K^{3l-1} \mathrm{d} K}{ (K^2)^{2l-1} },
\label{Eq_lec6:rewrite l loop}
\end{equation}
so that  the $l$-loop contribution to $\Sigma$ reads 
\begin{equation}
		\Sigma^{(l)} (p=0)
	\sim
		\frac{1}{m} \left( \frac{a}{\lambda^2} \right)^l
		\int_{\kappa_c}
			\frac{ \mathrm{d} K }{ K^{l-1} }
	\sim
		\frac{1}{m} 
		\left( \frac{a}{\lambda^2}               \right)^2
		\left( \frac{a}{\lambda^2} \frac{1}{\kappa_c} \right)^{l-2}.
\label{Eq_lec6:l loop result}
\end{equation}
 As anticipated, we are indeed confronted to a situation which is very reminiscent 
to one that we have already encountered. Namely, as $\kappa_c\to 0$ all terms are infrared divergent. I will show 
you later that there are good reasons to choose $\kappa_c$ of order $a/\lambda^2$. 
But if you do that, $a/\lambda^2 \cdot 1/\kappa_c$ is of order 1. That means 
that  all the diagrams in 
perturbation theory will be  of the same order of magnitude. 
 Perturbation theory breaks down. It cannot be used to 
calculate the shift of the critical temperature. But we know what to do. This is 
very analogous to what we have encountered earlier. And I have shown you in 
the case of the scalar field that we can develop other tools to handle this 
problem. One of them is effective field theory, to which I now turn. 


\begin{figure}
 \begin{tabular}{ccc}
  \begin{minipage}{0.33\textwidth}
   \begin{center}
    \includegraphics[scale=0.8,clip]{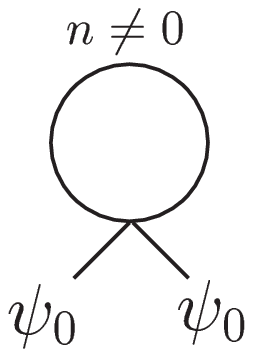}
   \end{center}
  \end{minipage} & 
  \begin{minipage}{0.33\textwidth}
   \begin{center}
    \includegraphics[scale=0.8,clip]{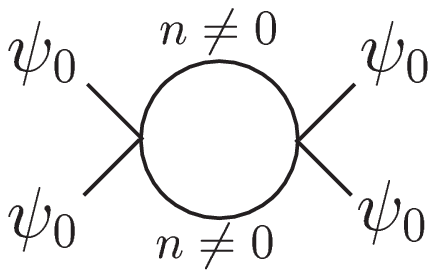}
   \end{center}
  \end{minipage} &
  \begin{minipage}{0.33\textwidth}
   \begin{center}
    \includegraphics[scale=0.7,clip]{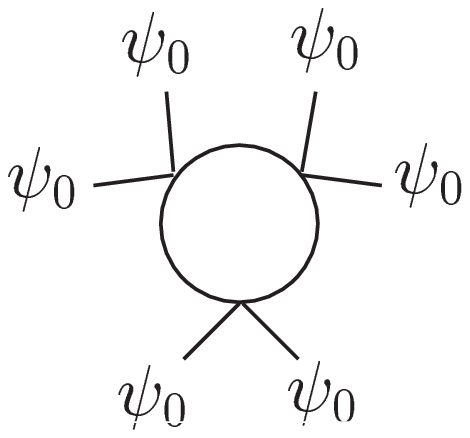}
   \end{center}
  \end{minipage} \\
  (a) &(b) &(c) \
 \end{tabular}
 \caption{Some diagrams contributing to the parameters of the effective theory}
 \label{Fig_lec6:246legs}
\end{figure}

\subsection{Effective theory}

Infrared divergences  occur in Feynman diagrams only when all the Matsubara frequencies $\omega _n$ are set equal to 0.
Let's then focus on this particular sector and recall what we did for the scalar field.  
The effective theory amounts to retain in the field expansion 
\begin{align}
\psi (\tau ,\r) =T \psi _0 (\r) +T\sum _{n\neq 0} e^{-i\omega _n \tau} \psi _n (\r),
\end{align}
only the component $\psi _0 (\r)$, which does not depend on the imaginary time. As you know from previous lectures the effective theory for 
$\psi _0 (\r)$ will be an effective theory for a field in 3 dimensions.
One dimension is lost because I have abandoned the imaginary time dependence.

If I just replace the field  $\psi (\tau,\r)$ by $\psi _0 (\r)$, I can perform trivially the integration of 
the imaginary time in the action, and get:
\begin{align}
{\cal Z} \sim \int \mathcal{D} \psi _0 \,e^{-\beta \int d^3 \r \left( \mathcal{H} (r)-\mu {\mathcal N}(r)\right)},
\end{align}  
where 
\begin{align}
\mathcal{H} (r)-\mu {\mathcal N}(r) 
 = \psi _0 ^* (r) \left( -\frac{\nabla ^2}{2m} -\mu \right) \psi _0 (r) +\frac{g}{2} \left( |\psi _0 (r)|^2 \right) ^2.
 \label{Eq_lec6:H-mN}
\end{align}
This is the leading order, where the effective action is just $\beta$ times the energy of the field configuration. This is  what  I called earlier the classical field approximation.  
We  know how to calculate the corrections to this simple approximation. 
Remember that I described to you the systematic procedure to do that. What we have to do is  to calculate Feynman diagrams where the external lines corresponds to  $\psi _0 (\r)$ or $\psi _0^* (\r)$
and the internal lines corresponds to the modes with non-vanishing Matsubara frequencies. 
Such diagrams are displayed in Fig.~\ref{Fig_lec6:246legs}. The diagram of Fig.~\ref{Fig_lec6:246legs}a is proportional to $a$, and it is a correction to the chemical potential $\mu$. 
We shall not need to evaluate this  correction because the chemical potential is 
eventually adjusted so that the system is  critical, or at the phase transition. 
But there are other corrections. 
Let me first look at the correction to the coupling constant, the diagram Fig.~\ref{Fig_lec6:246legs}b.  
This  correction is proportional to $a^2$. 
Since we are interested in leading order calculation, it can be ignored. 
The correction in Fig.~\ref{Fig_lec6:246legs}c represents a three-particle interaction, and 
 corresponds to a term not present in Eq.~(\ref{Eq_lec6:H-mN}). 
This is of order $a^3$, and can again be ignored.
The lesson of this brief analysis, which we have done more extensively in the case of the scalar field theory, is that if one is  interested in
the leading order effect of the coupling constant, one can just use, as an effective theory, the classical field approximation,  
whose energy density is given by Eq.~(\ref{Eq_lec6:H-mN}). 

However, we are a priori  somewhat stuck here, because with this effective theory perturbation theory cannot be used even though the coupling $g$ can as small as one wants.
Let me briefly review that issue, and recall why \textit{a priori} we should expect a problem with perturbation theory.
When we do perturbation theory, we are assuming in some way that, in Eq.~(\ref{Eq_lec6:H-mN}), the kinetic energy $\frac{1}{2m} (\nabla \psi _0 )^2$ 
is big compared to the potential energy $g\left( |\psi _0 |^2 \right) ^2$. 
Naively we expect that if $g$ is very small,  the potential energy is correspondingly small. But this is not always the case. 
Perturbation theory breaks down precisely  at that particular scale where all the terms 
in the effective action are of the same order of magnitude, that is, in particular, when (in average)
\begin{align}
\frac{1}{2m} (\nabla \psi _0 )^2 \sim g\left( |\psi _0 |^2 \right) ^2. \label{Eq_lec6:KvsP}
\end{align}
Let me repeat here an analysis that we have done in a more general context. Remember that we can separate layers of fluctuations corresponding to different momenta or different wavelengths. Consider the particular fluctuations of momentum $\kappa_c$, so that 
\begin{align}
(\nabla \psi _0 )^2 \sim \kappa_c ^2 |\psi _0 |^2.
\end{align}
The density  in the effective theory here is $\langle |\psi _0 |^2\rangle$. 
This is also given by 
\begin{align}
n \sim \langle |\psi _0 |^2\rangle 
  \sim \int \frac{d^3 p}{(2\pi )^3} \frac{1}{e^{(\epsilon _p -\mu)/T} -1}. \label{Eq_lec6:density1}
\end{align}
Close to condensation, I can ignore the chemical potential (or a constant self-energy), and replace $(\epsilon _p -\mu)/T$ by $p^2/2mT$.  For  long wavelength modes, $p^2/2mT$ is small and I can expand the exponential factor,
and rewrite Eq.~\eqref{Eq_lec6:density1} as
\begin{align}
n \sim \int \frac{d^3 p}{(2\pi )^3} \frac{2mT}{p^2}. \label{Eq_lec6:density2}
\end{align}  
(This approximation  is in fact in line with the classical field approximation.)
This  integral diverges at large momentum. But it is supposed to represent only the contribution to the density of long wavelength modes, with $p\lesssim \kappa_c$. 
With  $\kappa_c$ as ultraviolet cut-off, this gives $n\propto mT\kappa_c$. 
Let's then compare $\frac{\kappa_c ^2}{2m} |\psi _0|^2$ with $g\left( |\psi _0 |^2 \right) ^2$, or equivalently $\frac{\kappa_c ^2}{2m}$ and $g |\psi _0|^2$, with $\langle |\psi _0|^2\rangle\approx mT\kappa_c$. 
I can rewrite Eq.~\eqref{Eq_lec6:KvsP} as 
\begin{align}
\frac{k_c ^2}{2m} \sim gmT\kappa_c.
\end{align}
Remembering  that $g\sim a/m$ and $mT\sim 1/\lambda ^2$, 
one sees that the condition \eqref{Eq_lec6:KvsP} implies that
\begin{align}
\kappa_c \sim \frac{a}{\lambda ^2}.
\end{align}
This is the characteristic scale that we encountered before and which signals the breakdown of perturbation theory. We see here how this scale emerges  from a simple analysis of  the effective action.

Now, if we cannot expand, what can we do? 
We can still make progress, 
because we know that Eq.~\eqref{Eq_lec6:delta_nc} is an exact formula for $\Delta n_c$, and we  can deduce from it an equivalently ``exact''  formula for $\Delta n_c$ in the framework of the effective theory.
This is very easy to do.
We just ignore all the terms which have $n$ different from 0 in Eq.~\eqref{Eq_lec6:delta_nc}. 
This leads to 
\begin{align}
\Delta n_c = \int \frac{d^3 \p}{(2\pi )^3} \left \{ \frac{1}{\epsilon _\p ^0 +\Sigma _{cl} (\p) -\Sigma _{cl} (0)}
 -\frac{1}{\epsilon _\p ^0} \right \}. \label{Eq_lec6:Dn_c1}
\end{align}
where $ \Sigma _{cl}(\p)\equiv \Sigma(i\omega_n=0,\p)$. This formula is ``exact'' in the effective theory, in the sense that it does not involve any approximations, beyond those made to arrive at the effective theory. 

\begin{figure}
 \begin{center}
    \includegraphics[scale=0.25,clip]{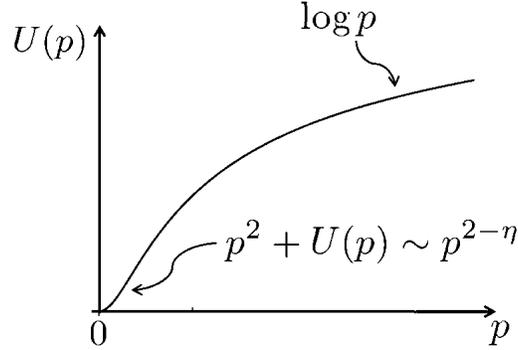}
 \end{center}
 \caption{General behavior of $U(p)$ as a function of $p$.}
 \label{Fig_lec6:Up}
\end{figure}

I'm going now to introduce a new notation. Let me set  
\begin{align}
U(\p) = 2m \left( \Sigma _{cl} (\p) -\Sigma _{cl} (0) \right).
\end{align}
Then, with a little algebra, I can rewrite the formula \eqref{Eq_lec6:Dn_c1} as 
\begin{align}
\Delta n_c = -\frac{2}{\pi \lambda ^2} \int _0 ^\infty dp\, \frac{U(p)}{p^2 +U(p)}. \label{Eq_lec6:Dn_c2}
\end{align}
This is still an exact formula (within the effective field theory). 
And if I know the exact expression for $U(p)$, then I can conclude something.
Now, I have to tell you a few more things about  the general behavior of $U(\p)$. 
Before, I have done the calculation with an infrared regulator: 
remember that when the momentum was smaller than the regulator,  $U(\p)$ was analytic,  $\sim p^2$. But when one removes the regulator, the behavior is actually logarithmic (in leading order), and more generally a power law,  $p^2 +U(p)\sim p^{2-\eta}$ where $\eta$, a small number, is the anomalous dimension (a small number of order 0.1). 
This power law behavior is characteristic of the scaling regime of a second order phase transition.
The main point at this stage is  that the integral in Eq.~\eqref{Eq_lec6:Dn_c2}, if  the function $U(p)$ 
 behaves as I indicated, is a  finite integral.

There is more that we can say, namely that  $U(p)>0$. I do not have a complete, analytical, proof to offer for this property. One may just observe that it is verified in the second order calculation, Eq.~(\ref{Eq_lec6:result of second order diagram}). The critical fluctuations correct the small momentum behavior ($p^2\to p^{2-\eta}$), but do not alter the fact that $U(p)$ is a growing function of  $p$ at small $p$. As for the large momentum behavior, it is correctly given by perturbation theory. Detailed calculations confirm that $U(p)$ is an increasing function of $p$, with the general shape displayed in Fig.~\ref{Fig_lec6:Up}. Now, if $U(p)$ is positive, so is the integral in Eq.~\eqref{Eq_lec6:Dn_c2}. 
Therefore $\Delta n_c$ is negative, which implies that $\Delta T_c$ is positive. 
This answers the question concerning the sign of  $\Delta T_c$.

The last thing we have to do is to show that  the shift of $\Delta n_c$ is linear in $a$. 
This can be done via a simple analysis. 
Let me first remark that the effective theory which I have written here in principle requires an ultra-violet cut-off.
You have seen that  already from the calculation of the density in Eq.~\eqref{Eq_lec6:density2}: 
if you want to calculate the density with a statistical factor of the form $\frac{2mT}{p^2}$,
you need an ultra-violet cut-off.
This is in line with the philosophy of the effective theory. 
Remember what we did in the scalar field theory: we introduced an intermediate scale, i.e., a separation scale $\Lambda$, 
and we integrated the modes above this scale $\Lambda$ to get an effective theory valid below that scale. 
Of course it may happen that the effective theory contains ultra-violet divergences,  and these divergences are 
in principle compensated by the cut-off dependence of the coefficients of the effective theory. 
Here you don't have to worry about such issues because, as I argued, 
the density enters mostly the correction to the chemical potential. 
Observe also that in the integral of Eq.~\eqref{Eq_lec6:Dn_c1} we are taking the difference of two contributions, and  this  difference is ultra-violet finite. 
But there are  cases where we need a cut-off $\Lambda$. 
What is the scale of this cut-off? To answer that question we need to ask ourselves first why we get ultra-violet divergences in the effective theory.
The answer to that is contained in Eq.~\eqref{Eq_lec6:density1}. The full statistical factor there, $1/(e^{(\epsilon _p -\mu)/T} -1)$, kills all the momenta which are bigger than the temperature. But we are using an approximation to the statistical factor that is valid only at small momenta, i.e., for $p\lesssim \sqrt{mT}$, so that the natural cutoff for the effective theory is of  order $\sqrt{mT}\sim 1/\lambda$.

Now  the effective theory  is a 3-dimensional theory, and  is  {\lq\lq}super renormalizable".
What it means is that the only ultraviolet divergence is the one that I have discussed.
It corresponds to a correction to the chemical potential (or to the mass,  in the language of field theory). 
As I have argued, such a correction to the chemical potential is innocuous 
because  the chemical potential is adjusted to be at criticality.
Let us then go through some dimensional analysis. 
As you've seen, there is a natural momentum scale in the problem which is $a/\lambda ^2$.
So let us  set $p=x\frac{a}{\lambda ^2}$ (with $x$ a dimension-less variable).
In principle, quantities calculated within the effective theory depend   also on an ultra-violet cut-off $\Lambda\sim 1/\lambda$. 
Thus I can write 
\begin{align}
U(p=x\frac{a}{\lambda ^2},\Lambda ) = \left( \frac{a}{\lambda ^2} \right) ^2 
 \sigma \left( x,\frac{a}{\lambda ^2} \frac{1}{\Lambda} \right),  \label{Eq_lec6:U}
\end{align}
where $\frac{a}{\lambda ^2} \frac{1}{\Lambda}\sim a/\lambda$ if $\Lambda\sim 1/\lambda$. In fact, since the theory is super renormalizable, I can let the cut-off $\Lambda$ go to infinity, and get a finite result. 
But the fact that I can let $\Lambda$ go to  infinity does  not guarantee that the results will not depend on $\Lambda$, if it is kept finite. Finite cutoff corrections will be truly negligible only when $a/\lambda$ is sufficiently small. 
In that case, I can replace Eq.~\eqref{Eq_lec6:U} by
\begin{align}
U(p=x\frac{a}{\lambda ^2},\Lambda ) \simeq \left( \frac{a}{\lambda ^2} \right) ^2 \sigma (x),
\end{align}
where $\sigma (x)$ is a universal function of $x$ (that is, independent of $a$).

Then I can rewrite $\Delta n_c$ as 
\begin{align}
\Delta n_c \simeq -\frac{2}{\pi \lambda ^2} \frac{a}{\lambda ^2} \int _0 ^\infty \frac{dx}{x}\, \frac{x\sigma (x)}{x^2 +\sigma (x)},
\label{Eq_lec6:Dn_c3}
\end{align}
where I have extracted the logarithmic integration measure to emphasize that the variations of the integrand are best visualized on a logarithmic scale (see Fig.~\ref{integrand} below). 
The integral is just a number, so that 
$\Delta n_c$ is proportional to $a$.
Basically, the dependence in $a$ just follows from dimensional considerations, once we have made sure that possible ultraviolet cutoff effects play no role (which, as we have argued, requires $a/\lambda$ to be small enough).

At this point, it is useful to briefly review the steps that led to this result.  
The first step is to recognize that perturbation theory doesn't work.
Naively one could think of expanding in powers of $a$, because $a$ is very small quantity.
But we have seen that this doesn't work because the calculation of  Feynman diagrams is plagued with infrared divergences. 
We have exploited the fact that these  divergences occur when all the Matsubara frequencies are vanishing.
Then, by relying on what we did earlier, I explained to  you how one can deal with all the Feynman diagrams at once, by constructing an effective theory.
In this particular case, the effective theory is extremely simple 
because it just amounts to replace in the original action the bosonic field, 
which is a function of three space coordinates and the imaginary time, 
by a field which is independent of time and depends  only on three spatial coordinates.
This leads to a three dimensional effective field theory, 
which in principal allows us to calculate $\Delta n_c$. 

The effective theory makes obvious the reasons for the breakdown of perturbation theory: there exists a typical scale of fluctuations at which kinetic energy and potential energy are   
of the same order of magnitude.  
Therefore, in order to calculate within this effective theory,  we have to use  tools other than perturbation theory, even when the coupling is small. 

But even without doing explicit calculations, we have been able to extract from  the effective theory   the 
 answers to   the questions that we are addressing. 
Namely  the fact that (a) $\Delta n_c$ is negative or $\Delta T_c$ is positive
and (b) $\Delta n_c$ is linear in $a$.

\begin{description}
\item{\bf Q}: You take some kind of classical field approximation somewhere?
\item{\bf A}: Yes. The effective theory that I have discussed is  what I called the classical field approximation. I have argued  that  Feynman diagrams are divergent when all Matsubara frequencies are vanishing.
Earlier in the lectures I have shown that  we can construct an effective field theory to handle systematically all these Feynman diagrams at once. This effective theory is in general an infinite series in local operators, such as 
 $\phi ^4$,  $\phi ^6$,  $\phi ^8$, $\phi^2(\nabla \phi )^2$, etc. An important point is that the coefficients can be calculated in perturbation theory. If you are interested in leading order, then the leading terms in the expansion are enough, and this leads to the classical $\phi^4$ theory in 3 dimension (with O(2) symmetry). 
So, in the particular context of the present problem, things are extremely simple because the effective theory is just the leading order term.  
Namely you just take the initial action, which is   
\begin{align}
S = \int _0 ^\beta d\tau \int d^3 \r \,\psi ^* (\r,\tau ) \left( -\frac{\nabla ^2}{2m} \right) \psi (\r,\tau ) + \cdots ,
\end{align}
and replace $\psi (\r,\tau )$ by $\psi _0 (\r)$, to get
\beq
S = \beta \int d^3 \r \,\psi _0 ^* (\r) \left( -\frac{\nabla ^2}{2m} \right) \psi _0 (\r) + \cdots .
\eeq
\item{\bf Q}:  Is that a classical field approximation? 
The Heisenberg $\hbar$ goes away by this procedure.
\item{\bf A}: Yes. If you wish. Another way to think about this approximation is as a high temperature (small $\beta$) approximation. 
In any case, this approximation allows me to do calculations  outside the framework of perturbation theory,
but still in leading order in $a$.
That is to say, all correction to the effective theory are order of $a^2$ or higher (these corrections may for instance induce corrections of order  $a^2 \log a$ to $\Delta T_c$).  
\item{\bf Q}:  In the  ratio of $\Delta T_c /T_c \propto n^{1/3}$,  both sides are dimensionless. But 
$\lambda$ depends on $T$. If you take the ratio, $\lambda$ disappears or not?
\item{\bf A}: If you go through the calculation you will see that the factor $n^{1/3}$ can be read as $1/\lambda$. And indeed $\lambda$ depends on $T$. But the temperature is fixed in the calculation of $\Delta T_c /T_c$.
That is,  $\lambda$ is here $\lambda _c$ which  can be estimated by taking the critical temperature in the absence of interaction. So this temperature dependence does not spoil the linear relation between $\Delta T_c /T_c$ and $a$.
\item{\bf Q}:  OK. Thank you.
\end{description}

Let me tell you a bit more before ending this lecture.
We have established the relation
\begin{align}
\frac{\Delta T_c}{T_c} = c\left( an^{1/3} \right),
\end{align}
in which  $c$ is a positive constant, given by an integral of the form 
\begin{align}
c \propto \int _0 ^\infty dx \frac{\sigma (x)}{x^2 +\sigma (x)}.
\end{align}
I remind you that $\sigma (x)$ is essentially  $\Sigma _{cl} (p)-\Sigma_{cl}(0)$.
So, provided you know how to calculate the self-energy and its momentum dependence, then in principle
we can calculate $c$. The problem is to calculate $\sigma (x)$. 

\begin{figure}
 \begin{center}
    \includegraphics[scale=0.5,clip]{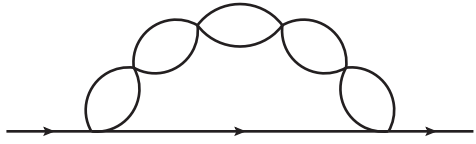}
 \end{center}
 \caption{Self-energy in the large $N$ expansion}
 \label{Fig_lec6:babble}
\end{figure}

To do so, we must use non perturbative techniques, of which there are not so many. 
The first technique that I want to discuss is the large $N$ approximation. 
The idea is to replace the field $\psi _0 (r)$, which is a complex field written as 
$\phi _1 (r) +i\phi_2 (r) =(\phi _1 ,\phi_2)$, by an $N$-component vector 
$\vec{\phi} (r)=(\phi _1 ,\phi_2 ,\cdots ,\phi _N )$.
Then we can write an action
\begin{align}
S = \int d^3 \r \left \{ \vec{\phi} (\r) \left( -\frac{\nabla ^2}{2m} \right) \vec{\phi} (\r)
 + \frac{g}{2} \left( \left( \vec{\phi} (\r) \right) ^2 \right) ^2 \right \},
\end{align}
which is invariant under $O(N)$ transformations. 
The usefulness of this strategy is that an analytic calculation is possible when $N \to \infty$. 
For the calculation of the self-energy, this amounts essentially  to resum the chain of bubbles  in Fig.~\ref{Fig_lec6:babble} (this is actually a $1/N$ correction; the leading order correction is the mean field correction that leads to a momentum independent self-energy). 
If you remember what I said in the last lecture, 
this chain of bubbles is what produces screening of the long wavelength density fluctuations. 
Because of this screening, the resulting calculation of the self-energy is infrared finite.
The calculation of the self-energy in this order can be done analytically, and yields  $c=2.3$. 

Another possible strategy is to do a ``brute force'' lattice calculation, as you do in QCD for instance. This leads to the most accurate determination of $c$. 
The results obtained almost simultaneously by two groups are 
$c = 1.32\pm 0.02$ \cite{latt2} , and $c= 1.29\pm 0.05$ \cite{latt1}.

After these results were obtained, the calculation of $c$ has become a  play ground for testing various approximation 
in field theory or many body physics (see for instance \cite{Blaizot:2008xx} for references). 

\begin{figure}
  \begin{center}
   \includegraphics[scale=1,clip]{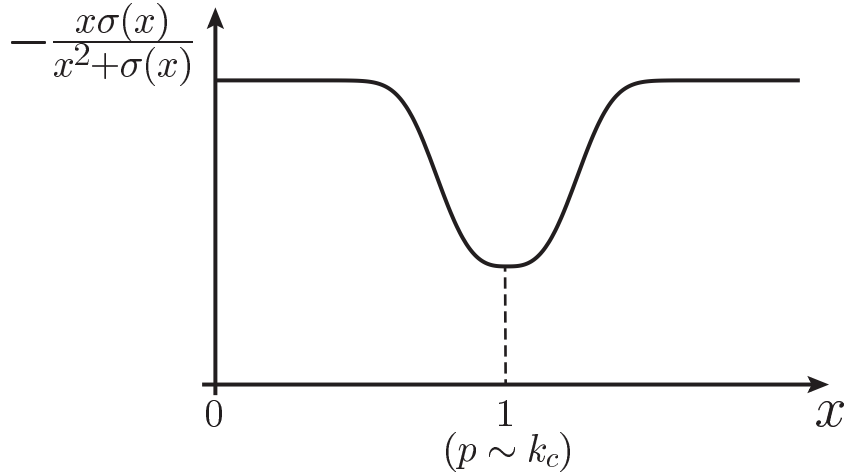}
  \end{center}
  \label{integrand}
  \caption{Schematic behavior of the integrand in Eq.~(271)}
\end{figure}

Perhaps before closing I should tell you one word about why it is hard to calculate $\sigma (x)$.
The reason can be understood from looking at the plot of the integrand in Eq.~(\ref{Eq_lec6:Dn_c3}). As can be seen on Fig.~\ref{integrand}, the integrand is peaked around $x\sim1$, that is at the momentum $p\sim \kappa_c\sim a/\lambda ^2$,
which delineates the frontier between  two very distinct regimes: 
At momenta smaller than $\kappa_c$, we have the critical regime, well described by the theory of critical phenomena which allows in particular a precise determination of the anomalous dimension $\eta$.  This is well under control. 
The other regime, that of large momenta, is also very well under control 
because in this regime  perturbation theory can be applied (perturbation theory is accurate at large momenta in a three dimensional scalar field theory).
But the quantity which we need is sensitive to  what happens at the border line between these two regimes.
What we need to get with accuracy is the precise point where this transition between the two regimes occurs. 
And this is hard.
That is why we need non-trivial non-perturbative techniques to do the explicit calculation (and why simple methods, like the large  $N$ expansion,   are off quantitatively by almost a factor two).

What was going to come after that was a discussion of this problem, from the  perspective of the non-perturbative (or ``exact'') renormalization group \cite{Blaizot:2008xx}.  But that will be for another occasion... as the time for these lectures is over.

%
%


\noindent {\bf Acknowledgements.}  I thank the Physics Department of the University of Tokyo for hospitality during the winter 2009 when these lectures where delivered. I am very grateful to Tesuo Matsui, for his invitation, and for taking the initiative of recording the lectures.  Special thanks are due to the students who produced the transcript of the recorded lectures.  I also gratefully acknowledge the contributions of  Professors  Hirotsugu
Fujii and Yusuke Kato  to the  pre-editing of
the transcribed text. Clearly, without all this  help these notes would not exist. I also gratefully acknowledge the hospitality of the Physics Department of Nanjing University, where I had a chance to give similar lectures in March 2011, and to work there on the final version of these notes.

\end{document}